\documentclass[a4paper,12pt, epsfig]{article}
\usepackage{epsfig}
\usepackage{amssymb,latexsym}
\usepackage{amsfonts}
\usepackage{amsmath}
\usepackage[colorlinks, linkcolor=blue, citecolor=blue, urlcolor=blue]{hyperref}
\usepackage{epstopdf}

\usepackage{CJKutf8}

\newskip\humongous \humongous=0pt plus 1000pt minus 1000pt

\newif\ifdtup

\catcode`\@=11

\@addtoreset{equation}{section}

\def\@normalsize{\@setsize\normalsize{15pt}\xiipt\@xiipt
\abovedisplayskip 14pt plus3pt minus3pt%
\belowdisplayskip \abovedisplayskip
\abovedisplayshortskip \z@ plus3pt%
\belowdisplayshortskip 7pt plus3.5pt minus0pt}

\def\small{\@setsize\small{13.6pt}\xipt\@xipt
\abovedisplayskip 13pt plus3pt minus3pt%
\belowdisplayskip \abovedisplayskip
\abovedisplayshortskip \z@ plus3pt%
\belowdisplayshortskip 7pt plus3.5pt minus0pt
\def\@listi{\parsep 4.5pt plus 2pt minus 1pt
      \itemsep \parsep
      \topsep 9pt plus 3pt minus 3pt}}

\relax

\catcode`@=12

\catcode`\@=11

\def\section{\@startsection{section}{1}{\z@}{3.5ex plus 1ex minus
    .2ex}{2.3ex plus .2ex}{\large\bf}}

\def\thesection{\arabic{section}}
\def\thesubsection{\arabic{section}.\arabic{subsection}}

\def\appendix{\setcounter{section}{0}
  \def\thesection{Appendix \Alph{section}}
  \def\thesubsection{\Alph{section}.\arabic{subsection}}
  \def\theequation{\Alph{section}.\arabic{equation}}}

\def\SymBoxes#1#2#3#4{\newdimen\un@t \un@t#3%
\raisebox{#1}{\rule{#2\un@t}{#4}\hskip-#2\un@t
\@tempdimb\un@t \advance\@tempdimb by-#4\@tempcntb#2\relax%
\@whilenum{\@tempcntb>0}\do{
\rule{#4}{\un@t}\hskip\@tempdimb \advance\@tempcntb by\m@ne}%
\hskip-#2\un@t \rule[\un@t]{#2\un@t}{#4}%
\rule[\un@t]{#4}{#4}\hskip-#4
\rule{#4}{\un@t}}\hskip-#4}                

\begin{document}


\newcommand{\dd}{\textrm{d}}

\newcommand{\beq}{\begin{equation}}
\newcommand{\eeq}{\end{equation}}
\newcommand{\bea}{\begin{eqnarray}}
\newcommand{\eea}{\end{eqnarray}}
\newcommand{\beas}{\begin{eqnarray*}}
\newcommand{\eeas}{\end{eqnarray*}}
\newcommand{\defi}{\stackrel{\rm def}{=}}
\newcommand{\non}{\nonumber}
\newcommand{\bquo}{\begin{quote}}
\newcommand{\enqu}{\end{quote}}
\newcommand{\tc}[1]{\textcolor{blue}{#1}}
\renewcommand{\(}{\begin{equation}}
\renewcommand{\)}{\end{equation}}
\def\de{\partial}
\def\Om{\ensuremath{\Omega}}
\def\Tr{ \hbox{\rm Tr}}
\def\rc{ \hbox{$r_{\rm c}$}}
\def\H{ \hbox{\rm H}}
\def\HE{ \hbox{$\rm H^{even}$}}
\def\HO{ \hbox{$\rm H^{odd}$}}
\def\HEO{ \hbox{$\rm H^{even/odd}$}}
\def\HOE{ \hbox{$\rm H^{odd/even}$}}
\def\HHO{ \hbox{$\rm H_H^{odd}$}}
\def\HHEO{ \hbox{$\rm H_H^{even/odd}$}}
\def\HHOE{ \hbox{$\rm H_H^{odd/even}$}}
\def\K{ \hbox{\rm K}}
\def\Im{ \hbox{\rm Im}}
\def\Ker{ \hbox{\rm Ker}}
\def\const{\hbox {\rm const.}}
\def\o{\over}
\def\im{\hbox{\rm Im}}
\def\re{\hbox{\rm Re}}
\def\bra{\langle}\def\ket{\rangle}
\def\Arg{\hbox {\rm Arg}}
\def\exo{\hbox {\rm exp}}
\def\diag{\hbox{\rm diag}}
\def\longvert{{\rule[-2mm]{0.1mm}{7mm}}\,}
\def\a{\alpha}
\def\b{\beta}
\def\e{\epsilon}
\def\l{\lambda}
\def\ol{{\overline{\lambda}}}
\def\ochi{{\overline{\chi}}}
\def\th{\theta}
\def\s{\sigma}
\def\oth{\overline{\theta}}
\def\ad{{\dot{\alpha}}}
\def\bd{{\dot{\beta}}}
\def\oD{\overline{D}}
\def\opsi{\overline{\psi}}
\def\dag{{}^{\dagger}}
\def\tq{{\widetilde q}}
\def\L{{\mathcal{L}}}
\def\p{{}^{\prime}}
\def\W{W}
\def\N{{\cal N}}
\def\hsp{,\hspace{.7cm}}
\def\hspp{,\hspace{.5cm}}
\def\bo{\ensuremath{\hat{b}_1}}
\def\bfo{\ensuremath{\hat{b}_4}}
\def\co{\ensuremath{\hat{c}_1}}
\def\cfo{\ensuremath{\hat{c}_4}}
\def\th#1#2{\ensuremath{\theta_{#1#2}}}
\def\c#1#2{\hbox{\rm cos}(\th#1#2)}
\def\s#1#2{\hbox{\rm sin}(\th#1#2)}
\def\cp#1#2#3{\hbox{\rm cos}^#1(\th#2#3)}
\def\sp#1#2#3{\hbox{\rm sin}^#1(\th#2#3)}
\def\ctp#1#2#3{\hbox{\rm cot}^#1(\th#2#3)}
\def\cpp#1#2#3#4{\hbox{\rm cos}^#1(#2\th#3#4)}
\def\spp#1#2#3#4{\hbox{\rm sin}^#1(#2\th#3#4)}
\def\t#1#2{\hbox{\rm tan}(\th#1#2)}
\def\tp#1#2#3{\hbox{\rm tan}^#1(\th#2#3)}
\def\m#1#2{\ensuremath{\Delta M_{#1#2}^2}}
\def\mn#1#2{\ensuremath{|\Delta M_{#1#2}^2}|}
\def\u#1#2{\ensuremath{{}^{2#1#2}\mathrm{U}}}
\def\pu#1#2{\ensuremath{{}^{2#1#2}\mathrm{Pu}}}
\def\meff{\ensuremath{\Delta M^2_{\rm{eff}}}}
\def\an{\ensuremath{\alpha_n}}
\newcommand{\Z}{\ensuremath{\mathbb Z}}
\newcommand{\R}{\ensuremath{\mathbb R}}
\newcommand{\rp}{\ensuremath{\mathbb {RP}}}
\newcommand{\vac}{\ensuremath{|0\rangle}}
\newcommand{\vact}{\ensuremath{|00\rangle}                    }
\newcommand{\oc}{\ensuremath{\overline{c}}}
\renewcommand{\cos}{\textrm{cos}}
\renewcommand{\sin}{\textrm{sin}}
\renewcommand{\cot}{\textrm{cot}}

\newcommand{\Vol}{\textrm{Vol}}

\newcommand{\half}{\frac{1}{2}}

\def\changed#1{{\bf #1}}

\begin{titlepage}

\def\thefootnote{\fnsymbol{footnote}}

\begin{center}
{\large {\bf
Advantages of Multiple Detectors for the Neutrino Mass Hierarchy Determination at Reactor Experiments
  } }

\bigskip

\bigskip

{\large \noindent Emilio
Ciuffoli$^{1}$\footnote{ciuffoli@ihep.ac.cn}, Jarah
Evslin$^{1,2}$\footnote{\texttt{jarah@ihep.ac.cn}}, Zhimin Wang$^{3}$\footnote{\texttt{wangzhm@ihep.ac.cn}},\\ Changgen Yang$^{3}$\footnote{\texttt{yangcg@ihep.ac.cn}}, Xinmin Zhang$^{2,
1}$\footnote{\texttt{xmzhang@ihep.ac.cn}} and Weili Zhong$^{3}$\footnote{\texttt{zhongwl@ihep.ac.cn}} }
\end{center}

\renewcommand{\thefootnote}{\arabic{footnote}}

\vskip.7cm

\begin{center}
\vspace{0em} {\em  { 1) TPCSF, IHEP, Chinese Acad. of Sciences\\
2) Theoretical physics division, IHEP, Chinese Acad. of Sciences\\
3)  Particle astrophysics division, IHEP, Chinese Acad. of Sciences\\
YuQuan Lu 19(B), Beijing 100049, China}}


\vskip .4cm


\end{center}

\vspace{1.3cm}

\noindent
\begin{center} {\bf Abstract} \end{center}

\noindent
We study the advantages of a second identical detector at a medium baseline reactor neutrino experiment.  A major obstruction to the determination of the neutrino mass hierarchy  is the detector's unknown nonlinear energy response, which even under optimistic assumptions reduces the confidence in a hierarchy determination by about 1$\sigma$ at a single detector experiment.  Various energy response models are considered at one and two detector experiments with the same total target mass.  A second detector at a sufficiently different baseline eliminates this $1\sigma$ reduction.   Considering the unknown energy response, we find the confidence in a hierarchy determination at various candidate detector locations for JUNO and RENO 50.  The best site for JUNO's near detector is under ZiLuoShan, 17 km and 66 km from the Yangjiang and Taishan reactor complexes respectively.  We briefly describe other advantages, including a more precise determination of $\theta_{12}$ and the possibility of a DAE$\delta$ALUS inspired program to measure the CP-violating phase $\delta$ using a single pion source about 10 km from one detector and 20 km from the other.  Two identical detectors provide a better energy resolution than a single detector, further increasing the confidence in a hierarchy determination.


\vfill

\begin{flushleft}
{\today}
\end{flushleft}
\end{titlepage}



\setcounter{footnote}{0}

\section{Introduction}
In 2002 Petcov and Piai \cite{petcovidea} proposed that the shape of the oscillated $\overline{\nu}_e$ spectrum observed at a medium baseline from a nuclear power plant can in principle be used to reconstruct the $\nu_e$ survival probability whose fine structure in turn can determine the neutrino mass hierarchy.  This fine structure consists of two beating modes of atmospheric oscillations with amplitude $\spp2213$.  Last year's discovery that $\theta_{13}$ is large \cite{dayabay,reno} implies that these oscillations are large enough to be observed.  As a result at least two experiments, JUNO \cite{yifangmarzo} and RENO 50 \cite{reno50}, have been proposed to measure the reactor neutrino spectrum using inverse $\beta$ decay and so determine the hierarchy.

These experiments will have to face some challenges which have never before been faced.  For example, as the largest and most transparent liquid scintillator detectors in history, the optical properties of the scintillator will need to be understood as never before.  Repeated Rayleigh scattering means that photons can scatter multiple times before being detected, smearing the pulse shapes which are usually used to discriminate $\alpha$ particle backgrounds.  This background rejection will be all the more difficult because the liquid scintillators will not contain gadolinium and neither experiment will be very deep.  

\subsection{Nonlinear Energy Response}

The present paper concerns a different challenge.  The fine structure of the spectra in the case of the two hierarchies differ in the relative energies of the oscillation peaks at high (about 6 MeV) and low (about 3 MeV) energies.  This difference, which must be observed to determine the hierarchy, is only about 1\%.  This means that the statistical fluctuations in the number of photoelectrons observed, which characterize the detector's energy resolution, must be at least a factor of two smaller than has ever been achieved.  Furthermore the systematic shifts in the relative energy response of the detector, which we call the nonlinear energy response, must be understood to an unprecedented level.

The nonlinear energy response arises from a variety of sources.  For example the electronics determines the sensitivity to multiple photoelectrons in close proximity.  This provides a nonlinear energy dependence.  The optical properties of the scintillator determine the number of photons that eventually arrive at the photomultiplier tubes from various locations in the detector, meaning that the nonlinear response also is position dependent.  Furthermore Cherenkov radiation by the positrons created in the inverse $\beta$ decay provides an energy dependence and a position dependence.   Liquid organic scintillator detectors in reactor neutrino experiments have historically had to deal with time dependence in the optical properties of the scintillators, which degrades the energy resolution but also further complicates this nonlinear response.

The light yield of the scintillator, which determines the energy response, depends on the particle being detected.  As the particles of interest are positrons, one would ideally calibrate this nonlinear response using monochromatic positron sources.  The trouble is that lepton number conservation implies that radioactive nuclei emit positrons together with neutrinos, resulting in a continuous positron spectrum.  Monochromatic positron sources are thus difficult to come by, and so calibration will be done largely using other particles.  While such a calibration can in principle determine the position dependence of the energy response of the detector with reasonable precision, the overall nonlinear shape of the response to positrons is not precisely determined by the response to any other particle.  Therefore in this note we will not consider the position dependence of the nonlinearity, assuming that it has been determined perfectly via calibration.

The best yet determination of this nonlinear energy response has been made by the KamLAND experiment.  The precision is about 2\%.  The JUNO experiment hopes to double this, determining the energy response at the 1\% level.

This leads one to ask, just how well does the nonlinearity need to be understood to determine the hierarchy?   Similarly one can ask, for a certain level of nonlinearity, just how strongly is the confidence in a hierarchy determination degraded?  In this note we will answer these questions for various models of the nonlinear response.  In particular we will find that, just as the use of identical detectors at distinct baselines appreciably reduces systematic errors due to flux uncertainties in Daya Bay \cite{dayabay} and RENO \cite{reno}, identical detectors at sufficiently distinct baselines also appreciably reduce correlated systematic errors due to energy uncertainties at JUNO and RENO 50 as had been suggested in Refs.~\cite{noisim,noi2rivel}.  Evidence for this effect has recently been obtained in Ref.~\cite{snowmass} which used only the ratio of the fluxes at the two detectors.  Our work instead analyses the full flux spectrum obtained at both detectors.

\subsection{Multiple Detectors}

While the construction of a second identical detector is likely to be beyond the funding limits of such experiments right now, the second detector can be constructed later.  This of course will make the second detector less identical, one will need to make the scintillator in the two detectors as identical as possible.  However, unlike $\theta_{13}$ experiments where it is necessary that the two detectors run simultaneously because of the time-dependent flux coming from the reactors, hierarchy determinations are independent of the flux normalization but only depend on the relatively time-independent shape of the fine structure of their energy spectra.  Therefore it is less essential that the near and far detectors run at the same time, the only important point is that they have the same energy response.

There are several other motivations for considering multiple identical detectors.  First of all, the uncertainty in the background flux is an obstruction to the secondary goals of these experiments of measuring $\theta_{12}$ and geoneutrinos.  A second detector breaks the degeneracy between the strengths of these signals and the backgrounds, allowing them to be more accurately measured.  

Multiple detectors are also likely to each be smaller.  In our study we have compared a single detector to two half-size detectors, however in practice to balance the cost the pairs of detectors may be even smaller.  Smaller detectors have a number of advantages.  First of all they are easier to build.  In particular the construction of the enormous spherical detector for JUNO promises difficult engineering.  More importantly, light does not need to travel so far to reach the photomultiplier tubes.  This means that more photons will be received from each event, improving the energy resolution.  Even a modest improvement in the energy resolution yields a large improvement in the confidence of the hierarchy determination.  Also the photons will arrive in a shorter timescale, facilitating the identification of $\alpha$ particle backgrounds via the 2\% of photons emitted in the scintillator's long decay modes \cite{alpha}. 

In the longer term, one is interested in the neutrino mass hierarchy to a large extent because a determination of the hierarchy allows the breaking of a degeneracy at NO$\nu$A and T2K which could then allow a determination of sin$(\delta)$ where $\delta$ is the CP-violating leptonic phase.  However with 2 detectors $\delta$ could be determined directly using a strategy similar to that of LBNE's  DAE$\delta$ALUS program \cite{dead1}.  This program requires multiple cyclotrons which serve as high intensity sources of stationary pions at various baselines which decay to $\overline{\nu}_\mu$, which in turn oscillate to $\overline{\nu}_e$ which is observed at the detector.   The uncertainty in the relative intensity of the pion sources translates to an uncertainty in $\delta$.  

On the other hand, if JUNO or RENO 50 has multiple detectors then only a {\it{single}} pion source is needed, which then will be at a different baseline from the two detectors.  Ref.~\cite{dead2} suggests that a single such cyclotron facility would cost between 25 and 100 million dollars.  Thus with a single cyclotron facility, both the experiment is cheaper and also the uncertainty from the relative calibration of the sources is removed, although of course it is important to calibrate the relative cross sections and efficiencies of the detectors.  This can in principle yield one of the most precise determinations of $\delta$ of any proposed experiment.  Another advantage of JUNO or RENO 50 detectors for such a program is that, like LENA \cite{lena}, they are liquid scintillator detectors and so efficiently detect $\overline{\nu}_e$ via inverse $\beta$ decay.  The antineutrino energy from the pion decay is above the reactor $\overline{\nu}_e$ energy range, so the hierarchy and CP-violation studies can occur simultaneously.

In Sec.~\ref{metsez} we will provide a general description of our assumptions and methods.  Then in Sec.~\ref{gensez} we will provide the results for a general study of the effect of the nonlinear energy response upon the confidence of the hierarchy determination from abstract detectors and pairs of detectors at various distances from a single reactor neutrino source.  Finally in Sec.~\ref{junosez} we will consider the relevant sites for the JUNO and RENO 50 experiments.

\section{Methods} \label{metsez}
JUNO and RENO 50 both will detect reactor $\overline{\nu}_e$ using inverse $\beta$ decay (IBD) in a liquid scintillator.  The scintillator used by JUNO will be based on LAB, which is 12-13\% hydrogen by mass.  We will assume that RENO 50 has the same IBD cross section per target mass.

The total flux of $\overline{\nu}_e$'s detected is normalized to be 50,000 per 20 kton of target mass per 6 years at a baseline of 58 km from 17.4 GW of thermal capacity of reactors.  This includes oscillations,  the IBD cross section and the reactor flux normalization.  It is chosen for easy comparison with the literature, but also is in reasonable agreement with both a scaled up value from Daya Bay and also with a first principles calculation \cite{noiteor}.    We also assume a fractional energy resolution of $3\%/\sqrt{\tilde{E}({\rm{MeV}})}$ where $\tilde{E}=E-0.8$ MeV is the prompt energy corresponding to a neutrino of energy $E$.

We use the value of $\m21$ from Ref.~\cite{pdb}, $\spp2212$ from Ref.~\cite{gando}, $\mn32$ determined by combining $\nu$ and $\overline{\nu}$ mass differences from Ref.~\cite{minosneut2012}  and $\spp2213$ from \cite{neut2012} 
\bea
&&\m21=7.5\times 10^{-5}{\mathrm{\ eV^2}}\hsp
\spp2212=0.857\nonumber\\
&&\mn32=2.41\times 10^{-3}{\mathrm{\ eV^2}}\hsp
\spp2213=0.089.\nonumber
\eea
We fix all of the parameters except for $\mn32$, which we fit as described below.

For each configuration we determine the statistic
\beq
\Delta\chi^2=\chi^2_I-\chi^2_N.
\eeq
Here $\chi^2_N$ ($\chi^2_I$) is the minimal $\chi^2$ value obtained by fitting a given dataset to the theoretical spectrum corresponding the normal (inverted) hierarchy with a penalty term corresponding to the pull parameter treatment for the nuisance parameters.  In particularly the pull parameters are minimized separately for the two choices of hierarchies.  

We have performed this procedure on both theoretical spectra and on simulated data, finding compatible results.  However as our statistical sample is small, in this note we will only report the fits to theoretical data corresponding to the Asimov data set.  In Ref.~\cite{stat} it was shown that, in such experiments, the Asimov $\Delta\chi^2$ agrees with the mean simulated $\Delta\chi^2$ to within 5 to 10 percent.  Furthermore formulae were provided relating $\Delta\chi^2$ to the confidence in the determination of the experiment corresponding to each quantile of statistical fluctuations with any given Bayesian prior\footnote{An alternate frequentist approach to this confidence was introduced in Ref.~\cite{xinoct,snowmass} reflecting the chance that an experiment yields the correct hierarchy.  Our Bayesian approach answers a different question, it provides the probability that a fit to the correct hierarchy yields the lowest $\chi^2$ given that the measured $|\Delta\chi^2|$ assumes its median value or more generally lies in any given quantile.}.  Until recently there has been no experimental preference for either hierarchy, suggesting a symmetric 50\% prior for each hierarchy.  However the most recent $\spp2213$-$\delta$ fit by the T2K collaboration \cite{T2K} is  incompatible with Daya Bay's result for $\theta_{13}$ in the case of the inverted hierarchy at a 2$\sigma$ level, and even $2\sigma$ compatibility is only obtained with maximal CP violation $\delta\sim -\pi/2$.  On the other hand $\delta\sim -\pi/2$ allows T2K and Daya Bay to be compatible at the 1$\sigma$ level in the case of the normal hierarchy.  This suggests that a roughly $1\sigma$ prior in favor of the normal hierarchy may be reasonable.  Nonetheless, in our studies we have used a symmetric $50\%$ prior.

The determination of the hierarchy is difficult because of the degeneracy between the hierarchy and $\mn32$, therefore in an evaluation of the sensitivity of an experiment to the hierarchy it is essential that $\mn32$ not be fixed.  As a result our code always chooses $\mn32$ so as to minimize $\chi^2$ for each hierarchy separately.  We do not introduce a penalty term in $\chi^2$ for $\mn32$, we simply choose the value which minimizes each $\chi^2$.  We have checked that fitting the other neutrino mass matrix elements and allowing them to vary in simulations only slightly affects the confidence in a hierarchy determination.

To elucidate the effect of the nonlinearity, we perform studies with and without nonlinearity.  
In each case we fit the nonlinearity using a 3-parameter model
\beq
E_{\rm{observed}}=a+(1+b)E_{\rm{true}}+cE^2_{\rm{true}}. \label{obs}
\eeq
$E_{\rm{true}}$ is the true energy while $E_{\rm{observed}}$ is the energy that the experimentalist concludes that the $\overline{\nu}$ must have had based on the number of photoelectrons observed, the results of his calibration campaign and the results of Monte Carlo simulations.  We do not assume that $E_{\rm{observed}}$ is proportional to the number of photoelectrons observed, on the contrary not only will this dependence be nonlinear after the completion of the calibration campaign but also it will depend upon the reconstructed location of the event.  Here $a$, $b$ and $c$ are pull parameters for which quadratic penalty terms are added in the definition of $\chi^2$ corresponding to their assumed standard errors.  These errors reflect the degree to which the experimentalist is confident in his calibrations.   In the results that we will present below we have taken the 1$\sigma$ uncertainty on these three parameters to be $2\times 10^{-2}$ MeV, $2\times 10^{-2}$ and $2\times 10^{-3}$ MeV${}^{-1}$ respectively.  As a consistency check on our results, we have found that when analyzing theoretical spectra that were generated with a perfect energy response, the best fit values of $a$, $b$ and $c$ are generally smaller than $10^{-5}$.

This fitting of the nonlinearity reflects the fact that the energy response of the detector affects the locations of known structure in the reactor neutrino spectrum, and that therefore this known structure can be used to calibrate the detector by choosing $a$, $b$ and $c$ so as to separately minimize  $\chi^2_N$ and $\chi^2_I$.    Realistically, the observed spectrum at JUNO and RENO 50 will include significant backgrounds, as have recently been estimated in Ref.~\cite{snowmass}.  The large uncertainties in these backgrounds lead to an uncertainty in the expected flux which can be degenerate with the effects of the nonlinear response.  As a result, backgrounds make this kind of calibration less effective.  In the current study we ignore backgrounds, and so our analysis of the consequences of the nonlinear energy response on the confidence in the hierarchy determination will to some extent be overoptimistic.

As the true energy response model is not known to the experimenter, we fit the unknown energy response using a fixed parametrization (\ref{obs}) which in two of the three cases considered below does not contain the model used to generate the spectrum.  Ref.~\cite{yifangmarzo} considered a form of the nonlinearity which is nearly constant in the critical regime between 3\% and 6\%.  The unknown energy response in this model is much milder than the target 1\% expected at JUNO and therefore yields a very optimistic prediction for the effects of nonlinearity.

The energy response of the detector is the number of photoelectrons that are observed as a function of the energy of the event and its location.  It can be determined by a simulation to within some precision.  But this precision is far short of the requirements to determine the hierarchy.  This is why such simulations need to be supplemented with intensive calibration campaigns, which determine $E_{\rm{observed}}$ in Eq.~(\ref{obs}).  The unknown nonlinearity is the difference between $E_{\rm{true}}$ and $E_{\rm{observed}}$.  It cannot be determined by any Monte Carlo or calibration, as these are already considered in $E_{\rm{observed}}$.  The result is that the form of the unknown nonlinear detector response is totally unknown.

So then what model of nonlinearity do we use?   The space of possible functions $E_{\rm{observed}}(E_{\rm{true}})$ is infinite-dimensional, it would be futile and useless to try them all.  Given any model of nonlinearity, the space of small perturbations of the nonlinearity generates a vector space.  A basis can be chosen for this vector space such that all of the directions except for one, $v$, leave $\Delta\chi^2$ fixed, although of course they will not leave the individual $\chi^2$ values fixed which is what allows the pull  parameters to be determined by $\chi^2$ fits.   The vector $v$ is important because a general small perturbation in the nonlinearity, corresponding to a vector $w$, will affect $\Delta\chi^2$ by a quantity proportional to $v\cdot w$.  Therefore instead of considering an infinite-dimensional space of perturbations, it suffices to identify $v$ and study its effects on $\Delta\chi^2$, and then any other model of nonlinearity will affect $\Delta\chi^2$ in proportion to its similarity to $v$.  Of course, since the effect on $\chi^2$ itself is model-dependent, other nonlinearities may be easier or harder to calibrate depending on their effect on known features of the spectrum, such as the 1.8 MeV minimum energy for inverse $\beta$ decay.

Which is the direction, $v$, in deformation space which varies $\Delta\chi^2$?  It is the deformation which interpolates between the spectra of the normal and inverted hierarchies.  As the distinction between the normal and inverted hierarchies is the energy difference between the high and low energy peaks \cite{noiteor}, this means that the most dangerous nonlinearity is one which contracts or expands the spectrum in the critical region between about 3 MeV and 6 MeV.  In Ref.~\cite{xinaug} an approximate form of this most dangerous nonlinearity was found, which we will call the worst case model.   This model is defined by
\beq
E_{\rm{observed}}=\frac{2\mn32{\rm{(eV^2)}}+4\cp212\m21{\rm{(eV^2)}}-\frac{\phi}{1.27}\frac{E_{\rm{true}}{\rm{(MeV)}}}{L{\rm(m)}}}{2\mn32{\rm{(eV^2)}}+\frac{\phi}{1.27}\frac{E_{\rm{true}}{\rm{(MeV)}}}{L{\rm(m)}}}E_{\rm{true}} \label{modellox}
\eeq
where
\beq
{\rm{sin}}(\phi)=\frac{\cp212{\rm{sin}}\left(\frac{2.54\m21{\rm{(eV^2)}} L{\rm{(m)}}}{E_{\rm{true}}{\rm{(MeV)}}}\right)}{\sqrt{1-4\sp212\cp212{\rm{sin}}^2\left(\frac{2.54\m21{\rm{(eV^2)}} L{\rm{(m)}}}{E_{\rm{true}}{\rm{(MeV)}}}\right)}} .
\eeq
Notice that $E_{\rm{observed}}$ is always less than $E_{\rm{true}}$ with a maximum fractional difference of 3\% near 3 MeV which is reduced to 1.5\% by 6 MeV.  As a result the best fit $\mn31$, which is determined from the energy near 3 MeV \cite{noiteor}, is reduced by about 3\%.  On the other hand, the spectrum near 6 MeV determines the effective mass \cite{parke2005}
\beq
\Delta M^2_{\rm{eff}}=\mn31\mp\sp212\m21=\cp212\mn31+\sp212\mn32. \label{meff}
\eeq
Therefore the best fit $\Delta M^2_{\rm{eff}}$ is reduced by about 1.5\%.  Medium baseline reactor neutrino experiments determine the hierarchy by using the fact that $\mn31-\Delta M^2_{\rm{eff}}$ is positive (negative) if and only if the hierarchy is normal (inverted).  The absolute value of this difference is $\sp212\m21$ which is about 1\% of $\mn31$.  The model (\ref{modellox}) reduces $\mn31-\Delta M^2_{\rm{eff}}$ by about 1.5\% of $\mn31$, changing the best fit hierarchy from normal to inverted.

Such an energy response will not only shift the fine structure of atmospheric oscillations, but will shift all of the expected features in the observed spectrum.  This can result in a large $\chi^2$ value, which allows the experimenter to calibrate to some extent for the nonlinearity even without adding additional sources to the detector.   There is no form of the nonlinearity capable of transforming the spectrum of one hierarchy into the other at all baselines.  The form~(\ref{modellox}) has this property at a particular baseline $L$, but at other baselines the transformation is imperfect, which is the reason that a second detector breaks the degeneracy between the energy response and the hierarchy.  $L$ is a parameter of the nonlinearity model which needs to be chosen in the simulation.  We will always choose $L$=55 km in what follows.

In our general study on nonlinearity in Sec.~\ref{gensez} we will consider 3 models of the unknown energy response.  First we will consider a quadratic shift in the energy
\beq
\tilde{E}_{\rm{observed}}=\tilde{E}_{\rm{true}}-\alpha \tilde{E}_{\rm{true}}^2 \label{quad}
\eeq
where $\alpha=0.0015\ {\rm{MeV}}^{-1}$ and again $\tilde{E}=E-0.8$ MeV is the prompt energy.  We will refer to this model as the quadratic model.  As this is an example of an energy response described by Eq.~(\ref{obs}), $\chi^2$ will be minimized when it is eliminated entirely.  We will see that therefore our best $\chi^2$ fit yields an observed spectrum and so a $\Delta\chi^2$ which is virtually identical to the case in which the energy response of the detector is known perfectly.  

Next we will consider a model similar to that of Ref.~\cite{yifangmarzo}
\beq
\tilde{E}_{\rm{observed}}=\left(\frac{\alpha+\delta  \tilde{E}_{\rm{true}}}{1+\beta \rm{e}^{-\gamma \tilde{E}_{\rm{true}}}}+(1-\alpha)\right) \tilde{E}_{\rm{true}}
\eeq
with parameters
\beq
\alpha=1.1057\hsp \beta=0.23\hsp \gamma=2.35\ {\rm{MeV}}^{-1}\hsp \delta=10^{-4}\ {\rm{MeV}}^{-1}.  \label{exp}
\eeq
We will call this the exponential model.  The corresponding unknown nonlinear response varies by only 0.13\% between 3 MeV and 6 MeV, therefore this model is quite optimistic.  Its main impact is below 3 MeV, where atmospheric oscillations in the spectrum are obscured by the finite energy resolution.  

Finally we will consider the worst case model with
\beq
\Delta E=E_{\rm{observed}}-E_{\rm{true}}
\eeq
scaled down by a factor of 3 so that
\beq
E_{\rm{observed}}=\frac{2\mn32{\rm{(eV^2)}}+\frac{4}{3}\cp212\m21{\rm{(eV^2)}}+\frac{\phi}{3.8}\frac{E_{\rm{true}}{\rm{(MeV)}}}{L{\rm(m)}}}{2\mn32{\rm{(eV^2)}}+\frac{\phi}{1.27}\frac{E_{\rm{true}}{\rm{(MeV)}}}{L{\rm(m)}}}E_{\rm{true}}. \label{terzo}
\eeq
We will call this the one third worst model.  The relative difference in the energy response between 3 MeV and 6 MeV is about one half of a percent, therefore the normalization of the nonlinear response in this model is about twice as good as JUNO's target 1\%.  This means that the results of our simulations employing the one third worst model are somewhat optimistic.


\section{General Study of the Nonlinear Energy Response} \label{gensez}

We have performed a systematic study of the effect of the nonlinearity on $\Delta\chi^2$.  We assumed that all of the reactor neutrinos are emitted by a single 36 GW thermal capacity reactor neutrino point source.  Then we considered configurations with one detector at various baselines with 240 kton years of exposure.  We also considered two detectors each with 120 kton years of exposure, one at 55 km and the other at various baselines.  We considered the generated theoretical spectra and also ran simulations using the three nonlinearity models above.  The statistics of the simulations were not sufficient to draw conclusions, but merely to test the consistency of our theoretical results.  We then performed similar studies with one half and one quarter of the events in the first study.

\begin{figure} 
\begin{center}
\includegraphics[width=6.5in,height=3in]{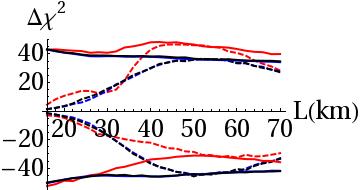}
\includegraphics[width=6.5in,height=3in]{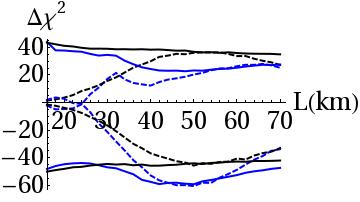}
\caption{In the upper panel $\Delta\chi^2$ is evaluated for the Asimov data assuming a perfectly understood energy response (black), a quadratic nonlinearity model  (blue) and an exponential model (red) and 240 kton years of target exposure.  The dashed curves correspond to single detectors at various baselines.  The solid curves correspond to one half size detector at 55 km and another at the given baseline.  Curves on the upper (lower) half of each panel correspond to theoretical spectra generated using the normal (inverted) hierarchy.  The quadratic nonlinearity yields the same confidence as no nonlinearity at all as it is fit perfectly by the assumed fitting function.  The exponential nonlinearity appreciably reduces the confidence in a hierarchy determination if the true hierarchy is normal in the case of 1 detector or 2 detectors at similar baselines.  However this effect is canceled entirely if 1 detector is placed at 55 km and another beneath about 30 km.  The lower panel is the same except the no nonunderstood nonlinearity case (black) is compared with the third worst model (blue).}
\label{sisfig40}
\end{center}
\end{figure}

\begin{figure} 
\begin{center}
\includegraphics[width=6.5in,height=3in]{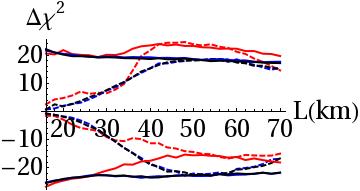}
\includegraphics[width=6.5in,height=3in]{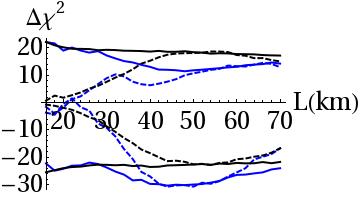}
\caption{As in Fig.~\ref{sisfig40} but now the total exposure is 120 kton years.}
\label{sisfig20}
\end{center}
\end{figure}

\begin{figure} 
\begin{center}
\includegraphics[width=6.5in,height=3in]{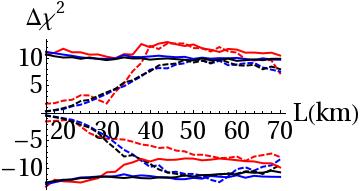}
\includegraphics[width=6.5in,height=3in]{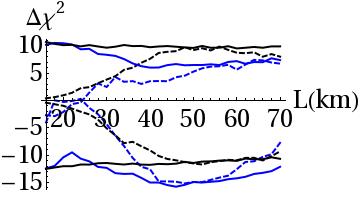}
\caption{As in Fig.~\ref{sisfig40} but now the total exposure is 60 kton years.}
\label{sisfig10}
\end{center}
\end{figure}

We chose $\mn32$ and also the pull parameters $a$, $b$ and $c$ so as to separately minimize $\chi^2_N$ and $\chi^2_I$ for fits to spectra obtained using the corresponding hierarchies.  Subtracting these yields $\Delta\chi^2$ which is plotted in Figs.~\ref{sisfig40}, \ref{sisfig20} and \ref{sisfig10} for all of the nonlinearity models with a total exposure of 240 kton years, 120 kton years and 60 kton years respectively.

The most obvious feature of these figures is that in the case of one detector the confidence $\Delta\chi^2$ goes to zero on the left whereas in the case of 2 detectors it does not.  The reason that the single detector confidences tend to zero is that for baselines shorter than 30 km \cite{noiteor} the reactor spectrum is only sensitive to $\meff$ and not to the hierarchy.  The hierarchy can only be determined by comparing $\meff$ with a different combination of the atmospheric splittings.  In the two detector case on the other hand, one detector is always at 55 km.  While the near detector measures $\meff$, the far detector measures other splittings, such as $\mn31$, whose comparison with $\meff$ yields the hierarchy whatever the location of the near detector.  This is why the two detector confidence does not go to zero at the left end of the figures.

One easy consistency check of the figures is that the dashed curves, corresponding to 1-detector configurations, and the solid curves, corresponding to 2-detector configurations, agree at 55 km.  This is because the far detector is always at 55 km and the total target volume is held fixed.  Thus 2 detectors both at 55 km are equivalent to a single detector at 55 km.  Of course in practice these situations are somewhat different because in the case of 2 detectors the photons must travel less distance to the photomultiplier tubes so more will arrive and the energy resolution will be improved and also in the 2 detector case there will be more cosmogenic muons detected, although the total number of cosmogenic muons that pass through any given volume will be the same.  However at the level of our calculations 1 detector at 55 km is indistinguishable from 2 half-sized detectors at 55 km and so the dashed and solid curves agree.


As can be seen, the quadratic nonlinearity model is indistinguishable from no nonlinearity at all.  This is because the nonlinearity model (\ref{quad}) is a special case of our fitting function (\ref{obs}).  On the other hand the exponential nonlinearity has a noticeable effect on the confidence of the hierarchy determination, at 55 km and 240 kton years reducing $\Delta\chi^2$ from -44 to -30 in the case in which the true hierarchy is inverted.  Using the statistical interpretation of Ref.~\cite{stat} this reflects a drop from over 6$\sigma$ of confidence in the median experiment to 5$\sigma$.  On the other hand, if one detector is placed at 55 km and another at 30 km then in the case of exponential nonlinearity one still obtains a $\Delta\chi^2$ value of -45.

One may be tempted to claim that the unknown energy response is not so bad, since it reduces the confidence by 1$\sigma$ in the case of the inverted hierarchy but increases the confidence by a similar amount in the case of the normal hierarchy.  The trouble is that this preference is the result of a sign choice in the model parameters (\ref{exp}).  For the other choice, there would be a 1$\sigma$ reduction in the confidence of the normal hierarchy.  As the true parameters in the unknown energy response are of course unknown, it will not be known to the experimenters just which confidence is correct.  The resulting confidence intuitively is the average of the $p$-values corresponding to the confidences with both signs, which is essentially just the confidence of the worst case.  In other words, when interpreting Figs.~\ref{sisfig40}, \ref{sisfig20} and \ref{sisfig10} the relevant information at each energy is the curve whose absolute value is smallest, reflecting the most pessimistic of the two hierarchies.  As a result, in the analysis that follows we will identify the confidence with the lowest $|\Delta\chi^2|$ statistic of the two hierarchies.

Nonetheless this 1$\sigma$ degradation of the signal due to the nonlinearity may seem small.  However the exponential nonlinearity model is extremely optimistic.  While below 2.2 MeV it represents a nonlinearity above 1\%, by 2.5 MeV that has been reduced to 0.5\% and about 3 MeV it is about than 0.13\%.  Due to the limited energy resolution, the fine structure of atmospheric oscillations are strongly suppressed below about 3 MeV.  

The third worst model, while still optimistic, seriously affects the confidence in the hierarchy determination.  Consider for example Fig.~\ref{sisfig40} reflecting 240 kton years.  In the case of a single detector the unknown energy response causes the confidence $\Delta\chi^2$ to drop from 36 to 24, losing more than 1$\sigma$.  On the other hand, a near detector at 30 km leads to a confidence of $\Delta\chi^2=39$, even better than the 1 detector case with no nonlinearity at all.

What about the case of shorter exposures?  Consider 120 kton years as shown in Fig.~\ref{sisfig20}, corresponding to 6 years of running of the current JUNO proposal, although without the interference effects between baselines which will be included in Sec.~\ref{junosez}.  Now with no nonlinearity at 55 km one may expect a confidence of $\Delta\chi^2=18$\ $(3.7\sigma)$, but the third worse model reduces this to $\Delta\chi^2=12$\ $(2.8\sigma)$, a reduction of nearly 1$\sigma$.   On the other hand two half-sized detectors, one at 55 km and one at 30 km, yield $\Delta\chi^2=18$, again as good as a single detector before the correction for the nonlinearity was applied.  

Finally in Fig.~\ref{sisfig10}, 60 kton years of running is considered, corresponding to 3 years of JUNO, 3.3 years of RENO 50 at Munmyeongsan or 6.6 years of RENO 50 at Guemsong.  Now the third worst nonlinearity model makes $\Delta\chi^2$ fall from 9.5 $(2.4\sigma)$ to 6 $(1.7\sigma)$, while a second detector at 30 km leads to a partially recovered confidence of 8.5 $(2.2\sigma)$.  A yet lower baseline for the near detector restores the confidence even more, with about 25 km necessary to fully restore the original confidence.  More generally we find that the advantage of a second detector is reduced for smaller numbers of kton years of exposure.  The reason for this is that in such cases the hierarchy determination is limited by statistics, and not by systematics such as the nonlinear energy response.  For example in the case of JUNO and 60 kton years of total exposure, the 2 detector scenario corresponds to 3 years of a 10 kton detector at 55 km and a 10 kton near detector.  Only the far detector is sensitive to $\mn31$, in fact only the low energy (3 MeV) part of its spectrum is sensitive to $\mn31$.  An accurate determination of $\mn31$ is necessary to determine the hierarchy at a reactor experiment, but 30 kton years of exposure at 55 km presents too few neutrinos in the required range to robustly measure this quantity.  For this reason, the second detector proposal is most advantageous for long running experiments, which are limited by systematic errors, like those of Fig.~\ref{sisfig40}.

\section{JUNO and RENO 50} \label{junosez}

\subsection{The Sites Considered}

We have also considered the effect of the one third worst model of nonlinearity on candidate sites for the JUNO and RENO 50 experiments.   The far detector of the JUNO experiment will be in Jiangmen county, at the DongKeng site of Ref.~\cite{noisim}.  It will use flux from two reactor complexes, Yangjiang and Taishan.  Currently no reactors are operational in either complex.  However 20.8 GW of thermal capacity of reactors is under construction and another 15 GW are planned, although it remains to be seen how the recent cancellation of a reactor fuel production plant in Jiangmen will affect these plans.  We have identified two sites for a near detector.  First, it may be built under the 350m hill LuGuJing about 40 km from each reactor complex.  At this baseline it will require about 600 meters of overhead rock to block atmospheric muons responsible for various backgrounds.  Second, it may be built under the 750 meter mountain ZiLuoShan which is 17 km from Yangjiang but 66 km from Taishan.  Therefore Taishan would provide a background.   However the proximity to Yangjiang means that it will enjoy even more reactor neutrino flux than a detector under LuGuJing.  These sites are summarized in Table~\ref{disttabdaya}.

The preferred site for the far detector of RENO 50 is deep under the 450 m hill GuemSeong.  As was observed in Ref.~\cite{noisim}, the  perpendicular location of GuemSeong with respect to the Hanbit reactor complex means that it is the same 47.4 km from all of the reactors and so there is negligible interference between baselines.   A corresponding near detector may be placed under Jangamsan which is 23 km from Hanbit.  Alternately a far detector could be placed under Munmyeongsan or Munsusan receiving flux from the Hanul complex.  While this complex currently has the same power has Hanbit, more reactors are under construction.  When these reactors are completed, in 2018 when RENO 50 is scheduled to begin taking data, the total flux will be almost 50\% greater.  Even more reactors are planned, perhaps doubling the thermal capacity of the site by 2022.  In addition these mountains are appreciably higher than GuemSeong, and so such a site would not require vertical digging.  The disadvantages are that it could not receive neutrinos from the J-PARC beam and also there would be significant interference from reactors 150 km away.  This interference is less at Munsusan, but in exchange there is interference between the various Hanul reactors at this location.  Both Hanul sites could use a near detector at a location that we have called Buncheon-ri.  These sites are summarized in Table~\ref{disttab}.

\begin{table}[position specifier]
\centering
\begin{tabular}{c|l|l|l|l|l|}
Reactor&Status&Th. Cap&DongKeng(\begin{CJK}{UTF8}{gbsn}东坑\end{CJK})&LuGuJing(\begin{CJK}{UTF8}{gbsn}碌古径\end{CJK})&ZiLuoShan(\begin{CJK}{UTF8}{gbsn}紫罗山\end{CJK})\ \ \      \\
\hline\hline
Latitude&&&$22.12^{\circ}$N&$21.90^{\circ}$N&$21.84^{\circ}$N\\
\hline
Longitude&&&$112.52^{\circ}$E&$112.59^{\circ}$E&$112.35^{\circ}$E\\
\hline
Elevation&&&450 m&350 m&750 m\\
\hline
Relief&&&450 m&350 m&750 m\\
\hline
YangJiang 1&Under Cons&2905 MW&52.75 km&40.55 km&17.43 km\\
\hline
YangJiang 2&Under Cons&2905 MW&52.84 km&40.6 km&17.52 km\\
\hline
YangJiang 3&Under Cons&2905 MW&52.42 km&40.2 km&17.09 km\\
\hline
YangJiang 4&Under Cons&2905 MW&52.51 km&40.3 km&17.18 km\\
\hline
YangJiang 5&Planned&2905 MW&52.12 km&39.9 km&16.79 km\\
\hline
YangJiang 6&Planned&2905 MW&52.21 km&40.0 km&16.88 km\\
\hline
Taishan 1&Under Cons&4590 MW&52.76 km&40.55 km&65.84 km\\
\hline
Taishan 2&Under Cons&4590 MW&52.63 km&40.4 km&65.7 km\\
\hline
Taishan 3&Planned&4590 MW&52.32 km&40.1 km&65.5 km\\
\hline
Taishan 4&Planned&4590 MW&52.20 km&39.9 km&65.4 km\\
\hline
DB-LA&Operational&17430 MW&215 km&215 km&242 km\\
\hline
\end{tabular}
\caption{JUNO sites}
\label{disttabdaya}
\end{table}

\begin{table}[position specifier]
\centering
\begin{tabular}{c|l|l|l|l|l|l|l|}
Reactor&Status&Th. Cap&GuemSeong&Jangam&Munmyeong&Munsusan&Buncheon\ \ \      \\
\hline\hline
Latitude&&&$35.05^{\circ}$N&$35.26^{\circ}$N&$36.82^{\circ}$N&$36.98^{\circ}$N&$36.93^{\circ}$N\\
\hline
Longitude&&&$126.70^{\circ}$E&$126.58^{\circ}$E&$128.92^{\circ}$E&$128.81^{\circ}$E&$129.09^{\circ}$E\\
\hline
Elevation&&&450 m&480 m&870 m&1180 m&900 m\\
\hline
Relief&&&420 m&440 m&690 m&900 m&520 m\\
\hline
Hanbit 1&Operational&2787 MW&47.4 km&22.9 km&274.6 km&277.8 km&294.2 km\\
\hline
Hanbit 2&Operational&2787 MW&47.4 km&22.9 km&274.4 km&277.6 km&293.9 km\\
\hline
Hanbit 3&Operational&2825 MW&47.4 km&22.8 km&274.1 km&277.3 km&293.7 km\\
\hline
Hanbit 4&Operational&2825 MW&47.4 km&22.8 km&273.8 km&277.1 km&293.4 km\\
\hline
Hanbit 5&Operational&2825 MW&47.5 km&22.8 km&273.6 km&276.8 km&293.2 km\\
\hline
Hanbit 6&Operational&2825 MW&47.5 km&22.8 km&273.3 km&276.6 km&292.9 km\\
\hline
Hanul 1&Operational&2785 MW&331.6 km&323.9 km&51.5 km&52.0 km&31.9 km\\
\hline
Hanul 2&Operational&2775 MW&331.6  km&323.9 km&51.5 km&52.1 km&31.9 km\\
\hline
Hanul 3&Operational&2825 MW&331.6  km&323.9 km&51.5 km&52.2 km&31.9 km\\
\hline
Hanul 4&Operational&2825 MW&331.6  km&324.0 km&51.5 km&52.3 km&31.9 km\\
\hline
Hanul 5&Operational&2815 MW&331.6  km&324.0 km&51.6 km&52.4 km&32.0 km\\
\hline
Hanul 6&Operational&2825 MW&331.6  km&324.0 km&51.6 km&52.5 km&32.0 km\\
\hline
Wolseong 1&Operational&2061 MW&262.7 km&267.0 km&133.0 km&153.2 km&139.7 km\\
\hline
Wolseong 2&Operational&2061 MW&262.6 km&266.9 km&133.1 km&153.3 km&139.8 km\\
\hline
Wolseong 3&Operational&2061 MW&262.5 km&266.8 km&133.2 km&153.4 km&140.0 km\\
\hline
Wolseong 4&Operational&2061 MW&262.3 km&266.7 km&133.3 km&153.5 km&140.1 km\\
\hline
Shin Wol. 1&Operational&2825 MW&263.1 km&267.3 km&132.2 km&152.4 km&138.9 km\\
\hline
Kori 1&Operational&1729 MW&237.8 km&246.0 km&170.2 km&189.9 km&180.0 km\\
\hline
Kori 2&Operational&1882 MW&237.9 km&246.1 km&170.2 km&189.9 km&180.0 km\\
\hline
Kori 3&Operational&2912 MW&238.2 km&246.4 km&170.3 km&189.9 km&180.0 km\\
\hline
Kori 4&Operational&2912 MW&238.4 km&246.6 km&170.3 km&190.0 km&180.1 km\\
\hline
Shin Kori 1&Operational&2825 MW&238.8 km&246.9 km&169.7 km&189.3 km&179.3 km\\
\hline
Shin Kori 2&Operational&2825 MW&238.8 km&246.9 km&169.6 km&189.2 km&179.2 km\\
\hline
Shin Kori 3&Under Cons&3983 MW&239.8 km&247.9 km&168.7 km&188.4 km&178.3 km\\
\hline
Shin Kori 4&Under Cons&3938 MW&239.9 km&247.9 km&168.6 km&188.3 km&178.1
 km\\
\hline
Shin Wol. 2&Under Cons&2825 MW&263.2 km&267.4 km&132.1 km&152.3 km&138.8 km\\
\hline
Shin Hanul 1&Under Cons&3938 MW&331.4 km&323.8 km&51.5 km&52.6 km& 31.9 km\\
\hline
Shin Hanul 2&Under Cons&3938 MW&331.4   km&323.8 km&51.5 km&52.7 km&31.9 km\\
\hline
Shin Hanul 3&Planned&3938 MW&331.4 km&323.8 km&51.5 km&52.9 km&32.0 km\\
\hline
Shin Hanul 4&Planned&3938 MW&331.3  km&323.8 km&51.5 km&52.9 km&32.0 km\\\hline
\hline
\end{tabular}
\caption{Reno 50 sites}
\label{disttab}
\end{table}

We consider single detectors of 20 kton for JUNO and 18 kton for RENO 50, as well as two detector scenarios with 10 kton and 9 kton of target mass per detector respectively.  Our results are summarized in two tables.  In Table~\ref{ristab} we present the expected $\Delta\chi^2$ assuming no nonlinearity and also the third worst model assuming that all reactors which are under construction or planned have already been built.  Note that reactor construction takes about 4 years in both China and South Korea, and so the planned detectors are not expected to be operational when these experiments begin taking data.  Therefore in Table~\ref{noplantab} we report the results using only those detectors which are operational or currently under construction, thus not including detectors which are only in their planning stages.  Due to the convergence properties of our code, the values of $\Delta\chi^2$ reported in these tables are stable to within an error of about $0.5$.

As in our general study, the fact that the model of the unknown energy response chosen here~(\ref{terzo}) increases the confidence in a determination of the inverted hierarchy and decreases that in the normal hierarchy reflects an arbitrary choice.  The sign of the systematic bias in energy will never be known.  Therefore the correct confidence in a hierarchy determination is better approximated by the entry of a given row of Table~\ref{ristab} and \ref{noplantab} which is most pessimistic for a given model.  For example, for the single detector JUNO site DongKeng, the third worst hierarchy leads to a $\Delta\chi^2=8.2$ confidence in the case of the normal hierarchy but $-21.5$ in the case of the inverted hierarchy. Since the true nonlinear response could just as well have the opposite bias, no result should be given more than $|\Delta\chi^2|=8.2$ of confidence, reflecting about $2.1\sigma$.   In particular, we claim that given an unknown nonlinear response about as large as the third worst model, JUNO will be able to determine the mass hierarchy with a confidence of about $2\sigma$ with a single 20 kton detector and 6 years of running. This is consistent with the conclusions of Ref.~\cite{snowmass}, whose statistical interpretation of $\chi^2$ agrees with ours in this regime.

\subsection{Greater Sensitivity to the Inverted Hierarchy}

Perhaps the most obvious trend in Tables~\ref{ristab} and \ref{noplantab} is that both experiments are more sensitive to the inverse hierarchy than the normal hierarchy.  The reason for this is as follows.  While the determination of $\meff$ at a reactor experiment is relatively easy, to determine the hierarchy, one needs to observe the peaks in the spectrum corresponding to neutrinos that have oscillated at least 13 times \cite{noiteor}.  The more times they have oscillated, the stronger the hierarchy dependence and so the stronger the signal.  For example, the energy of the $\mn31/(2\m21)$th oscillation determines $\mn31$, and the difference between $\mn31$ and $\meff$ is of order 1\% with a sign that determines the hierarchy.  If $\mn31$ is increased with $\m21$ held fixed then $\mn31/(2\m21)$ will increase and so one needs to observe more oscillations to measure $\mn31$ and so determine the hierarchy.  This means that higher values of $\mn31$ make the hierarchy more difficult to determine \cite{noiint}.

So how do we know $\mn31$?  In our study, we have fixed $\mn32$ to the value given by MINOS in Ref.~\cite{minosneut2012}.  So $\mn32$ is fixed.  If the hierarchy is normal then, for the same value of $\mn32$, $\mn31$ will be higher than it would be for the inverted hierarchy.  Then, as a result of the argument in the previous paragraph, fixing $\mn32$, if the hierarchy is normal then JUNO and RENO 50 will determine it with less confidence, reproducing the results of the tables.

Actually these results exaggerate the dependence of the sensitivity on the true hierarchy for two reasons.  First of all, MINOS does not really measure $\mn32$.  It combines accelerator data, which measures a combination of $\mn31$ and $\mn32$ called the atmospheric mass difference in \cite{parke2005}, with atmospheric neutrino data which measures different combinations for different angles.  In general these combinations are closer to $\mn32$ than is $\meff$, and therefore fixing the mass measured by MINOS, medium baseline reactor experiments will nonetheless be more sensitive to the inverted hierarchy than the normal hierarchy.  The mass differences measured by MINOS lie between $\mn31$ and $\mn32$, and so by naively identifying them with $\mn32$, as we have done, one exaggerates the effect.  

The second reason that the higher sensitivity of the inverted hierarchy is overstated here is that while presently MINOS provides the most precise atmospheric mass difference, already preliminary results by Daya Bay \cite{yifanggiugno} rival this precision.  However reactor experiments like Daya Bay measure $\meff$, which is closer to $\mn31$ than to $\mn32$.  Fixing $\meff$ instead of the atmospheric mass splitting, the medium baseline reactor experiments have similar sensitivities to the two hierarchies.  In the future Daya Bay and RENO will continue to improve measurements of $\meff$ while accelerator experiments like NO$\nu$A and T2K will improve the measurement of the atmospheric mass difference.  Thus the sensitivity to the hierarchy will continue to be greater in the case of the inverse hierarchy, but by less than is reported here.

\begin{table}[position specifier]
\centering
\begin{tabular}{c|l|l|l|l|}
Site&NH:No Nonlin&IH: No Nonlin&NH: Worst/3&IH: Worst/3\\
\hline\hline
DongKeng&14.1&-17.0&8.2&-21.5\\
\hline
DongKeng+LuGuJing&13.2&-16.2&7.8&-21.4\\
\hline
DongKeng+ZiLuoShan&13.5&-16.1&13.9&-15.3\\
\hline
GuemSeong&6.2&-7.7&3.3&-10.0\\
\hline
GuemSeong+Jangamsan&5.6&-6.6&5.3&-7.0\\
\hline
Munmyeong&11.8&-13.6&6.7&-18.3\\
\hline
Munmyeong+Buncheon-ri&11.5&-13.6&9.4&-16.4\\
\hline
Munsusan&9.4&-11.7&5.9&-16.1\\
\hline
Munsusan+Buncheon-ri&10.3&-12.0&8.6&-14.6\\
\hline
\end{tabular}
\caption{$\Delta\chi^2$ obtained with a perfect energy response and with the one third worst model at various JUNO and RENO 50 sites after 6 years of running.  Single detectors of 20 kton (18 kton) and pairs of 10 kton (9 kton) detectors are used for JUNO (RENO 50).  Flux is considered from all reactors which are operational, under construction and planned.}
\label{ristab}
\end{table}

\begin{table}[position specifier]
\centering
\begin{tabular}{c|l|l|l|l|}
Site&NH:No Nonlin&IH: No Nonlin&NH: Worst/3&IH: Worst/3\\
\hline\hline
DongKeng&8.4&-11.2&5.4&-13.4\\
\hline
DongKeng+LuGuJing&9.2&-10.6&5.1&-14.0\\
\hline
DongKeng+ZiLuoShan&9.2&-10.4&9.2&-8.8\\
\hline
GuemSeong&6.5&-7.2&2.8&-9.6\\
\hline
GuemSeong+Jangamsan&5.8&-6.7&5.5&-8.3\\
\hline
Munmyeong&8.2&-9.7&5.3&-12.1\\
\hline
Munmyeong+Buncheon-ri&8.2&-10.0&6.8&-11.3\\
\hline
Munsusan&6.9&-8.9&4.2&-11.1\\
\hline
Munsusan+Buncheon-ri&7.6&-9.0&7.2&-10.8\\
\hline
\end{tabular}
\caption{As in Table~\ref{ristab} but flux is considered from all reactors which are operational or under construction, not from those that are only planned, reflecting the flux expected when these experiments begin taking data.}
\label{noplantab}
\end{table}

\subsection{Consequences of the Nonlinear Energy Response}

It is clear from these tables that the nonlinear energy response will be a serious challenge both at JUNO and at RENO 50.  Even the optimistic unknown energy response model considered here decreases the expected sensitivity by about 1$\sigma$.   In particular in the case of JUNO it decreases the confidence from $\Delta\chi^2=14.1$ $(3.1\sigma)$ to $8.2$\ $(2.1\sigma)$ whereas for the three locations considered for RENO 50 it decreases the confidence from $\Delta\chi^2=6.2$ $(1.7\sigma)$, 11.8 $(2.8\sigma)$ and 9.4 $(2.4\sigma)$ to 3.3 $(1\sigma)$, 6.7 $(1.8\sigma)$ and 5.9 $(1.6\sigma)$ respectively.  

In some cases all of this decrease can be eliminated by dividing the target mass into two detectors.  However this strategy is only successful if the baselines of the two detectors differ significantly.  For example, the LuGuJing near site for JUNO has a baseline which is too close to that of DongKeng to break the degeneracy between the hierarchy and the unknown energy response of the detector.  We do not observe any advantage for a second detector placed at LuGuJing if the total target mass is fixed.  On the other hand the near site ZiLuoShan does have a sufficiently different baseline and so breaks this degeneracy quite well, restoring the confidence in the hierarchy to $\Delta\chi^2=13.9$ $(3.1\sigma)$, essentially equal to the one detector value before the effects of nonlinearity were considered.  Therefore we conclude that a two 10 kton (20 kton) detector proposal for JUNO can yield a more than 3$\sigma$ confidence determination of the hierarchy in 6 (3) years.

What about RENO 50?  Despite the  backgrounds from the reactor complexes at Kori and Wolseong, the highest $\Delta\chi^2$ can be achieved at Munmyeongsan.   Without considering the unknown energy response it can achieve $\Delta\chi^2=11.8$ corresponding to about $2.8\sigma$ in 6 years as compared with $\Delta\chi^2=6.2$ and $9.4$ at Guemseong and Munsusan respectively.  The unknown energy response reduces the confidence at all three sites drop by between $0.7\sigma$ and $1\sigma$.   More than half of this drop in confidence is recovered with the addition of a second detector.  As a result the best confidence achievable at RENO 50 with a total target mass of 18 kton and 6 years of running appears to be $\Delta\chi^2=9.4$ corresponding to $2.4\sigma$ at Munmyeongsan with an identical near detector at Buncheon-ri.  While these confidences are appreciably lower than what may be achieved at JUNO, at all three locations it seems as though RENO 50 is planning to have more overhead burden than the 700 meters foreseen by JUNO, therefore the backgrounds will be lower.  In this study we have not included the effects of the backgrounds on our results.

In Table~\ref{noplantab} we have considered the case in which reactors which have been planned at the Yangjiang, Taishan and Hanul complexes are not built.  In the case of JUNO this results in a reduction of $\Delta\chi^2$ to 5.4 $(1.5\sigma)$ and 8.8 $(2.3\sigma)$ in the 1 detector and 2 detector cases.  In the case of RENO 50 this has no effect on the preferred GuemSeong site, but it does reduce the confidence at the other two sites.  In particular the confidence at Munmyeong is reduced to $\Delta\chi^2=5.3$ $(1.5\sigma)$ and 6.8 $(1.9\sigma)$ in the 1 and 2 detector cases.  Nonetheless, the confidence at both Hanul sites remains higher than that at the Hanbit site GuemSeong.

\section{Conclusions}

In this note we have investigated the effect of the systematic unknown nonlinear energy response of the detectors upon the sensitivity to the mass hierarchy which can be expected at medium baseline reactor experiments like JUNO and RENO 50.  The size of this unknown depends on details of the calibration scheme which these experiments eventually adopt and so cannot be reliably determined at this time.  However a target of 1\% has been set for JUNO which is twice as small as the current record holder, KamLAND.   

It is clear that this nonlinear response will pose one of the greatest challenges at these experiments and it will reduce the confidence with which the hierarchy has been determined.  As had been suggested in Refs.~\cite{noisim,noi2rivel} and further confirmed in Ref.~\cite{snowmass}, a second identical detector at a significantly different baseline greatly reduces this effect as the relative energies of various features of the spectrum seen at the two detectors is independent of the energy response and is sufficient to determine the hierarchy.  

In this work we considered three models of detector energy response.  All three were optimistic.  In particular the first was a subset of our nonlinearity fitting function and so was removed entirely by the fitting procedure.  The second, the exponential model, is only appreciable in the very low energy regions of the spectrum where the finite resolution means that atmospheric oscillations are invisible.  This part of the spectrum is not useful in a hierarchy determination.  Above about 3 MeV, where the spectrum is sensitive to the neutrino mass hierarchy, this energy response model only shifts the energy by about $0.1\%$ and so is an order of magnitude better than JUNO's target.  Despite this we observed that the effect on the confidence of the hierarchy determination is already noticeable.

Our final model of the nonlinear energy response is the worst case model of Ref.~\cite{xinaug} scaled down by a factor of three.  Once it is scaled down by a factor of three, its effect on the relative energies is about 0.5\% and so it is still about twice as optimistic as JUNO's target, providing a four times more precise determination of the neutrino energies than has ever been provided in such an experiment.  Thus this third worst model is still quite optimistic, although less optimistic than the first two models.  This optimistic model leads to a reduction in the confidence of the hierarchy determination of roughly $1\sigma$ in the case of our one detector configurations.  However, two detectors with the same target mass and sufficiently different baselines are much less strongly affected by the unknown energy response, losing between 0 and $0.4\sigma$ of confidence.

While our study shows a reasonable advantage for two detectors over one detector with the same total target mass, it nonetheless underestimates this advantage in several key ways.  First of all, the energy response models considered were all very optimistic in that the unknown energy response at useful energies was in general limited to 0.5\%.  A larger unknown energy response is more problematic for the one detector scenarios, and therefore increases the advantage of a second detector.  

Second of all, in the two detector scenario, the second detector is either the same size as in the one detector scenario or else it is smaller.  If it is the same size, this means that although engineering considerations limit the mass of a spherical detector of this kind to about 20 kton, it would be possible to have 40 kton of target mass.  As seen in our general study, the greatest advantage of the 2 detector setup is in this high statistics range, and indeed it seems to provide the only way to obtain a robust 3$\sigma$ or more confidence in the hierarchy determination.  

Alternately, one may choose a two detector configuration in which each detector is smaller.  This also leads to several advantages.  First of all, a smaller detector means a smaller distance between the events and the photomultiplier tubes.  This means more photoelectrons will arrive per MeV, improving the energy resolution.  The confidence in a hierarchy determination is extremely sensitive to small changes in the energy resolution \cite{snowmass}.  The smaller distance that a photon must travel also makes the pulse shape narrower, which helps in the rejection of background and also in the determination of the location of the event.  As the response of the detector strongly depends on the location already at Daya Bay \cite{calib}, the determination of this location is necessary to control statistical errors in the number of photoelectrons so as to improve the energy resolution.

Finally, our study has underestimated the advantages of a second detector because we have considered individual experiments, which are statistically limited.  Indeed our general study reveals that the greater the target mass, the better the statistics, the greater the advantage in splitting a detector into two at distinct baselines.  In practice this statistical limitation can somewhat be overcome by coming data from RENO 50 and JUNO, and indeed even some data from 1 km baseline reactor experiments.

In summary, while the results of this paper show a clear advantage for the two detector proposal, this study underestimates this advantage in several key ways.  It may well be, if the unknown nonlinear response is truly of order 1\%, that a single detector proposal will not determine the hierarchy with even a single $\sigma$ of confidence.   In such a case a second detector would be necessary to make the resulting hierarchy determination credible.



\section* {Acknowledgement}

\noindent
JE is supported by the Chinese Academy of Sciences
Fellowship for Young International Scientists grant number
2010Y2JA01. EC and XZ are supported in part by the NSF of
China.  


\end{document}

\begin{table}[position specifier]
\centering
\begin{tabular}{c|l|l|l|l|}
Site&NH:No Nonlin&IH: No Nonlin&NH: Worst/3&IH: Worst/3\\
\hline\hline
DongKeng&14.0&-16.9&18.8&-8.8\\
\hline
DongKeng+LuGuJing&13.4&-16.2&20.1&-9.1\\
\hline
DongKeng+ZiLuoShan&13.6&-16.3&11.6&-17.0\\
\hline
GuemSeong&6.3&-7.3&8.9&-3.3\\
\hline
GuemSeong+Jangamsan&5.7&-6.5&5.5&-5.3\\
\hline
Munmyeong&11.8&-13.5&17.8&-6.9\\
\hline
Munmyeong+Buncheon-ri&11.6&-13.4&14.9&-10.3\\
\hline
Munsusan&9.4&-11.6&13.3&-6.0\\
\hline
Munsusan+Buncheon-ri&10.3&-11.9&11.3&-9.5\\
\hline
\end{tabular}
\caption{$\Delta\chi^2$ obtained with a perfect energy response and with the one third worst model at various JUNO and RENO 50 sites.  Single detectors of 20 kton (18 kton) and pairs of 10 kton (9 kton) detectors are used for JUNO (RENO 50).  Flux is considered from all reactors which are operational, under construction and planned.}
\label{ristab}
\end{table}

\begin{table}[position specifier]
\centering
\begin{tabular}{c|l|l|l|l|}
Site&NH:No Nonlin&IH: No Nonlin&NH: Worst/3&IH: Worst/3\\
\hline\hline
DongKeng&8.5&-11.0&11.5&-6.1\\
\hline
DongKeng+LuGuJing&9.4&-10.6&13.2&-5.8\\
\hline
DongKeng+ZiLuoShan&9.2&-10.4&7.6&-9.9\\
\hline
GuemSeong&6.6&-7.2&9.5&-3.5\\
\hline
GuemSeong+Jangamsan&5.6&-6.6&5.8&-5.9\\
\hline
Munmyeong&8.1&-9.8&12.2&-5.4\\
\hline
Munmyeong+Buncheon-ri&8.2&-9.8&10.3&-7.9\\
\hline
Munsusan&6.9&-8.9&9.9&-4.4\\
\hline
Munsusan+Buncheon-ri&7.6&-8.9&8.1&-6.8\\
\hline
\end{tabular}
\caption{As in Table~\ref{ristab} but flux is considered from all reactors which are operational or under construction, not from those that are only planned, reflecting the flux expected when these experiments begin taking data.}
\label{noplantab}
\end{table}

\begin{table}[position specifier]
\centering
\begin{tabular}{c|l|l|l|l|l|l|}
Reactor&Status&Th. Cap&DongKeng(\begin{CJK}{UTF8}{gbsn}东坑\end{CJK})&LuGuJing(\begin{CJK}{UTF8}{gbsn}碌古径\end{CJK})&JiZaiNao(\begin{CJK}{UTF8}{gbsn}鸡仔脑\end{CJK})&ZiLuShan(\begin{CJK}{UTF8}{gbsn}紫罗山\end{CJK})\ \ \      \\
\hline\hline
Elevation&&&450 m&350 m&450 m&750 m\\
\hline
YangJiang 1&Under Construction&2905 MW&52.75 km&40.55 km&39.48 km&17.43 km\\
\hline
YangJiang 2&Under Construction&2905 MW&52.84 km&40.6 km&39.6 km&17.52 km\\
\hline
YangJiang 3&Under Construction&2905 MW&52.42 km&40.2 km&39.1 km&17.09 km\\
\hline
YangJiang 4&Under Construction&2905 MW&52.51 km&40.3 km&39.2 km&17.18 km\\
\hline
YangJiang 5&Planned&2905 MW&52.12 km&39.9 km&38.8 km&16.79 km\\
\hline
YangJiang 6&Planned&2905 MW&52.21 km&40.0 km&38.9 km&16.88 km\\
\hline
Taishan 1&Under Construction&4590 MW&52.76 km&40.55 km&41.65 km&65.84 km\\
\hline
Taishan 2&Under Construction&4590 MW&52.63 km&40.4 km&41.5 km&65.7 km\\
\hline
Taishan 3&Planned&4590 MW&52.32 km&40.1 km&41.2 km&65.5 km\\
\hline
Taishan 4&Planned&4590 MW&52.20 km&39.9 km&41.0 km&65.4 km\\
\hline
DB-LA&Operational&17430 MW&215 km&215 km&216 km&242 km\\
\hline
\end{tabular}
\caption{JUNO sites}
\label{disttabdaya}
\end{table}

\noindent
Last year the Daya Bay \cite{dayabay,neut2012} and RENO \cite{reno} experiments demonstrated beyond any reasonable doubt that $\theta_{13}$ is as much as an order of magnitude larger than had been suspected several years ago.  
This large value of $\theta_{13}$ implies that 1-3 reactor neutrino oscillations may be observed at medium baselines, which we define to be 40-80 km.  The medium baseline neutrino spectrum may then be used to determine the neutrino mass hierarchy \cite{petcovidea}.  Such experiments are now not only practical but indeed they will be performed within the next decade \cite{caojunseminario,renonuturn,yifangseminario}.  

How does this determination work?  The reactor emits $\nu_e$'s and $\overline{\nu}_e$'s.  The $\overline{\nu}_e$'s are detected via inverse beta decay upon their interaction with free (not bound to other nucleons) protons in a detector.   Some of the $\overline{\nu}_e$'s oscillate into other flavors, providing an energy-dependent reduction of the flux which depends on the leptonic mixing angles $\theta_{12}$ and $\theta_{13}$, on the neutrino mass differences and in particular on the neutrino mass hierarchy.  In all, the $\overline{\nu}_e$ survival probability is 
\bea
P_{ee}&=&\sp413+\cp412\cp413+\sp412\cp413+\frac{1}{2}(P_{12}+P_{13}+P_ {23})\nonumber\\
P_{12}&=&\spp2212\cp413\cos\left(\frac{\m21L}{2E}\right)\hsp \label{sopraviv}
P_{13}=\cp212\spp2213\cos\left(\frac{\mn31L}{2E}\right)\nonumber\\
P_{23}&=&\sp212\spp2213\cos\left(\frac{\mn32L}{2E}\right).\nonumber
\eea
The largest contribution to the depletion is caused by the $P_{12}$ term in Eq.~(\ref{sopraviv}).  At a medium baseline this corresponds to a single, broad dip in the measured neutrino spectra.  On the other hand $P_{13}$ and $P_{23}$, which we refer to collectively as 1-3 oscillations, provide a fine structure of small oscillations in the observed spectrum.  Of these, $P_{13}$ has twice the amplitude of $P_{23}$, so $P_{23}$ provides a perturbation to the $P_{13}$ oscillations, which on their own would have been periodic in $1/E$.  As a result of $P_{23}$, the fine structure is not quite periodic in $1/E$ \cite{parke2007}.   

This deviation from periodicity in the fine structure of the observed spectrum determines the hierarchy.   More quantitatively, let $E_n$ be the energy of the peak in the oscillated neutrino spectrum corresponding to neutrinos which have oscillated $n$ times between the reactor and the detector.  Note that higher values of $n$ correspond to lower values of energy.  It was shown in Ref.~\cite{noiteor} that the inverse energies of the first 10 peaks are indeed periodic to within experimental error and indeed are well approximated by 2 flavor neutrino oscillation with an effective mass difference of~\cite{parke2005}
\beq
\Delta M^2_{\rm{eff}}=\cp212\mn31+\sp212\mn32. \label{meff}
\eeq
On the other hand, by the 16th peak $P_{13}$ and $P_{23}$ are in phase, and so at energies as low as $E_{16}$ the peak locations are instead roughly those of 2-flavor oscillation with a mass of $\mn31$, which is greater (less) than $\meff$ if the hierarchy is normal (inverted).  If the hierarchy is normal then $\mn31$ will be greater than $\meff$ and so the energies $E_n$ of the low energy peaks, corresponding to $n$ well above 10, will be higher than would be obtained from a simple periodic extrapolation of the high energy ($n\leq 10$) peaks.  For example, fixing $M^2_{\rm{eff}}$, $E_{16}$ would be about 2\% higher in the case of the normal hierarchy.

Clearly such an experiment needs to be able to measure the energy with a precision much better than 2\%.  The energy of the neutrino cannot be measured directly, but the energy of the positron resulting from IBD  is determined by counting photoelectrons in a photomultiplier.  At low energies, not many of these photoelectrons are detected and so statistical fluctuations in the number of photoelectrons limit the energy resolution.  As a result, the low energy peaks, which anyway are closer together, are smeared.  This means that for $n$ greater than about 17, the identification of an individual peak is hopeless.  On the other hand, the high energy peaks which determine $\Delta M^2_{\rm{eff}}$ can be reliably measured.  As a result, such experiments can easily measure $\Delta M^2_{\rm{eff}}$, the main difficulty in the determination of the hierarchy comes from the low energy measurement of $\mn31$.  In particular, since all of the peaks $n\leq 10$ measure the same quantity $\meff$, little is gained by considering the peaks in the high energy tail of the reactor neutrino spectrum.

All experimental analyses that determine the hierarchy solely from reactor neutrinos rely upon the breakdown in periodicity described above.  Two kinds of analysis have been studied extensively in the literature.  First, one may perform a $\chi^2$ fit to the observed spectra assuming both hierarchies, and conclude that the hierarchy is the one which minimizes $\chi^2$.  This method suffers from the fact that there are many  nuisance parameters which need to be considered in the determinations of the spectra, and it would be impractical to extremize $\chi^2$ with respect to all of them.  As a result, a simpler method has been proposed in Ref.~\cite{hawaii}, in which one considers a Fourier transform of the observed spectrum and identifies several hierarchy-dependent quantities associated to the transformed spectrum which are reasonably independent of some of these nuisance parameters.   More such properties were identified in Refs.~\cite{caojun,noiteor} and applied to simulated data in Ref.~\cite{caojun2}.


In Ref.~\cite{oggi}, the authors observed that the hierarchy-dependent quantities, contrary to their original motivation, are extraordinarily sensitive to the neutrino mass differences and also to the model of the reactor spectrum.  We will now will explain the origin of this sensitivity.



As the dependences of the various quantities are virtually indistinguishable, for brevity we will consider only~\cite{caojun}
\beq
RL=\frac{R-L}{R+L}
\eeq
which is the fractional difference between two minima $R$ and $L$ of the Fourier cosine transform of the neutrino spectrum 
\beq
F_c(k)=\int d\left(\frac{L}{E}\right)\frac{E^2}{L} \frac{\Phi(E)\sigma(E)}{4\pi L^2}  P_{ee}\left(\frac{L}{E}\right) \cos\left(\frac{kL}{E}\right) \label{fc}
\eeq
where $E$ is the neutrino's energy and the tree level neutrino inverse $\beta$ decay cross section is~\cite{sezionedurto}
\beq
\sigma(E)=0.0952\times 10^{-42}\mathrm{cm}^2\frac{E_e\sqrt{E_e^2-m_e^2}}{\mathrm{MeV}^2}\hsp
E_e=E-m_n+m_p.
\eeq
A $3\%/\sqrt{(E_e+m_e)/\mathrm{MeV}}$ energy resolution is included by convoluting the observed energy spectrum with
\beq
\rm{exp}\left(-\frac{(E-E\p)^2}{0.0018(E_e+m_e)MeV}\right).
\eeq
The masses of the electron, proton and neutron are $m_e$, $m_p$ and $m_n$.  We use the neutrino mass matrix parameters of Ref.~\cite{caojunseminario}.

As was demonstrated in Ref.~\cite{noiteor}, the minima whose difference defines $RL$ lie just on either side of $k=\mn31/2$.  These minima arise from the Fourier transform of $P_{13}$ which is independent of the hierarchy, but the contribution of $P_{23}$  provides a perturbation which makes the right (left) minimum deeper for the normal (inverted) hierarchy.   Thus the authors concluded that the hierarchy can be well determined by simply observing which of the two minima is deeper.

\begin{figure} 
\begin{center}
\includegraphics[width=5.2in,height=2in]{interferenza.eps}
\caption{The cosine transforms of the unoscillated reactor flux (black dashed curve) and the full $P_{13}+P_{23}$ oscillated flux (red solid curve) are shown.  Note that the amplitude of the reactor flux oscillations is not much smaller than the difference between the depths of the two minima of $P_{13}+P_{23}$, which yields the hierarchy signal $RL$.  For the value of $\meff$ drawn here, the reactor flux attains a maximum at the shallowest (left side) minimum of $P_{13}+P_{23}$, and so its contribution enhances $RL$, artificially inflating the confidence of the hierarchy determination.}

\label{vecflufig}
\end{center}
\end{figure}

In Ref.~\cite{oggi} the authors described a problem with this approach.  They observed that, depending upon the reactor flux model used, the transform of the unoscillated reactor flux $\Phi(E)\sigma(E)E^2/L^3$ itself may contribute near $k=\mn31/2$, interfering with  $P_{13}+P_{23}$ and so affecting $RL$.  The unoscillated reactor flux contains no information about neutrino masses or the hierarchy, this contribution therefore contaminates the hierarchy-dependent observable $RL$.    Thus $RL$ does indeed depend upon the reactor flux model.  But in Ref.~\cite{oggi} the authors also observed that this dependence is highly sensitive to the mass splitting, why is this?

While the cosine transform of the unoscillated flux $\Phi(E)\sigma(E)E^2/L^3$ is itself independent of the neutrino mass splittings, the locations of the peaks of $P_{13}+P_{23}$ are proportional to $\meff$.  This means that the relative phase between the Fourier transform of the unoscillated spectrum and that of $P_{13}+P_{23}$ depends on the precise value of $\meff$.    For example, for some values of $\meff$ the maximum of the Fourier transformed unoscillated reactor flux is coincident with the left (right) minimum of the transform of $P_{13}+P_{23}$, so this contribution increases (decreases) RL.  As a result the oscillations in the Fourier transform of $\Phi(E)\sigma(E)$ lead to an $\meff$-dependence in the quantity $RL$ just of the kind observed in Ref.~\cite{oggi} using old reactor flux models.  

In fact, using the $\u35$ flux from Ref.~\cite{uflusso}, the $\pu39$ and $\pu41$ fluxes from \cite{pflusso} and the Gaussian approximated $\u38$ flux from Ref.~\cite{quadflusso} with the isotope ratios of Ref.~\cite{caojun} we find an oscillation in the unoscillated spectrum term in Eq.~(\ref{fc}).  Using this old model of the reactor flux, in Fig.~\ref{vecflufig} we compare the Fourier transform of the unoscillated term with that of the $P_{13}+P_{23}$ term, which is sensitive to the hierarchy.  One can see that the unoscillated term is periodic with the same wavelength as was observed in Fig. 4 of Ref.~\cite{oggi}, and thus the interference between these two terms oscillates as $\meff$ varies, shifting the $P_{13}+P_{23}$ peaks and so reproducing the effect reported in that note.

\begin{figure} 
\begin{center}
\includegraphics[width=5.2in,height=2in]{cinquespettri.eps}
\caption{The cosine transforms of the unoscillated flux is shown for numerically interpolated fluxes from the 1980's \cite{uflusso,pflusso,quadflusso} (black dotted curve), for a quadratic fit to fluxes  from the 1980's \cite{quadflusso} and for quintic fits of the new fluxes of Ref.~\cite{huber}.  The latter two are shown with cutoffs of 8.5 MeV (dashed curves) and 12.8 MeV (solid curves).  The blue dashed curve corresponds to the quadratic fit flux.  The red and green solid curves, corresponding to 12.8 MeV cutoffs, are close to zero.  Therefore the interference effect is present if the cutoff is at 8.5 MeV and but not if the fits are naively extrapolated to 12.8 MeV.  This demonstrates that $RL$ is sensitive to the neutrino spectrum above 8.5 MeV.}
\label{cutofffig}
\end{center}
\end{figure}

\begin{figure} 
\begin{center}
\includegraphics[width=5.2in,height=2.6in]{Con_e_Senza_Cut.eps}
\caption{Simulated average values of $RL+PV$ obtained from 100,000 neutrinos observed at a baseline of 58 km for various reactor flux models and mass differences $\mn32$.   The black dotted curve corresponds to the numerically interpolated fluxes from the 1980's \cite{uflusso,pflusso,quadflusso}, the dotted (solid) blue curve corresponds to a quadratic fit to fluxes  from the 1980's \cite{quadflusso} cut off at  8.5 MeV (12.8 MeV) and the dotted (solid) red curves to the quintic fits of the new fluxes of Ref.~\cite{huber} cut off at 8.5 MeV (12.8 MeV) .  Notice that an unphysical extrapolation to 12.8 MeV eliminates the oscillations, therefore the oscillations result from the high energy part of the spectrum.}
\label{rlpvfig}
\end{center}
\end{figure}

Ref.~\cite{oggi} concludes that this strong dependence of $RL$ upon the reactor flux means that a precise knowledge of this flux is desirable to determine the neutrino mass hierarchy at a short baseline experiment.  Our conclusions differ, we claim that this apparent dependence on the reactor model and the mass splittings is merely a dependence upon the high energy tail of the spectrum.  This can clearly be seen to be the case in Fig.~\ref{cutofffig}, in which we plot the cosine transforms of the unoscillated reactor flux for several different reactor models with various high energy cutoffs.   The oscillations are large for {\it{every}} reactor flux model with a cutoff of 8.5 MeV, but for none with a cutoff at 12.8 MeV.  Thus, the large oscillations of the Fourier transform of the unoscillated spectrum are caused almost entirely by the very high energy tail of the spectrum, above 8.5 MeV, where there are few events and essentially no information regarding the hierarchy.


To demonstrate that these fluctuations in the cosine transform of the unoscillated reactor spectrum are indeed the culprit behind the spurious mass splitting dependence in Ref.~\cite{oggi}, in our Fig.~\ref{rlpvfig} we reproduce their Figure 4.  This figure shows $RL$ determined from simulated data with various mass splittings, reactor models and cutoffs.   We find that the spurious dependence on the mass splitting in fact exists for {\it{all}} flux models, but for each model it goes away with a naive extrapolation of the corresponding flux fitting function up to 12.8 MeV.   Indeed, we see that even for the old quadratic fits the oscillation can be removed for such a naive extrapolation.  

This does not imply that the problem observed in Ref.~\cite{oggi} can simply be eliminated by not cutting off the spectrum at high energies.  The problem with this approach is that the flux fitting functions extrapolated up to 12.8 MeV in many previous studies do not provide good approximations to the reactor flux at these high energies.  On the contrary, as was noted in Ref.~\cite{quadflusso} the extrapolation of the quadratic flux fit to 12.8 MeV yields 2 to 3 times more flux than is observed above about 8.5 MeV, and so therefore is unphysical.  Thus although the real reactor neutrino flux does not exhibit a hard cut off at 8.5 MeV, a hard cut off at 8.5 MeV may nonetheless provide as good of an approximation to the true spectrum as no cut off at all.  The quintic fit, as it is based on a fit to data at energies below 8.5 MeV, has a similar problem.   

From this analysis we learn two lessons.  First of all, as the real reactor neutrino flux above 8.5 MeV is not well approximated by the simple fitting functions which work at lower energies, one can expect that the Fourier transform of the true spectrum will exhibit the oscillations described above and so contaminate $RL$ and $RL+PV$ in the manner observed in Ref.~\cite{oggi}.  Second, this effect is missed in a simulation which naively extrapolates these fitting functions well beyond 8.5 MeV, perhaps explaining why it had not been observed in earlier studies.


{\it{As $RL$ depends strongly on the spectrum between 8.5 and 12.8 MeV, which in turn is independent of the hierarchy, this high energy tail provides a nuisance parameter for the determination of the hierarchy using $RL$.}}  The solution suggested in Ref.~\cite{oggi} is to determine the spectrum precisely, however so few neutrinos are observed in this range that such a determination would be difficult, indeed the spectrum is not understood at the required precision even at the energies with high fluxes \cite{reattoreanom}.  Even if such a measurement were possible, then $RL$ would still depend upon $\meff$ with a higher sensitivity than the mass determination at MINOS, making a determination of the hierarchy at a medium baseline more challenging.

\begin{figure} 
\begin{center}
\includegraphics[width=5.2in,height=2.5in]{Medie_Totali_PesiDef.eps}
\caption{Simulated average values of $RL+PV$ obtained from 100,000 neutrinos observed at a baseline of 58 km assuming the numerically interpolated reactor spectra from the 1980's \cite{uflusso,pflusso,quadflusso}.  The black dotted curve uses an unweighted Fourier transform and the red solid curve uses the weight  exp$(-0.04E^2/\rm{MeV}^2)$.}
\label{conosenzafig}
\end{center}
\end{figure}


Our solution is to replace $RL$ and $PV$ with quantities that are insensitive to the high energy neutrino spectrum, by providing an energy-dependent weight $w(E)$ on the neutrino spectrum in the Fourier transform.  As we saw in Fig.~\ref{cutofffig}, a simple cutoff in the Fourier transform will amplify the spurious dependence.  The weight needs to cut off the high energies gradually, with derivative scales much longer than $\mn31$, so as to not itself introduce spurious peaks in the critical part of the Fourier transforms.  One such choice of weight which we have found works quite well is a Gaussian
\beq
F_c(k)=\int d\left(\frac{L}{E}\right)e^{-\frac{0.04E^2}{\rm{MeV}^2}}\frac{E^2}{L} \frac{\Phi(E)\sigma(E)}{4\pi L^2}  P_{ee}\left(\frac{L}{E}\right) \cos\left(\frac{kL}{E}\right).
\eeq
The same weight serves well in both the sine transform and also the nonlinear transforms of Ref.~\cite{noiteor} which determine the hierarchy more reliably than $RL+PV$ at baselines below about 55 km \cite{noisim}.   In Fig.~\ref{conosenzafig} we use simulated data to compare the value of $RL+PV$ obtained from the ordinary Fourier transform with that obtained using the weighted Fourier transform.  The oscillations almost disappear in the weighted case.

As can be seen by comparing the unweighted and weighted cosine transforms in Figs.~\ref{vecflufig} and~\ref{xformpesfig}, not only does the weighting procedure preserve $RL$, but given enough detected neutrinos it actually increases the difference in the peak sizes between the normal and inverted hierarchies.  Thus this solution to the dependence upon the high energy neutrino tail not only removes the spurious dependence, for any high energy reactor spectrum, but it can even increase the chance of success of the determination of the hierarchy.  This benefit is greatest when a large number of neutrinos is detected, since the weighting effectively reduces the statistics at high energies.  In Ref.~\cite{noisim} we use the weighted Fourier transform to analyze simulated data.

\begin{figure} 
\begin{center}
\includegraphics[width=5.2in,height=2in]{pesati.eps}
\caption{Here we see the theoretical weighted Fourier transform of the spectrum without oscillations (blue solid curve) and with oscillations in the case of the normal (black solid curve) and inverted (red dashed curve) hierarchies.  One can see that the solid, blue unoscillated curve is very close to zero.  We have checked that this curve is essentially independent of the cutoff and so the reactor spectrum no longer affects $RL$.   Comparing with Fig.~\ref{vecflufig} one can see that the difference $RL$ between the depths of the minima is even greater in this weighted case, allowing for a better determination of the hierarchy than was possible with an unweighted Fourier transform.}
\label{xformpesfig}
\end{center}
\end{figure}

\section* {Acknowledgement}

\noindent
JE is supported by the Chinese Academy of Sciences
Fellowship for Young International Scientists grant number
2010Y2JA01. EC and XZ are supported in part by the NSF of
China.  


\end{document}

\bibitem{minos}
R. Nichol,
``Final MINOS Results," presented at Neutrino 2012 in Kyoto.
Available at http://neu2012.kek.jp/neu2012/programme.html.
P.~Adamson {\it et al.}  [MINOS Collaboration],
  ``Measurements of atmospheric neutrinos and antineutrinos in the MINOS Far Detector,''
  arXiv:1208.2915 [hep-ex].

\bibitem{nuovoflusso}

\bibitem{noiunokm}

\end{document}

However at these baselines, due to a degeneracy in the high energy neutrino spectrum \cite{parke2007,oggi,noi}, a determination of the hierarchy requires a measurement of 1-3 oscillations at low neutrino energies $E$.  As a result of the finite energy resolution of the detector and various interference effects \cite{noi} these low energy peaks are difficult to identify individually at medium baselines $L$.  Nonetheless if the nonlinear energy response of a detector is well understood then one may measure {\it{the sum of the peaks}} by studying the $k\sim\mn31/2$ region of the  $L/E$-Fourier transform of the neutrino spectrum \cite{hawaii}.  The most popular variables for such a determination are the fractional difference $RL$ between the deepest minima of the Fourier cosine transform and the difference $PV$ between the deepest minimum and the highest peak of the Fourier sine transform \cite{caojun,caojun2}.  Although these two variables are somewhat degenerate, an improvement may be obtained by considering their sum $RL+PV$.

A serious obstruction to this analysis, and thus to plans to measure the neutrino mass hierarchy at medium baselines, has been described in Ref.~\cite{oggi}.  The authors observed that the combination $RL+PV$ is very sensitive to the choice of model of the reactor neutrino flux $\Phi(E)$ and to variations of $\mn32$ which are smaller than the precision to which this mass difference has been determined by MINOS \cite{minos}.  While the observed shift appears to depend upon both $\mn32$ and the hierarchy, from Fig. 4 of Ref.~\cite{oggi} it can be seen that the shift depends only upon the effective mass difference \cite{parke2005,noi}
\beq
\meff=\cp212\mn31+\sp212\mn32.
\eeq
The neutrino flux from reactors is known poorly.  The theoretical normalization has recently increased by about 3\% \cite{nuovoflusso} and the 6+ MeV flux has increased by an additional 3\% \cite{huber}.  The flux beyond about 8 MeV is not known at all due to its strong dependence upon decays of exotic isotopes \cite{nuovoflusso}.  Even worse, all of these theoretical fluxes are about 6\% above the observed fluxes at very short \cite{reattoreanom} and 1 km \cite{noiunokm} baselines.  Thus the large sensitivity of $RL+PV$ upon the poorly known fluxes and $\meff$  appreciably reduces the probability that a medium baseline reactor experiment can correctly determine the neutrino mass hierarchy.

\section* {Acknowledgement}

\noindent
JE is supported by the Chinese Academy of Sciences
Fellowship for Young International Scientists grant number
2010Y2JA01. EC and XZ are supported in part by the NSF of
China.  


\end{document}

\section{Introduction}

\section{The Monte Carlo}

\subsection{The Parameters}

This paper builds upon the simulations of medium baseline neutrino oscillation in Ref. \cite{caojun2}.  For ease of comparing the results of that study and ours, we have adopted many of the same for the neutrino mass matrix
\beq
\m21=7.6\times 10^{-5}\rm{\ eV}^2\hsp
\mn22=2.4\times 10^{-3}\rm{\ eV}^2\hsp
\sp212=0.32
\eeq
and for the absolute detector energy resolution
\beq
\sigma_E=.03\sqrt{E_ {pr}\rm{(MeV)}}
\eeq
where the prompt energy $E_{pr}$ is related to the neutrino energy $E$ and the positron energy $E_e$ by
\beq
E_{pr}=E_e+m_e=E-m_n+m_p+m_e\sim E-780\ \mathrm{keV}\hsp
\eeq

While these parameters have changed very little over the past 3 years, the motivation for our analysis is the large increase in $\theta_{13}$ \cite{dayabay,reno,doublechooz}.  We will use Daya Bay's most recent result \cite{neut2012}
\beq
\spp2213=0.089.
\eeq 
This increase is critical to the determination of the hierarchy at a neutrino disappearance experiment, as the hierarchy manifests itself via energy-dependent shifts of the 1-3 oscillation peaks \cite{petcovidea,parke2007}.

\subsection{The Simulation}

The overall reactor flux normalization is fixed to be that of Ref \cite{caojun2}.  It is asserted that the Daya Bay plus Ling Ao I and II reactor complexes, which together have 17.4 GW of thermal capacity, lead to the observation of 25,000 neutrinos at a 20 kton detector with a baseline of 60 km.  To be precise this does not fix the reactor flux normalization, but rather the overall normalization of the product of the reactor flux, the antineutrino cross section of the target, the detector efficiency and the effect of neutrino oscillations.  This condition, together with the tree level inverse $\beta$ decay cross section in Ref. \cite{sezionedurto}
\beq
\sigma(E)=0.0952\times 10^{-42}\mathrm{cm}^2(E_e\sqrt{E_e^2-m_e^2}/\mathrm{MeV}^2)
\eeq
and the oscillation probability $P_{ee}$
\bea
P_{ee}&=&\sp413+\cp412\cp413+\sp412\cp413+\frac{1}{2}(P_{12}+P_{13}+P_ {23})\nonumber\\
P_{12}&=&\spp2212\cp413\cos\left(\frac{\m21L}{2E}\right)\hsp
P_{13}=\cp212\spp2213\cos\left(\frac{\mn31L}{2E}\right)\nonumber\\
P_{23}&=&\sp212\spp2213\cos\left(\frac{\mn32L}{2E}\right)\nonumber
\eea
are then used to normalize the effective reactor flux $\Phi(E)$ per time per GW of thermal capacity.  This reactor flux is effective in the sense that it is already multiplied by the efficiency of the detector.

With this normalization in hand, the average number density of antineutrinos at energy $E$ from a given reactor which without oscillation would be observed during 3 years at a 20 kton detector at a baseline $L$ is then
\beq
N(E)P_{ee}(L/E)=\frac{\Phi(E)\sigma(E)P_{ee}(L/E)}{4\pi L^2}.
\eeq
In each 60 kton year experiment we simulate precisely $N(E)$ neutrinos per unit energy from each reactor.  We will recreate 120 and 300 kton year experiments by simply summing the neutrinos from pairs or quintuplets of 60 kton year experiments.  To minimize the relative statistical errors between experiments with different exposure times,120 and 300 kton year experiments will be created from 60 kton experiments reported on the same tables.


Unlike Ref. \cite{caojun2,caojunseminario} our total neutrino flux in each experiment depends on $L$.  In fact it increases faster than $1/L^2$ as $L$ is decreased from 58 km because there is less loss due to 1-2 neutrino oscillations.  This means that at short baselines, such as 40-50 km, our simulations allow a precise determination of the high energy peaks, which are hardly depleted by 1-2 oscillation.  The energies of these peaks $2\pi/\Delta M^2_{\rm{eff}}$ determine the effective mass splitting \cite{parke2005,noi}
\beq
\Delta M^2_{\rm{eff}}=\cp212\mn31+\sp212\mn32. \label{meff}
\eeq
As a result we will see that our simulations favor shorter baselines than those preferred in Refs~\cite{caojun2,caojunseminario}.  We will also see that shorter baselines are preferred as they reduce the fractional backgrounds from distant reactors, backgrounds not included in previous studies.

Once we have fixed the numbers and energies of the neutrinos arriving at our detector, the finite resolution $\sigma_E$ of the detector is applied.   With infinite luminosity the observed neutrino spectrum $P_{obs}(E)$ would be the theoretical spectrum convoluted with a Gaussian of width $\sigma_E$.  To implement this effect at a finite luminosity we shift the energy of a neutrino by a random variable  $\delta$ with Gaussian probability density $\rm{exp}(-\delta^2/2\sigma_E^2)/\sqrt{2\pi\sigma_E}$.  

Thus our simulation correctly accounts for statistical errors in the determination of the energy due to the finite number of photoelectrons detected.  However it does not take into to account for a systematic nonlinear error in the determination of energy.  While linear errors are harmless to the Fourier analysis that will be performed later, they simply shift the points, even a small linear uncertainty can alter \cite{petcov2010} or destroy \cite{oggi} the effectiveness of such an analysis.  In practice the energy can be calibrated, at least at some energies and at some points inside of the detector, by inserting radioactive sources.  A peak by peak analysis of the spectrum can be performed just using peaks at these energies \cite{noi} but this requires more flux than a Fourier analysis.

\subsection{Improvements over Previous Simulations}

Although the simulation does not account for the unknown nonlinear response of the detector, it does include an interference effect \cite{noi} which has so far not been simulated in the literature.  The reactor complex at Daya Bay is separated by more than a kilometer from the reactor complexes Ling Ao I and Ling Ao II.  While this set of reactors is in general an ideal choice for medium baseline experiments, given the high flux and the existence of 8 short baseline detectors, in general the 1-3 oscillations from Ling Ao and Daya Bay will be out of phase at low energies.  The measurement of 1-3 oscillations at these low energies are essential to break a degeneracy which prohibits the determination of the neutrino mass hierarchy \cite{parke2007,oggi,noi}.  Thus such interference poses a severe problem for the determination of the hierarchy.  However, since the Daya Bay, Ling Ao I and Ling Ao II complexes, like RENO, lie along a line, it is possible to evade this interference effect if the detector is roughly orthogonal to the line.  The price of this choice is that one cannot benefit from the flux at the proposed HuiDong reactor complex, which would no longer be equidistant from the detector.  

Our simulation also includes the flux from many distant reactors, of which TaiShan and YangJiang will begin to come online next year.  We will see that these backgrounds have a nonnegligible effect on the probability of determination of the neutrino mass hierarchy.   They provide both more backgrounds but also potentially more usable information for the determination of $\theta_{12}$.

\section{Data Analysis}

\subsection{Fourier Analysis and Reactor Flux Models} \label{flusssez}

In this note we will not consider a $\chi^2$ or peak energy analysis of the data, although these require an understanding of the nonlinear response of the detector only in limited regions.  Instead we will consider a Fourier analysis \cite{hawaii}, which has the advantage that it combines neutrinos from multiple peaks and so requires less flux, despite the fact that it requires an unprecedented understanding of the nonlinear response of the detector over a wide range of energies. Following Refs.~\cite{caojun,caojun2} we will decompose this transform into a real and a complex part
\bea
F_c(k)&=&\int d\left(\frac{L}{E}\right)\frac{E^2}{L} N(E)  P_{ee}\left(\frac{L}{E}\right) \cos\left(\frac{kL}{E}\right)\nonumber\\
F_s(k)&=&\int d\left(\frac{L}{E}\right)\frac{E^2}{L} N(E)  P_{ee}\left(\frac{L}{E}\right) \sin\left(\frac{kL}{E}\right) \label{fcos}
\eea
where $N(E)$ is the number of neutrinos in the energy bin centered at energy $E$ in a given experiment and the factor $E^2/L$ is the Jacobian factor which transforms a neutrino density in $E$ into a density in $L/E$.

We will be interested in the form of the Fourier transforms near the 1-3 oscillation peak $k=\mn31/2$.  At these wavenumbers it is commonly believed that gross features of the spectrum are independent of the reactor flux and $P_{12}$, but this is not quite true.  Recently Ref.~\cite{oggi} found that the quantities (\ref{rlpv}) used in the analysis of Refs.~\cite{caojun,caojun2} depend strongly upon the reactor flux model used and for some flux models even upon $\mn32$, with variations as large as a factor of 5.  The authors concluded that a precise determination of the reactor spectrum is necessary to determine the hierarchy.  This is also not quite true.  

Instead the effect discovered in Ref. \cite{oggi} is the strong dependence of the quantities (\ref{rlpv}) upon the high energy part of the neutrino flux spectrum, in particular above about 8 MeV.  Even though the neutrino flux at such high energies is very small, it can be leading contribution to the Fourier transform of the unoscillated neutrino flux $\Phi(E)$ at $k\sim\mn31/2$.  For example a cutoff even as high as 8.5 MeV can lead to fluctuations in the Fourier transform of the reactor flux $\Phi(E)$ which are as large as the fluctuations in $P_{13}+P_{23}$ which determine the quantities (\ref{rlpv}) and therefore the hierarchy.  These oscillations in the Fourier transform of $\Phi(E)$ in the presence of a cutoff have just the same wavelength as those reported in Ref. \cite{oggi}.  The $\mn32$-dependence observed in Fig. 4 of that reference arises from the $\meff$ dependence of the $P_{13}+P_{23}$ peaks, which determines the interference with the $\mn32$-independent transform of the reactor flux.  This poses a real problem as the $\mn32$-dependence is large on scales within the MINOS error on the determination of $\mn32$, and so it cannot be simply corrected during the data analysis.

How can this problem be resolved?  Ref. \cite{oggi} suggests an accurate reactor flux determination, and to back up this claim shows that the effect is minimized using the new fluxes of Refs. \cite{reattoreanom,huber2011}.  However this approach only appeared to work because the fit to the neutrino flux of Refs. \cite{reattoreanom,huber2011} is of a form such that, when extrapolated to high energies, the contribution to the $k\sim\mn31/2$ oscillations is small.  This is not because those fluxes are correct at such high energies, indeed above 6.5 MeV they begin to become unreliable as multistage decays of very excited isotopes become important \cite{reattoreanom}.  It is simply a feature of the extrapolation of a particular fitting function beyond its range of validity.  In fact the Gaussian fitting functions used in Refs. \cite{caojun,caojun2,noi} can be extrapolated to arbitrarily high energy and give no discernible contribution at all near $k=\mn31/2$.  

The real high energy neutrino spectrum at these high energies is unknown, but there is no reason to believe that it will be close to a function of a special form with a compactly supported Fourier transform.   Thus even if at some future date the high energy reactor spectrum can be measured precisely, the strong $\mn32$-dependence in the Fourier transform quantities will still be present and so it will not be useful in the determination of the hierarchy.  

In fact the general arguments in Refs. \cite{parke2007,noi} make it obvious that the high energy spectrum cannot contribute to a determination of the hierarchy.  Neutrinos at such high energies have not oscillated enough times to depend observably on the hierarchy.  Thus we conclude that an accurate determination of the high energy neutrino spectrum, in contradiction with the claims in Ref. \cite{oggi}, is unlikely to resolve to strong $\mn32$-dependence which they observed reduces the chance of successfully determining the hierarchy.

This problem greatly reduces the probability of success of a determination of the hierarchy using the variables (\ref{rlpv}), but it does not affect the positions of the peaks, which after all contain all of the information concerning the hierarchy.  The problem is created by a contribution to the analysis parameters (\ref{rlpv}) by neutrinos whose energy is so high that they are not sensitive to the hierarchy.  This is similar to the contribution of unoscillated high energy neutrinos to RENO's determination of $\theta_{13}$ \cite{reno}.  Just as that problem was solved in Ref. \cite{renofrancia} by a cutoff eliminating the spurious signal from high energy neutrinos, here it may be solved by an energy-dependent weight in the Fourier transform which reduces the impact of the high energy part of the spectrum.  We have checked that this can be arranged by simply inserting a scale factor in the Fourier transform, such as exp$(-0.08 E^2/{\rm{MeV}}^2)$, which suppresses the high energies near the cutoff by more than the lower energies which are sensitive to the hierarchy.  

\subsection{Observables}

Now that we have inserted a scale factor in the Fourier transform, the qualitative features of the transformed spectrum depend primarily upon $P_{13}$, which gives a symmetric $F_c(k)$ with a central peak and damped oscillations with a wavenumber of order $L\langle 1/E\rangle$ where $\langle 1/E\rangle$ is the average value of $1/E$.  At medium baselines, of order 40-80 km, this wavenumber is of order $\m21$.  Similarly the $P_{13}$ contribution to the sine transform $F_c(k)$ is antisymmetric about $k=\mn31/2$, with a maximum at higher $k$ and a minimum at lower again.  Again, as one varies $k$ from $k=\mn31/2$ one finds a series of ever shrinking oscillations separated by a characteristic distance of order $\m21$.  At $L=58$\ km analytic approximations to these features are derived in Ref.~\cite{noi}.  As the Fourier transforms are linear, they add the transform of the $P_{23}$ oscillations to that of the larger $P_{13}$ oscillations.

The hierarchy can be determined from the way in which the resulting asymmetric perturbation.  It was observed in Ref.~\cite{caojun} and derived in Ref.~\cite{noi} that in the case of the normal (inverted) hierarchy the $P_{23}$ oscillations render the first minimum $R$ to the right of the global maximum of $F_c$ larger (smaller) than its mirror image $L$ and similarly render the maximum $P$ of $F_{s}$ just to the right of $k=\mn31/2$ larger (smaller) than the minimum just to its left $V$.    Ref~\cite{caojun} introduced two parameters which characterize these effects
\beq
p_1=\frac{R-L}{R+L}\hsp
p_2=\frac{P-V}{P+V} \label{rlpv}
\eeq
finding that positive (negative) values of these two parameters tend to indicate the normal (inverted) hierarchy.

In Ref.~\cite{noi} two more parameters were suggested.  First, the value $\phi$ of $P_s$ at the maximum of $P_c$, which is positive (negative) for the normal (inverted) hierarchy.  Second, the difference in values $m_\pm$ of the maxima of hierarchy-dependent nonlinear Fourier transforms
\beq
F_n^\pm(k)=\int d\left(\frac{L}{E}\right)\frac{E^2}{L}\Phi(L/E) P_{ee}(L/E) g(L/E) \cos\left(k\frac{L}{E}\pm 2\pi\alpha\left(\frac{k}{2\pi}\frac{L}{E}\right)\right).
\eeq
These can be encoded into the normalized observables
\beq
p_3=\frac{\phi}{P+V}\hsp
p_4=\frac{m_+-m_-}{m_-+m_+}.
\eeq

In our analysis we have also introduced one new parameter.  We observed that statistical errors in Fourier space tend to have correlation lengths which are longer that the $\m21$-scale wavelength of oscillations in the Fourier transform of $P_{13}+P_{23}$.  Thus if one peak is erroneously too low or too high, its nearest neighbors are also likely to be too low or too high.  It was shown in Ref. \cite{noi} that as $k$ varies from the center point $\mn31/2$, the analytic (no statistical error) heights of the peaks shrinks exponentially, so the secondary peaks and valleys are much smaller than the primary peaks and valleys, which we define to be those closest to the center.  

The observables above all depend upon the heights of the primary peaks and valleys.  But this correlation means that when a primary peaks or valleys is too high (low) due to a statistical error, the corresponding secondary peak or valley should also be too high (low).  Thus one can reduce the statistical errors by using the size of the secondary oscillations to correct those of the primary oscillations.  This can be implemented automatically by defining a secondary analog of $p_1$
\beq
p_5=\frac{L_2-R_2}{R+L}
\eeq
where $R_2$ and $L_2$ are the depths of the secondary valleys, which are defined to be the minima within a distance $2\m21$ to the right and left of the corresponding primary valleys respectively.  Thus an indicator which is a linear combination of $p_1$ and $p_5$ can be expected to be less susceptible to statistical errors than $p_1$ alone.  Of course secondary analogs of the other indicators can also be introduced, but we expect them to be largely degenerate with $p_5$.

\subsection{Neural Network}

In the previous subsection we identified five quantities $p_i$ which may be used to determine the hierarchy.  The sign of any one of the first four is already a good indicator of the hierarchy, in fact they are quite degenerate.  The fifth indicator $p_5$ alone is an inferior indicator of the hierarchy, but it characterizes the direction of the statistical errors of the others.

Our goal is to determine the best indicator, that which, given a dataset, has the highest probability of correctly determining the mass hierarchy.  As these indicators are not precisely degenerate, the best indicator will not be a single $p_i$ but rather some combination of all 5.  For simplicity we will consider linear combinations
\beq
I=c_0+\sum_{i=1}^{5}c_i p_i \label{secondo}
\eeq 
where $c_i$ are 6  constants.  The constants will be chosen such that, given a set of experiments in which the neutrino mass hierarchy is normal in as many as it is inverted, the chance of success is the highest.  The chance of success is defined to be the average of the percentage of normal hierarchy experiments for which $I$ is positive with the number of inverted hierarchy experiments for which $I$ is negative.  Clearly the overall scale of $c_i$ is irrelevant, but the optimal choice of $c_i$ in general depends on the baseline, the reactors that are operational, the geometry of the reactors with respect to the detector and even the time for which the experiment has run.

We will optimize $c_i$ given a dataset, in general of 500 experiments of each hierarchy, by beginning at $c_i=0$ with a fixed temperature, and then at each step both cooling and moving the vector $c_i$ a temperature-dependent distance in a random direction.  If the probability of success is improves, the new location is the starting point for the next recursion.  After a fixed number of steps the current value of $c_i$ will be fixed.  It will then be applied to another set of 500 experiments of each hierarchy to determine the chance of success.  It is important that a separate set of experiments is used for this determination to avoid overfitting the data, which would lead to an artificially inflated chance of success.  Then the best $c_i$ will be found with the second set of data and the result applied to the first, to obtain a second chance of success.  The chance of success that we will report will be the average of these two numbers.

In general we have found that this 1-layer neural network method leads to an improvement over the $p_1+p_2$ quantity in Ref. \cite{caojun} of 1-2 percent.  This can be increased yet further with a second layer, applied before first layer.  We have seen in Subsec. \ref{flusssez} that in order to avoid the spurious dependence of $p_i$ upon the high energy neutrino spectrum observed in Ref. \cite{oggi} one may include energy dependent weights $w(E)$ in the functionals
\bea
\tilde{F}_c[w(E)](k)&=&\int d\left(\frac{L}{E}\right)w(E)\frac{E^2}{L} N(E)  P_{ee}\left(\frac{L}{E}\right) \cos\left(\frac{kL}{E}\right)\nonumber\\
\tilde{F}_s[w(E)(k)&=&\int d\left(\frac{L}{E}\right)w(E)\frac{E^2}{L} N(E)  P_{ee}\left(\frac{L}{E}\right) \sin\left(\frac{kL}{E}\right).
\eea
Here $E$ is the observed energy, which due to the finite energy resolution of the detector is not precisely equal to the true neutrino energy.

We noted that $w(E)=exp(0.08 E^2/\rm{MeV}^2)$ removes the spurious high energy dependence of Ref. \cite{caojun}.  What about other choices of weights?  In principle, for each Fourier transform one can consider weights to be an function of some arbitrary parameters.  One can then try to optimize these parameters together with the $c_i$, but optimizing so many parameters at once is not possible with the small number of experiments that we have generated.  However, in a first layer of the neural network, for each $i$, except for $i=5$ as $p_5$ does not determine the hierarchy well on its own, one can optimize the parameters in the weight function so as to maximize the chance of success of that $p_i$.  Then one can optimize the $c_i$ for these new and improved $p_i$.

For example, one may choose the weight functions $w(E)$ so as to deemphasize the spurious high energy neutrinos, to accentuate the neutrinos around the 1-2 oscillation maximum or to accentuate the lowest energy neutrinos whose signals are the most strongly hierarchy-dependent.  In realistic detectors in which the nonlinear response is only known in some region, where it has been calibrated by radioactive decays, one may wish to weight those regions more heavily.  This can be done, for example, by using a weight function which contains a Gaussian distribution centered upon the energy range to be accentuated.   The only constraint on the weight function is that its characteristic derivative scale be smaller than $\mn31$, so that it does not itself produce a spurious contribution to the Fourier transform near $k=\mn31/2$ and so to the quantities $p_i$.

One such choice of weight coefficients for the cosine transform is
\beq
F_c\p(k)=\tilde{F}_c[1](k)+a_1\tilde{F}_c[e^{-E/(3\rm{\ MeV})}](k)+a_2\tilde{F}_c[e^{-E^2/(12\rm{\ MeV}^2)}](k). \label{primo}
\eeq
Here the first weighted transform $\tilde{F}_c[1]$ is just the unweighted transform defined in Eq. (\ref{fcos}) while the second weights lower energy neutrinos more heavily and the third is that of Subsec. \ref{flusssez}.  A given $p_i$ can then be computed with the new $F\p$ and the weight coefficients $a_j$ can be optimized for each value of $i$ separately.

Summarizing, for each value of $i$ from 1 to 4 the first layer of the neural network optimizes $a_1$ and $a_2$ of Eq. (\ref{primo}) so as to maximize the probability of success of $p_i$ alone.  Then the second layer optimizes the $c_i$ in Eq. (\ref{secondo}) so as to maximize the probability of success of the indicator $I$.

\section{Optimizing the Baseline and Geometry}

\subsection{The Optimal Baseline}

\begin{table}[position specifier]
\centering
\begin{tabular}{c|l|l|l}
Baseline&$N_{\overline{\nu}_e}$ at 60 ky&Hierarchy at 60 ky&Hierarchy at 120 ky\\
\hline\hline
42 km&64,860&86.5\%\ (80.8\%)&\%\ (91.4\%)\\
\hline
44 km&55,578&86.6\%\ (82.6\%)&\%\ (91.2\%)\\
\hline
46 km&48,030&87.0\%\ (82.3\%)&\%\ (91.6\%)\\
\hline
48 km&41,900&87.8\%\ (86.0\%)&\%\ (94.2\%)\\
\hline
50 km&36,931&88.2\%\ (86.7\%)&\%\ (95\%)\\
\hline
52 km&32,915&88.0\%\ (87.1\%)&\%\ (94\%)\\
\hline
54 km&29,679&87.9\%\ (87.4\%)&\%\ (95\%)\\
\hline
56 km&27,080&\%\ (83.9\%)&\%\ (94\%)\\
\hline
58 km&25,000&\%\ (80.6\%)&\%\ (92\%)\\
\hline
60 km&23,342&83.0\%\ (82.3\%)&\%\ (93\%)\\
\hline
\end{tabular}
\caption{The observed $\overline{\nu}_e$ flux from a single 17.4 GW reactor and the probability of determining the hierarchy as a function of baseline and number of kton years determined using the neural network.  The neural networks trained on the neighboring baselines, and the values of the coefficients $c_i$ used to determine the chance of success are the averages of the coefficients at the available neighboring baselines.  Probabilities in parenthesis are those using $p_1+p_2$ as in Ref.~\cite{caojun}.}
\label{disttab}
\end{table}

As a first application of our simulation we have attempted to determine the optimal baseline in an idealized situation in which all 17.4 GW of thermal capacity are located at a single reactor, at a fixed baseline.  This eliminates the interference effects of Ref. \cite{noi} which will be investigated in Subsec. \ref{intsez}.  The probability of success then depends only upon the baseline, the number of ktons of the detector multiplied by the number of years of observation and the method of data analysis.  We will consider a Fourier transform analysis with two indicators, $p_1+p_2$ which was introduced in Ref. \cite{caojun} and also our neural network.  The results are displayed in Table \ref{disttab}.

\subsection{Interference Between Reactors in the Same Complex} \label{intsez}

No single reactor produces enough flux to determine the neutrino hierarchy in a reasonable amount of time.  Thus it is inevitable that complex of reactors need to be used, and often the distances between these reactors is considerable.  For example in Ref. \cite{weihai} the proposed cite for Daya Bay II is 3.5 km closer to the reactor complexes at Daya Bay and Ling Ao than to that planned at HuiDong.  These different baselines mean that for some energies the neutrinos from one reactor will arrive at the 1-3 oscillation maximum while neutrinos from the other will arrive at the 1-3 oscillation minimum, greatly reducing the amplitude of the 1-3 oscillations whose observation is necessary to determine the hierarchy \cite{noi}.  Note that the neutrinos arriving from different reactors are not coherent, the interference effect results from the addition of probabilities and not wavefunctions.

Such an interference effect is present using just the reactors at Daya Bay and Ling Ao alone, because Daya Bay and Ling Ao I are separated by 1.1 km and Ling Ao II is 400 meters further.   Fortunately these reactors all lie more or less along a line, so a medium baseline detector perpendicular to this line will be the same distance from each reactor, eliminating the interference effect.  In Table \ref{inttab} we have determined the effect of this interference on the probability of success for a detector as a function of its distance from the center of mass of Daya Bay and Ling Ao, its angle with respect to this line and the number kton years of observations.  To illustrate the effect of the angle, the interference from other, more distant, reactors is not considered.  These will be included in the full simulations reported in Sec. \ref{rissez}.

\begin{table}[position specifier]
\centering
\begin{tabular}{c|l|l|l|l|l|l|l|l}
Baseline&kt yrs&$0^\circ$&$15^\circ$&$30^\circ$&$45^\circ$&$60^\circ$&$75^\circ$&$90^\circ$ \\
\hline\hline
50 km&60&\%&\%&\%&\%&\%&\%&\%\\
\hline
50 km&120&\%&\%&\%&\%&\%&\%&\%\\
\hline
58 km&60&\%&\%&\%&\%&\%&\%&\%\\
\hline
58 km&120&\%&\%&\%&\%&\%&\%&\%\\
\hline
\end{tabular}
\caption{The probability of determining the hierarchy as a function of baseline and number of kton years determined using the neural network.  The baseline is the distance from the center of mass of the Daya Bay, Ling Ao I and II reactor complexes considering the distances between these complexes.  The angles are measured with respect to the line that nearly passes through all three complexes.}
\label{inttab}
\end{table}

\section{Comparing Detector Locations Near Daya Bay} \label{rissez}

\subsection{Locations of Reactors and Detectors}

China's Guangdong province is among the best locations for a medium baseline reactor experiment because it contains the powerful reactor complex consisting of 2 reactors at Daya Bay and 4 at Ling Ao and yet it is free from the large reactor neutrino backgrounds caused by many smaller complexes in France and, modulo tsunami induced shutdowns, in Japan.  In the next few years a number of new reactor complexes will come on line in Guangdong.  Two distant complexes, TaiShan and YangJiang, will see their first reactors become operational already next year.  Two closer complexes, HuiDong and LuFeng, have already passed several critical steps in the approval process and may be built.  The relevant reactors are listed in Table \ref{reactortab}.

\begin{table}[position specifier]
\centering
\begin{tabular}{l|l|l|l|l}
Name&Status&Latitude&Longitude&Thermal Power\\
\hline\hline
Daya Bay&Operational&$22^\circ$ 35'  53" N&$114^\circ$ 32' 35" E& 5.8 GW\\
\hline
Ling Ao I&Operational&$22^\circ$ 36' 19" N& $114^\circ$  33' 4" E& 5.8 GW\\
\hline
Ling Ao II&Operational&$22^\circ$ 36' 31" N& $114^\circ$  33' 14" E& 5.8 GW\\
\hline
TaiShan I&Under Construction&$21^\circ$ 55' 9" N& $112^\circ$  58' 57" E& 9.2 GW\\
\hline
TaiShan II&Planned&$21^\circ$ 55'  N& $112^\circ$  59' E& 9.2 GW\\
\hline
YangJiang I&Under Construction&$21^\circ$ 42' 29" N& $112^\circ$  15' 32" E& 5.8 GW\\
\hline
YangJiang II&Under Construction&$21^\circ$ 42' 36" N& $112^\circ$  15' 41" E& 5.8 GW\\
\hline
YangJiang III&Planned&$21^\circ$ 43'  N& $112^\circ$  16'  E& 5.8 GW\\
\hline
HuiDong&Planned&$22^\circ$ 42'  N& $115^\circ$  0'  E& 17.4 GW\\
\hline
LuFeng&Planned&$22^\circ$ 45'  N& $115^\circ$  49'  E& 17.4 GW\\
\hline
\end{tabular}
\caption{Reactors in Guangdong}
\label{reactortab}
\end{table}

Depending on a detector location the new reactors may help or hinder the determination of the neutrino hierarchy.  They will help at each detector which is equidistant from two reactors,  as the fluxes add without interference.  This is the case in the proposed location of Ref \cite{yifangseminario} and also at our proposed location DongKeng.  If the two reactors are nearly at the same distance, as in the proposal of Ref. \cite{weihai} and our proposed location XiKengDing, the addition of the second reactor will increase the flux but interference between the detectors will decrease the amplitudes of the 1-3 oscillations.   

However in general the additional reactors will be so distant that the 1-3 oscillation peaks will be too close together to be distinguishable by a detector with resolution $\sigma_E$.  As a result, the additional reactors will simply supply a background which impedes the measurement of the hierarchy.

A medium baseline detector experiment may also be used to measure $\theta_{12}$.  This can be done by comparing the flux at the $1-2$ oscillation minima and maxima.  In the new reactor is twice as far away as the desired reactor, as is the case for most of the positions that we will consider, then the distant reactor will be at its 1-2 minimum at the same energy range in which neutrinos from the near reactor arrive at their 1-2 maximum.  This means that at the 1-2 minimum, where sensitivity to the hierarchy and to $\theta_{12}$ is maximized, the neutrino flux will be dominated by noise from the distant reactor \cite{noi}.  In general this makes the determination of both the hierarchy and $\theta_{12}$ more difficult.  However the 1-2 oscillations of distant reactors do lie within the resolution of the detector and so these can be used to gain more information about $\theta_{12}$.  Furthermore, if multiple detectors are considered, then they can break the degeneracy between $\theta_{12}$ and flux from distant reactors, allowing a more precise determination of both.

\begin{table}[position specifier]
\centering
\begin{tabular}{l|l|l|l|l|l|l|l|l}
Name&Altitude&Latitude (N) &Longitude (E) &DB&TS&YJ&HD&LF\\
\hline\hline
BaiYunZhang&1000 m&$22^\circ$ 53'  52"&$114^\circ$ 15' 14"& 44.5&170.5&244.1&79.5&161.6\\
\hline
ShiYaTou&500 m&$22^\circ$ 52'  14"&$114^\circ$ 17' 28"& 39.7&171.6&245.7&75.5&157.6\\
\hline
ShuangFeiJi&700 m&$22^\circ$ 54'  19"&$114^\circ$ 10' 0"& 51.6&164.2&237.0&88.3&170.6\\
\hline
SanJiaBi&600 m&$22^\circ$ 54'  8"&$114^\circ$ 10' 41"& 50.6&164.9&237.8&87.3&171.6\\
\hline
XiangTouShan&800 m&$23^\circ$ 15'  24"&$114^\circ$ 21' 0"& 75.4&205.0&275.3&90.4&160.9\\
\hline
BaiMianShi&400 m&$23^\circ$ 6'  27"&$114^\circ$ 37' 2"&  56.3&214.1&288.0&59.4&129.7\\
\hline
DongKeng&200 m&$22^\circ$ 6'  4"&$112^\circ$ 31' 9"& 216.5&51.8&51.0&264.1&347.3 \\
\hline
HuangDeDing&500 m&$22^\circ$ 5'  23"&$112^\circ$ 29' 55"& 218.9 &53.3&48.8&266.5&349.7\\
\hline
\end{tabular}
\caption{Potential locations for Daya Bay II detectors and distances to reactors in km. DB~indicates the distance to the weighted center of Daya Bay and Ling Ao I and II.}
\label{rivelatoretab}
\end{table}

In Table \ref{rivelatoretab} we list the detector positions that we propose together with BaiMianShi, which was proposed in Ref. \cite{weihai}.  These points are chosen to be at a medium baseline, to minimize interference effects, to be underneath mountains to minimize backgrounds and when possible to be a similar distance from multiple reactors to enhance the useful neutrino flux.  No site that we have found simultaneously achieves all of these goals.

\subsection{Monte Carlo Results}

The simulation of planned reactors is impeded by the fact that we do not know the configurations of the reactors.  These configurations are important when the reactors are used as sources of flux because it determines the reduction in the resulting 1-3 peaks due to interference.  It is not important when the reactors provide a background source.  While in general these configurations cannot be predicted, one natural guess is that the reactors will follow the coast, which is in general a sufficiently strong assumption to determine the resulting interference effect.

\begin{table}[position specifier]
\centering
\begin{tabular}{l|l|l|l|l|l|l|l|l}
Name&60 HD LF&60 HD&60 LF&60&120 HD LF&300 HD LF\\
\hline\hline
BaiYunZhang&77.8\%&&&&&\\
\hline
HS+XTS&&&&&&\\
\hline
SanJiaBii&77.0\%&&&&&\\
\hline
ShuangFeiJi&\%&&&&&\\
\hline
BaiMianShi&54.5\%&&&&&\\
\hline
BMS Ideal&79.4\%&&&&&\\
\hline
DongKeng&&&&&& \\
\hline
XiKengDing&&&&&&\\
\hline
\end{tabular}
\caption{Potential locations for Daya Bay II detectors and chances of success for various numbers of kton years of flux and choices of active reactors.  In all columns the reactors at Daya Bay, Ling Ao I and II, TaiShan and YangJiang.  HS+XTS refers to two detectors, one under HuaShan and one under XiangTouShan, each of which is half as large as indicated in the corresponding column.  BMS ideal is a mountainless point 2 km from BaiMianShi which is equidistant from the center of mass of the Daya Bay/Ling Ao complex and HuiDong.}
\label{risultatitab}
\end{table}

\section{Conclusions}

\section* {Acknowledgement}

\noindent
JE is supported by the Chinese Academy of Sciences
Fellowship for Young International Scientists grant number
2010Y2JA01. EC and XZ are supported in part by the NSF of
China.  


\end {document}

10 years ago Petcov and Piai suggested that a 20-25 km baseline neutrino detector may be able to determine the neutrino mass hierarchy if the solar mass splitting $\m21$ is greater than about $10^{-4}\ \mathrm{eV}^2$ \cite{petcovidea}.  It turned out that $\m21$ is smaller than this threshold, meaning that the difference between the normal and inverted hierarchy cannot be seen at a baseline below about 40 km.  The low fluxes at these long baselines led various groups \cite{hawaii,caojun} to conclude that the individual peaks in the neutrino spectrum would be difficult to resolve, and so the only hope for a hierarchy determination would be to sum them via a Fourier transform.  Even in this case the necessary detectors were extremely large and the experiments slow.

This all changed with the recent determination of $\theta_{13}$ \cite{daya,reno} confirming hints last year \cite{doublechooz,globale} that $\spp2213$ is up to 10 times higher than had been considered by the authors of Refs. \cite{hawaii,caojun}.  The larger mixing angle increases the sizes of the oscillations used to determine the hierarchy by an order of magnitude.  This means both that a reactor experiment to determine the hierarchy is now practical and also that the analysis of the optimal baseline and experimental configuration must now be redone.  Indeed, such a reanalysis is urgent as such an experiment will be built soon \cite{caojunseminario,renonuturn,yifangseminario}.

We will perform this updated analysis in two papers, mirroring the structure of Ref. \cite{caojun}.  While the second paper will include a description of an the results of our simulations, indicating for example the optimal baseline, in this first paper we will analyze a medium baseline reactor experiment analytically.  We will derive old observations relating observables of the Fourier transformed spectrum to the neutrino mass hierarchy and we will also provide new methods of determining various combinations of the neutrino masses and the hierarchy.  The optimal method will be a combination of the old and the new to be determined by simulations.  We will also discuss problems which have so far escaped attention in the literature.  

In particular we will discuss the fact that, since reactors are typically separated by of order 1 km in a reactor array, the baselines of neutrinos from different reactors are different.  As a result the oscillations of low energy neutrinos, which perform half an oscillation traveling from one reactor to the next, will be invisible at the detector as the neutrinos from one reactor arrive at the oscillation minimum while the other is at its maximum.  We will show that, due to a new degeneracy between the hierarchy and an effective mass difference, a determination of the neutrino mass hierarchy is impossible in an experiment which cannot resolve these low energy peaks.  Thus the angle between the detector and the lines extending between reactors is an essential variable in the determination of the optimal detector location.  For example, for a linear array of reactors like RENO or Daya Bay plus Ling Ao, this effect can be eliminated if the detector is placed orthogonal to the array.

The greatest sensitivity to the hierarchy arises near the 1-2 oscillation minimum, at about 60 km.  However the flux from a distance reactor at the 1-2 oscillation maximum of 120 km will dominate over the flux of the near reactor in this region.  In fact, this is will be the case for a detector placed 60 km away from Daya Bay and Ling Ao in the orthogonal direction, as the proposed Haifeng reactor will be near the 1-2 maximum.  Also if a detector is placed equidistant from Daya Bay and Haifeng at the position suggested in Ref. \cite{caojunseminario} then the 1-2 minimum neutrinos from Daya Bay and Haifeng will correspond to the 1-2 maximum for neutrinos from the proposed reactor at Lufeng \cite{yifangseminario}.  This not only makes the determination of the hierarchy more difficult, but also is detrimental to the sensitivity to $\theta_{12}$.  The ideal solution to this problem is to use two detectors at different distances, say 40 and 70 km.  However a more economical solution is to keep the baselines short so that the flux from the desired reactor complex dominates over the fluxes from others, for example one can consider a baseline of 45-50 km. 

We will begin in Sec. \ref{teorsez} with a review of standard results on 3-flavor neutrino oscillations and the electron antineutrino survival probability.  We describe the interference between the 1-3 and 2-3 oscillations \cite{petcovidea} which leads to beats and we numerically find the combined peaks.  Then in Sec. \ref{masssez} we describe how the positions of various peaks can be used to obtain various combinations of the neutrino mass differences.  Each peak provides a different combination.  And we will see that the first 10 peaks do not allow a determination of the mass hierarchy, but the next 5 do, which is why short baselines are not sufficient for a determination of the hierarchy.  In Sec. \ref{vincsez} we discuss the consequences of the finite energy resolution and the finite neutrino flux.  Both are provide obstacles to locating the $n>10$ peaks, and so for determining the hierarchy.  In Sec. \ref{fouriersez} we discuss analyses of the Fourier transform of the neutrino spectrum.  These have the advantage that, when the nonlinearity of the detector response is well understood, they sum the peaks together, and so render the signal stronger. We rederive three old ways in which the hierarchy can be determined from this transformed spectrum and add two new methods to the list. 

Finally in Sec. \ref{intsez} we discuss the consequences of the fact that not all of the neutrinos detected traveled the same distance.  The distances may differ by of order a kilometer because the individual reactors in an array are not coincident, leading to an interference effect which greatly diminishes the amplitudes of the low energy, high $n$, peaks. We will see that this interference can be avoided if the detector is placed perpendicular to the array.   Also, while it has long been known \cite{hawaii} that neutrinos from reactors at the 1-2 oscillation minimum baseline are the most useful for determining the hierarchy, we will see that this signal can be overwhelmed by neutrinos from distant reactors at the 1-2 maximum, and we see that this is indeed the case if a detector is placed at the 1-2 minimum orthogonal to the Daya Bay, Ling Ao reactor array.  This problem, as well as the related error in a determination of $\theta_{12}$, can be reduced by shortening the baseline or, if possible, adding another detector at a different baseline.

\section{The electron neutrino survival probability} \label{teorsez}

\subsection{Short and long oscillations}

The electron neutrino weak interaction eigenstate $|\nu_e\rangle$ is not an energy eigenstate $|k\rangle$, but it can be decomposed into a real sum of energy eigenstates
\beq
|\nu_e\rangle=\c12\c23|1\rangle+\s12\c13|2\rangle+\s13|3\rangle.
\eeq
In the relativistic limit, after traveling a distance $L$, the survival probability of a coherent electron (anti)neutrino wavepacket with energy $E$ can be expressed in terms of the mass matrix $\mathbf{M}$
\bea
P_{ee}&=&|\langle\nu_e|\mathrm{exp}\left(i\frac{\mathbf{M}^2L}{2E}\right)|\nu_e\rangle|^2\label{pee}\\
&=&\sp413+\cp412\cp413+\sp412\cp413+\frac{1}{2}(P_{12}+P_{13}+P_ {23})\nonumber\\
P_{12}&=&\spp2212\cp413\cos\left(\frac{\m21L}{2E}\right)\hsp
P_{13}=\cp212\spp2213\cos\left(\frac{\mn31L}{2E}\right)\nonumber\\
P_{23}&=&\sp212\spp2213\cos\left(\frac{\mn32L}{2E}\right)\nonumber
\eea
where $\m{i}{j}$ is the mass squared difference of mass eigenstates $i$ and $j$.  Notice that the survival probability is a sum of cosines and so its cosine Fourier transform with respect to the variable $L/E$ is just a sum of delta functions whereas its sine transform vanishes.

The three cosines in the survival probability (\ref{pee}) identify two characteristic frequencies of the $L/E$ oscillations.  The $P_{12}$ term oscillates at a low frequency 
\beq
\frac{\m21}{2}\sim 3.8\times 10^{-5}\ \mathrm{eV}^2\sim 0.17\ \mathrm{MeV/km} .
\eeq
Therefore the maximum 1-2 oscillation occurs at
\beq
\frac{L}{E}=\frac{\pi}{\m21/2}\sim 18 \ \mathrm{km/MeV}. \label{unodue}
\eeq
A medium baseline reactor experiment, with a baseline of under 100 km,  can observe at most one or two such oscillations.  Instead such experiments will focus on shorter oscillations, characterized by the $P_{13}$ term, which have frequency
\beq
\frac{\mn31}{2}\sim 1.2\times 10^{-3}\ \mathrm{eV}^2\sim 5.5\ \mathrm{MeV/km} 
\eeq
corresponding to a wavelength of
\beq
\Delta\left(\frac{L}{E}\right)=\frac{2\pi}{\mn31/2}\sim 1.1 \ \mathrm{km/MeV}. \label{corto}
\eeq
At a medium baseline reactor one may hope to see 5 to 15 such oscillations.

What about the $P_{23}$ term  in (\ref{pee})?  The frequency is $\mn32/2$, which is about 3\% more or less than $\mn31/2$ depending on the hierarchy.  However the amplitude is less than that of $P_{13}$ by a factor of
\beq
a=\tp212\sim 0.5.
\eeq
As the frequencies of the two short oscillations are similar but $P_{23}$ has a smaller amplitude, the total short distance oscillation $P_{13}+P_{23}$ is a deformation of $P_{13}$ alone.  However the 2-3 oscillations serve to slightly displace the 1-3 peaks and shift the amplitudes with a pattern which repeats at the beat frequency $\m12/2$.

More precisely, while the $n$th maximum of $P_{13}$ is at $L/E=4\pi n/\mn31$, the $n$th peak of $P_{13}+P_{23}$ is at
\beq
\frac{L}{E}=\frac{4\pi}{\mn31}(n+\an) \label{pichi}
\eeq
for a vector $\an$ which is determined entirely by the neutrino mass matrix.  As the derivatives of the neutrino spectrum and the 1-2 oscillations are small, the peaks of the total survival probability $P_{ee}$ and even its product with no-oscillation neutrino spectrum, which is the observed neutrino spectrum, are roughly located at the values given in Eq. (\ref{pichi}).

From Eq. (\ref{pichi}) it is possible to see the main obstruction to determinations of the hierarchy from near reactors.  The position of the $n$th peak is only sensitive to the mass combination
\beq
\Delta M^2_{eff}=\frac{\mn31}{1\pm\an/n}.
\eeq
As we will see, at low $n$, corresponding to peaks visible at relatively short baselines, the values of $\an$ are roughly linear.  Thus in this regime the positions of all of the peaks are sensitive to the same effective mass and so are independent of the hierarchy so long as that effective mass applies. {\textbf{The hierarchy can only be determined from the nonlinearity of $\an$.}}

\subsection{Finding $\an$}

The vector $\an$ encodes the effect of the neutrino mass matrix on the location of the survival probability peaks.  Therefore a knowledge of this function together with a measurement of the peaks allows one to reconstruct some elements of the mass matrix.

The values of $\an$ are determined from (\ref{pichi}) by the extrema of $P_{13}+P_{23}$
\bea
0&=&\frac{\partial}{\partial E}(P_{13}+P_{23})\propto \frac{\partial}{\partial E}\left[\cos\left(\frac{\mn31L}{2E}\right)+\tp212\cos\left(\frac{\mn32L}{2E}\right)\right]\\
&\propto&\sin\left(2\pi\an\right)+\left(1\pm\epsilon\right)\tp212\sin\left(2\pi[\an
\pm\epsilon(n+\an)]\right)\hsp
\epsilon=\frac{\m21}{\mn31}\nonumber
\eea
where the $-$ sign applies to the normal hierarchy and the $+$ sign to the inverse hierarchy.

As $\epsilon<<1$ and $\an<<n$ we may approximate
\beq
0\sim \sin(2\pi\an)+\tp212\sin(2\pi[\an\pm\epsilon n]). \label{alfeq}
\eeq
This can be expanded in a power series in $n$.  The highest energy peaks occur at small values of $n$, where the linear term in this expansion suffices
\beq
0\sim 2\pi(1+\tp212)\an\pm 2\pi\tp212\epsilon n
\eeq
which is easily solved for $\an$
\beq
\an\sim\mp\epsilon\sp212 n\sim \mp 0.011 n.
\eeq

Thus the n$th$ peak, for $n$ sufficiently small, lies at
\beq
\frac{L}{E}\sim\frac{4\pi}{\mn31}(1\mp\epsilon\sp212) n=\frac{4\pi n}{\mn31\pm\sp212\m21} \label{degen}
\eeq
where the lower sign corresponds to the normal hierarchy.  

The basic problem facing shorter baseline experiments, which are only sensitive to peaks at small $n$, is already apparent in Eq. (\ref{degen}).  The mass difference $\mn31$ is degenerate with the hierarchy
\beq
\mn31(\mathrm{direct})= \mn31(\mathrm{inverse})+2\sp212\m21 .
\eeq
Therefore any experiment with such a short baseline that all observable peaks have $n$ in the regime in which $\alpha$ is linear are, alone, incapable of determining the hierarchy.  They can, however, determine the combination 
\beq
\Delta M^2_{eff}=\mn31\pm\sp212\m21=\cp212\mn31+\sp212\mn32. \label{meff}
\eeq

Therefore reactor experiments can determine the hierarchy in only two ways.  Either an accurate determination of $\Delta M^2_{eff}$ can be combined with an accurate determination of another combination of the mass differences obtained from another experiment, a difference which needs to be known more precisely than the atmospheric mass difference measured by MINOS, or else the experiment needs to be sensitive to peaks at large enough $n$ that the linear approximation breaks down.  So just how large does $n$ need to be?

\subsection{Cubic terms is $\an$} \label{cubicosez}
To see where interference between 2-3 and 1-3 oscillations push the peaks of $P_{13}+P_{23}$ at larger values of $n$, we need to expand \an to cubic order
\beq
\an\sim  \mp\epsilon\sp212 n+b n^3. \label{cubico}
\eeq
and to substitute this expansion into Eq. (\ref{alfeq}).   The linear term in $\an$ already solves this equation at linear order.  At cubic order it yields
\beq
0\sim 2\pi(1+\tp212)b \pm\frac{4\pi^3}{3}\epsilon^3\sp212(\sp212-\cp212)       
\eeq
and so
\beq
b\sim\mp\frac{2\pi^2}{3}\epsilon^3\sp212\cp212(\sp212-\cp212) \sim \pm 2\times 10^{-5} .
\eeq

The linear approximation to $\an$ is reliable when the cubic term in Eq. (\ref{cubico}) is much smaller than the linear term
\beq
n<<\sqrt{\frac{\epsilon\sp212}{b}}\sim 20.
\eeq
For example, at the 10th peak the contribution of the cubic term to the energy is only one quarter of that of the linear term.  The linear term we have seen shifts the effective mass by about 1\%, thus the cubic term, which is the leading hierarchy-dependent term, only shifts the energy of the tenth peak by about one quarter of a percent.  The other hierarchy would lead to a shift of a quarter percent in the other direction, so overall the difference in the energies of the 10th peaks in the normal and inverted hierarchy is about one half percent.

Therefore if only the first ten peaks can be measured at a given baseline, the detector will need to be able to determine the position of the tenth peak with a precision of a half percent for only a one sigma determination of the hierarchy, making a determination of the hierarchy using such an experiment alone quite unlikely.

\begin{figure} 
\begin{center}
\includegraphics[width=5in,height=2in]{alfa.pdf}
\caption{$\an$ for the normal hierarchy}
\label{anfig}
\end{center}
\end{figure}

The cubic expansion is no longer reliable for higher peaks, but $\an$ can be determined numerically.  As seen in Fig. \ref{anfig} in the case of the normal hierarchy, it is periodic moduli $1/\epsilon$ and is zero at every multiple of $1/2\epsilon$.  But the main problem is that it is nearly linear for $n<10$, which is why the hierarchy is so nearly degenerate with the shift in the mass differences at all of these peaks.

A $2/k$ percent precision measurement of the peak energy of the the 14th peak would also give a $k$ sigma indication of the hierarchy.  The 14th peak can only be seen if it occurs at a high enough energy that the neutrino flux and detector resolution are sufficient to discern it.  For example if one requires it to appear at at least 3 MeV, so that the detector resolution may be better than 2\%, then using Eq. (\ref{corto}) the minimum baseline is about
\beq
L_{\mathrm{min}}\sim 45\ \mathrm{km}.
\eeq

\section{Determining mass differences from peak positions} \label{masssez}

\subsection{The reactor neutrino energy spectrum}

In the previous section we saw that the locations of the aperiodicity of the peaks is determined by the neutrino mass spectrum.  In particular, in $L/E$ space the peak positions are periodic modulo $1/\epsilon\sim 32$ and each set of $1/2\epsilon$ peaks is displaced about 1 percent either left or right depending on the hierarchy.  Thus, to determine the hierarchy, it suffices to measure the positions of enough peaks to within 2 percent precision.

The trouble with such a procedure is that no single experiment has access to all of the peaks, as one effectively has access to only a single baseline per detector and the energy spectrum is limited to that which is produced by a nuclear reactor, which effectively leads to a maximum usable neutrino energy.  Not even all of these energies are accessible as each type of neutrino detector has a minimum energy which it is able to detect.   To determine which peaks may be seen by a particular experiment, one must combine both of these constraints.

The neutrino flux from a reactor results almost entirely from decays of just 4 isotopes: \u35,\ \pu39,\ \pu41\ and \u38.  The flux $\phi_i(E)$ of neutrinos from each isotope $i$ is traditionally approximated as the exponential of a polynomial in the neutrino energy $E$ \cite{vogelengel}
\beq
\phi_i(E)=\mathrm{exp}\left(\sum_{k=1}^{m} a_{ki}E^{k-1}\right) \label{polinom}
\eeq
Theoretical errors on these fluxes are often claimed to be near the 2-3\% level, although recent theoretical fluxes \cite{nuovoflusso} appear to be about 6\% higher than the fluxes measured at very short baseline experiments \cite{reattoreanom} and at 1 kilometer experiments \cite{noiunokm}.  

The precision of such a phenomenological law depend on the degree $n$ of the fit polynomial.  For neutrinos with between 2 and 7.5 MeV, which will be the ones of interest in this note, it was shown in Ref. \cite{huber2004} that a quadratic ($m=3$) fit tends to introduce errors of order 2-3\% whereas a 6 parameter ($m=6$) fit introduces errors of order 1\%, well below the error in the theoretical flux.  The differences between these parameterizations are oscillations over a characteristic scale of 1-2 MeV, which are likely too broad to give a false signal for a 1-3 oscillation peak, but may well disguise the depth of 1-2 oscillations and so affect the measured value of $\theta_{12}$.  

The most recent theoretical estimate of the flux is Fig 53 of Ref. \cite{cartabianca}, which shows a systematic 3\% excess over last year's estimates \cite{nuovoflusso} at energies about 6 MeV and a one percent deficit beyond 4 MeV, which is well within the theoretical errors of the calculations.  This correction again will affect a single detector determination of $\theta_{12}$.

Not all of the neutrinos which are generated by the reactor will be measured.   The maximum number of neutrinos which can be measured at a given energy $E$ is the product of the produced flux with the fraction of neutrinos at that energy which can be measured.  For example, if neutrinos are measured via the inverse $\beta$ decay reaction
\beq
\overline{\nu}_e+p\rightarrow n+e^+
\eeq
then the maximum number of neutrinos detected is the flux/area at the baseline $L$ multiplied by the inverse $\beta$ decay cross section which at tree level is \cite{sezionedurto}
\beq
\sigma(E)=0.0952\times 10^{-42}\mathrm{cm}^2(E_e p_e/\mathrm{MeV}^2) \label{albero}
\eeq
where the positron energy and momenta are, ignoring the neutron recoil, given in terms of the neutron, proton and electron rest masses
\beq
E_e=E-m_n+m_p+m_e\sim E-780\ \mathrm{keV}\hsp
p_e=\sqrt{E_e^2-m_e^2}. \label{posenergia}
\eeq

\begin{figure} 
\begin{center}
\includegraphics[width=5in,height=2in]{Spettro_NO.pdf}
\caption{The theoretical reactor neutrino spectrum as measured with inverse beta decay.}
\label{flussoorig}
\end{center}
\end{figure}

Combining the reactor flux (\ref{polinom}), with the coefficients of Ref. \cite{vogelengel}, with the tree level cross section (\ref{albero}), one obtains the theoretical reactor flux, depicted in Fig. \ref{flussoorig}.  While inverse $\beta$ decay is kinematically forbidden if $E<m_n+m_e-m_p\sim 1.8$\ MeV, it can be seen in Fig. \ref{flussoorig} that the flux/energy is maximized at 3.6 MeV, falls to one third of its maximum by 6 MeV and to 10\% of its maximum by 7.5 MeV.  Thus useful information can only be obtained about the spectrum for energies within a factor of 2, which for each detector corresponds to values of $L/E$ within about a factor of 2.

\subsection{Effective masses at various baselines}

When neutrino oscillations are included, the theoretical flux $\Phi(E)$ from a reactor at a distance $L$ is then multiplied by the survival probability $P_{ee}$ given in Eq. (\ref{pee})
\beq
\Phi(E)=\sum_i c_i \phi_i(E)\sigma(E) P_{ee}(L/E)
\eeq
where $c_i$ is the quantity of each isotope in the reactor.

Using $\m21$ and $\spp2212$ from Ref. \cite{gando}, $\mn32$ determined by combining neutrino and antineutrino mass differences from Ref.\cite{minosneut2012}  and $\spp2213$ from \cite{noiunokm} 
\beq
\m21=7.50\times 10^{-5}\hspp
\mn32=2.41\times 10^{-3}\hspp
\spp2212=0.857\hspp
\spp2213=0.096
\eeq
where $\mn31$ is determined using the normal and inverted hierarchies, this total neutrino flux is given in Fig. \ref{tuttiflussi} at baselines of 40, 50, 60 and 70 km.

\begin{figure} 
\begin{center}
\includegraphics[width=3.2in,height=2in]{Plot_40_no_shift.pdf}
\includegraphics[width=3.2in,height=2in]{Plot_50_no_shift.pdf}
\includegraphics[width=3.2in,height=2in]{Plot_60_no_shift.pdf}
\includegraphics[width=3.2in,height=2in]{Plot_70_no_shift.pdf}
\caption{Theoretical neutrino fluxes, including 3 flavor oscillation, for both hierarchies as seen at 40, 50, 60 and 70 km.}
\label{tuttiflussi}
\end{center}
\end{figure}

The numbers $n$ of the local maxima can be read from (\ref{pichi}) by approximating $\mn31\sim\mn32$ and setting $\an=0$
\beq
n\sim\frac{4\pi}{\mn32}\frac{L}{E}\sim 0.9 \frac{L/\textrm{km}}{E/\textrm{MeV}}. \label{nvalore}
\eeq
As the error on $\mn32$ is of order 3\% and the difference between $\Delta M^2_{eff}$ and $\mn32$ is also of order 2\%, one may expect an error of 3-5\% in Eq. (\ref{nvalore}).  The fractional energy difference between the $n$th and $(n+1)$st peak is $1/n$.  Therefore an optimal detector can determine $n$ given a single peak only if $n$ is less than about 20, whereas an ordinary detector, ideally calibrated by radioactive decays close to the energy in question, can reliably determine the value $n$ of a peak when $n\leq 10$. 

Although the values of peaks in Fig \ref{tuttiflussi} are strongly dependent upon the hierarchy at every baseline, this does not mean that a measurement of the spectra can actually allow one to determine the hierarchy.  The problem, as was described in Sec. \ref{teorsez}, is that the energies of the first 10 or so peaks only determine the mass difference $\Delta M^2_{eff}$ of Eq (\ref{meff}).  

\begin{figure} 
\begin{center}
\includegraphics[width=3.2in,height=2in]{Plot_40_sovrapposto.pdf}
\includegraphics[width=3.2in,height=2in]{Plot_58_sovrapposto.pdf}
\caption{Theoretical neutrino fluxes, including 3 flavor oscillation, at 40 and 58 km for both hierarchies with the same value of $\Delta M^2_{eff}$.  In the left panel, at 40 km, the hierarchies are difficult to distinguish because $\an$ is nearly linear at the visible peaks.  This degeneracy is broken by the higher $n$ peaks visible at 58 km, seen in the right panel.}
\label{degenfig}
\end{center}
\end{figure}

Thus if the baseline is low enough so that only these peaks may be reliably measured, then there will be a value of $\Delta M^2_{eff}$ that reproduces the peaks for both hierarchies, as seen at 40 km in the first panel of Fig. \ref{degenfig}.   Here the peak $n=5$ can barely be seen at 6.6\ MeV, whereas $n=6$ at 5.5 MeV is clearly discernible.  The largest peaks are $n=7,\ 8,\ 9$ amd $10$.  However the energies of these peaks are independent of the hierarchy at constant $\Delta M^2_{eff}$.  The hierarchy-dependence becomes somewhat larger at the 11th peak, which is located at about 3.35 MeV in the case of the normal hierarchy and 3.30 MeV in the case of the inverted hierarchy.  Thus if $\Delta M^2_{eff}$  peaks can be determined at better than 1\% from the low $n$ peaks and then the location of the 11th peak can be determined with a precision of better than 1\%, a determination of the hierarchy would be barely possible.  The lower energy peaks are more smaller, but more hierarchy dependent.  For example, the 15th peak would be at 2.50 MeV with the normal hierarchy, but 2.42 MeV with the inverted hierarchy.  This 3\% difference is well within the resolution of the proposed detectors of Refs. (\cite{caojun}), and so when combined with an accurate measurement of $\Delta M^2_{eff}$ one may could potentially determine the hierarchy at the 1-2$\sigma$ level.



In the second panel one can see the electron neutrino survival probability at 58 km with both hierarchies and the same value of $\Delta M^2_{eff}$.  At this long baseline one can see maxima up to $n=20$, where the nonlinearity in $\an$ is appreciable and so, as one can see, the low and mid energy peaks are hierarchy-dependent at the 1-2\% level.  

\subsection{The 1-2 oscillation minimum}
It is clear from Fig. \ref{degenfig} that if $\Delta M^2_{eff}$ is determined from the low $n$ peaks then the locations of the peaks at the 1-2 oscillation minimum
\beq
n=\frac{1}{2\epsilon}\sim 15\hsp
E=\frac{L}{18\mathrm{\ km}}\textrm{MeV}
\eeq
are hierarchy-dependent, and so one may hope to use their locations to determine the hierarchy.  This can be done if the peak locations allow for a combination of the mass differences which is distinct from $\Delta M^2_{eff}$.  In fact, two such combinations can be determined, one from the location of the peaks and one from the distance between them.

To derive these two combinations, we will need to find $\an$  near the 1-2 oscillation minimum.  This is easily obtained by expanding (\ref{alfeq}) about $n=1/2\epsilon$
\beq
0\sim \sin(2\pi\an)-\tp212\sin\left(2\pi\left[\an\pm\epsilon \left(n-\frac{1}{2\epsilon}\right)\right]\right)
\eeq
where again the positive sign applies to the inverted hierarchy.  Linearly expanding the sine function yields
\beq
0\sim (1-\tp212)\an\mp\epsilon \left(n-\frac{1}{2\epsilon}\right)
\eeq
and so for $n\sim 1/2\epsilon$
\beq
\an\sim\pm\epsilon \frac{n-\frac{1}{2\epsilon}}{ 1-\tp212}.  \label{anminimo}
\eeq

At $n=1/2\epsilon$, $\an=0$.  Of course $n$ must be an integer, so it can never be precisely equal to $1/2\epsilon$.  Nonetheless, $\an$ will be over order $0.01$ when $n$ is at the closest integral value, leading to a less than 0.1\% contribution to the energy.  When $\an=0$, the energy of the peak is 
\beq
E=\frac{4\pi n L}{\mn31}=\frac{2\pi L}{\m21}
\eeq
allowing for a precise determination of $\m21$.  Of course, to know that one is at the 1-2 minimum by counting peaks, one must know $\epsilon$ and so this is related to a measurement of $\mn31$.  Alternately, one can find the 1-2 minimum by looking at the large oscillations in the flux due to 1-2 mixing and then count peaks to determine $n$ at the minimum, which allows for a determination of $\epsilon$ and so $\mn31$ from $\m21$.

The distance between the peaks near the 1-2 minimum allows for a determination of the mass differences which is independent of precise knowledge of the 1-2 oscillation parameters.  Inserting (\ref{anminimo}) in (\ref{pichi}) one finds that the distance between two peaks is
\beq
\Delta\left(\frac{L}{E}\right)=\frac{4\pi}{\mn31}\left(1\pm\frac{\epsilon}{1-\tp212}\right)
\eeq
and so it determines the effective mass
\beq
\Delta M^2_{min}=\left(1\mp\frac{\epsilon}{1-\tp212}\right)\mn31=\frac{2-\tp212}{1-\tp212}\mn31-\frac{1}{1-\tp212}\mn32 .
\eeq 

This effective mass is quite different from that defined in Eq. (\ref{meff}).  Approximating $\tp212=1/2$ one finds that the spacing between the first 10 peaks yields an effective mass
\beq
\Delta M^2_{eff}=\frac{2}{3}\mn31+\frac{1}{3}\mn32 \label{meffdue}
\eeq
while the peaks between $n=14$ and $n=18$ yield
\beq
\Delta M^2_{min}=3\mn31-2\mn32 . \label{mmindue}
\eeq
An detector at a baseline of less than 50 km can accurately determine the combination (\ref{meffdue}) while one with a baseline between 45 and 70 km, as it sees higher $n$ peaks, may more easily measure the combination (\ref{mmindue}).   The normal mass hierarchy is equivalent to the second mass being larger than the first, and so by comparing these masses one can determine the hierarchy.  

The difference between these masses is quite large, about 7\%, however a 7\% measurement of $\Delta M^2_{min}$ requires a 7\% precision in the measurement of the difference between the the peaks in the linear regime near to the 1-2 oscillation minimum $n=1/2\epsilon$.   As can be seen in Fig \ref{anfig}, this linear regime includes about 8 peaks, corresponding to a 40\% variation in energy.  Therefore a 7\% precision in the distance between the peaks requires a 4\% precision in the energy of the peaks, much less than was required at shorter baselines.  Therefore a comparison of the low $n$ peak positions and the 1-2 minimum peak separations is a promising test of the hierarchy, so long as the statistics are sufficient for the peaks to be observed.

\section{Resolution and flux constraints} \label{vincsez}

\subsection{Energy resolution}

The energy of the neutrino $E$ is determined from the positron energy $E_e$ by subtracting 780 keV (\ref{posenergia}).   The positron energy is determined by counting photoelectrons in a scintillator.  The number of photoelectrons is proportional to the positron energy, and so the energy resolution $\sigma_E$ is proportional to the square root of the positron energy.  For example, at Daya Bay II it has been suggested in Ref. \cite{caojun} that a resolution of
\beq
\sigma_E=0.03\sqrt{(E_e){\mathrm{MeV}}}.
\eeq
The observed positron  energy spectrum is then the convolution of the true spectrum with a Gaussian of width $\sigma$
\beq
P(E_e^{\textrm{observed}})=\int dE_e P(E_e) e^{-\frac{(E_e-E_e^{\textrm{observed}})^2}{2\sigma_E^2}}.
\eeq
Both spectra are displayed in Fig. \ref{smearfig}.

\begin{figure} 
\begin{center}
\includegraphics[width=3.2in,height=2in]{Smear_40.pdf}
\includegraphics[width=3.2in,height=2in]{Smear_50.pdf}
\includegraphics[width=3.2in,height=2in]{Smear_60.pdf}
\includegraphics[width=3.2in,height=2in]{Smear_70.pdf}
\caption{The true neutrino spectrum and the measured spectrum with a resolution of $3\%/E$ using the normal hierarchy at baselines of 40, 50, 60 and 70 km.  Notice that the lowest energy oscillations are smeared away and the energy threshold for this smearing increases with the baseline, such that at higher baselines the maximum $n$ observable increases slowly.}
\label{smearfig}
\end{center}
\end{figure}

Even the observed spectrum is never observed.  It is the best that can be hoped for, once the response of the detector is understood and with an infinite number of neutrinos.  Finite flux effects will be briefly discussed in Subsec. \ref{flussosez} and then discussed in detail in the companion paper on our simulation results.  A poorly understood detector response is fatal to the Fourier analysis that will be discussed in Sec. \ref{fouriersez}, but individual peaks may be analyzed where the response is understood.

Even in this idealized setting, it is clear from Fig. \ref{smearfig} that the low energy (high $n$) peaks cannot be resolved.  How high is $n$ for the biggest peak that can be resolved?  This depends on the baseline and the neutrino flux.  However a rough answer is obtained by asserting that the distance between the $n$th maximum and the adjacent minimum (\ref{corto}) 
\beq
\Delta E=0.9\frac{L/\mathrm{km}}{n}\mathrm{MeV}-0.9\frac{L/\mathrm{km}}{n+1/2}\mathrm{MeV}= 0.45 \frac{L/\mathrm{km}}{n^2}\mathrm{MeV}
\eeq
be greater than
\beq
2\sigma =0.06\sqrt{(E_e){\mathrm{MeV}}}=0.06\mathrm{\ MeV}\sqrt{\frac{L/\mathrm{km}}{n}-0.8}.
\eeq
These two quantities are equal when 
\beq
L/\mathrm{km}=8\times 10^{-3}n^3\left(1\pm\sqrt{1-\frac{225}{n^2}}\right). \label{lnecc}
\eeq
The equality with the positive sign yields the upper bound on observable $n$ for a given baseline $L$, whereas the other typically yields a value of $n$ at energies where no reactor neutrinos are observed.  

Alternately (\ref{lnecc}) provides the baseline $L$ necessary to observe the $n$th peak.  When $n\leq 15$ it provides no bound at all, so long as there is enough flux and the detector response is well understood, the energy resolution is not an obstruction for observing these peaks.  Of course at low baselines they may be difficult to observe because there simply are not many or any neutrinos observed at the corresponding energy.  At $n=16$, $17$ and $18$\ one finds minimum baselines of 45 km, 58 km and 72 km respectively.  Of course the true minimum depends on the neutrino flux.  But this rough estimate shows an essential point, that with a $3\%/\sqrt{E}$ fractional resolution any medium baseline, between about 40 and 70 km, is sufficient to observe the 1-2 oscillation minimum if there is enough neutrino flux.  Longer baselines only marginally extend the reach to higher peaks, although each of these peaks is in the 1-2 minimum region and so even just 1 or 2 more peaks can greatly enhance the possibility of determining the mass hierarchy.   

While maximum observable $n$ is reasonably independent of the baseline, this derivation shows that it is strongly dependent upon $\sigma_E$.  A resolution comparable to that of Daya Bay or RENO would lead to a maximum $n$ which is still in the linear region of $\an$, and so the hierarchy would be unobservable.

\subsection{How much neutrino flux is required?} \label{flussosez}

The neutrino flux that can be observed by a large, distant detector via inverse beta decay is still quite unknown, even the efficiency of the detector is difficult to predict.  A rough estimate can be made using the flux observed at Daya Bay, using the flux normalization-independent determination of $\t13$ to eliminate the loss due to 1-3 operation.  At Daya Bay 17.4 GW of thermal power yields 80 neutrinos/day at each 20 ton detector at a weighted baseline of 1600 meters.  Therefore the total measured flux/year from a $P$ GW reactor complex measured at a detector of mass $M$ at a baseline $L$ is roughly
\beq
\Phi=365\times 80\times \frac{P}{17.4\ \mathrm{GW}}\frac{M}{0.02\ \mathrm{ktons}}\frac{2.6\ \mathrm{km}^2}{L^2}=2.2\times 10^5\frac{(P/\mathrm{GW})(M/\mathrm{ktons})}{(L/\mathrm{km})^2}. \label{totflusso}
\eeq
For example, at a distance $L$ from the Daya Bay and Ling Ao reactors a 20 kton detector may observe $7.6\times 10^7/L^2$ neutrinos/year, where the length $L$ is measured in kilometers.

Let the energy width of a peak be $\Delta E$, the integrated flux within that energy range be $\phi$ and the peak be a fraction $A$ higher than the flux in that energy range from the reactor in question after 1-2 oscillations plus all other reactors.   In the absence of other reactors the fraction $f$ varies from $\spp2213$ at low $n$ to $4\spp2213$ at the 1-2 minimum.  The fractional error in the flux in that range will be $1/\sqrt{\phi}$.  Therefore the peak may be observed if $\sqrt{\phi} A>1$.  

Simply observing the peaks is useful for two reasons.  First of all, the distance between peaks is roughly $L\mn31/\pi$ and so by identifying peaks, one can check the consistency of the location of other peaks.  Second, recall that it is easier to determine the $n$ value of the well-separated, high energy, low $n$ peaks.  If one can observe the peaks from low to high $n$ then it is possible to count them and so determine the $n$ values of the low energy peaks.  While the hierarchy can be determined from the distance between the high $n$ peaks, as described in the previous subsection, without knowing the precise value of $n$, nonetheless if one knows the $n$ value of a high $n$ peak it can be used to determine a combination of the mass differences via (\ref{pichi}).  This determination has a precision of $1/n$, and so at high $n$ it leads to a precise determination.

However, to determine the hierarchy it is not enough to observe the peaks, one must determine their energies as precisely as possible.  How precisely may they be determined?  They may be determined within an energy $\delta E$ if the neutrino surplus can be seen in width $\delta E$ bands within the peak.  This requires
\beq
1<\sqrt{\phi}\frac{A\delta E}{\Delta E}=
\frac{\sqrt{\phi}Af}{n}
\eeq
where $f$ is the fractional energy precision desired.  This implies that the maximal fractional precision with which the energy of a peak can be measured by a detector with perfect resolution is
\beq
f>\frac{1}{An\sqrt{\phi}}.
\eeq
As mentioned above, with no unwanted backgrounds from other reactors, $A$ varies between $0.1$ at the 1-2 maxima to $0.4$ at the minima.  

Consider for example the tenth peak at a 40 km baseline, which lies at 3.6 MeV.  At this point $A\sim 0.2$.  The flux within the peak is about 5\% of the total flux (\ref{totflusso}), which each year at a 20 kton detector at 40 km from Daya Bay may be
\beq
\phi=0.05\times 7.6\times 10^7/(40)^2=2.4\times 10^3.
\eeq
Therefore, after $m$ years, the best resolution of an ideal detector would be
\beq
f_{min}=\frac{1}{0.2\times 10\sqrt{2.4m\times 10^3}}=\frac{1}{100\sqrt{m}}.
\eeq
Therefore an ideal detector can find the tenth peak energy to within $1/\sqrt{m}$ percent after 10 years.  As the peak width is greater than the resolution of Daya Bay or RENO, the high energy resolution proposed at a new medium baseline detector experiment in Ref. \cite{caojun} will not significantly alter the determination of this peak.  Therefore one may expect $\Delta M^2_{eff}$ to be determined to a precision significantly better than 1 percent at a 40 kilometer baseline experiment with no backgrounds from other reactors.

However, to determine the hierarchy, one also needs to determine another combination of the neutrino mass differences.  This requires the measurement of a higher $n$ peak.  Consider for example the $n=15$ peak at 2.5 MeV.  While it is in general a poor approximation to ignore the backgrounds provided by distant reactors, if one does ignore them then since since peak is near the 1-2 minimum, the relative peak height is $A=0.4$.   The total flux in the peak is only about 0.5
\beq
f_{min}=\frac{1}{0.4\times 15\sqrt{2.4m\times 10^2}}=\frac{1}{90\sqrt{m}}.
\eeq
As can be seen in Fig \ref{degenfig}, a 3\% precision required to determine the hierarchy.  As the width of the peak is about 3\%, a $3\%/\sqrt{E}$ resolution may reduce the amplitude by a factor of 2, still allowing for a determination of the hierarchy.

\section{The Fourier transform of the survival probability} \label{fouriersez}

While each peak provides some information regarding a combination of neutrino mass differences and therefore the hierarchy, it may well be that the fluxes are too weak or the backgrounds too small for the peaks to be reasonably well identified.  Complimentary information can be combined by combining the peaks.  As the electron survival probability is a sum of periodic cosine functions, they can be combined by a Fourier transform.  Even when individual peaks are hard to identify, the combination probed by the Fourier transform may well be visible and so may provide the best chance for determining the hierarchy \cite{hawaii}.

\subsection{The complex Fourier transform}

For simplicity we will approximate the observed electron antineutrino spectrum by a Gaussian distribution in $L/E$ space
\beq
\Phi\left(\frac{L}{E}\right)=e^{\left(\frac{L}{E}-L\langle\frac{1}{E}\rangle\right)^2/\sigma^2} \label{spec}
\eeq
where $\langle 1/E \rangle$ is the average $1/E$ of a neutrino which arrives.  While with only slightly more complicated equations the following could be avoided, we will make the crude approximation that (\ref{spec}) is the neutrino spectrum after 1-2 neutrino oscillations, and that the expectation value of the inverse energy is therefore taken with respect to the 1-2 oscillated spectrum, which depends upon $L$. 

1-3 oscillations affect this spectrum by introducing a modulation equal to $\Phi(L/E) P_{13}(L/E)$.  The Fourier transform of this modulation is
\bea
F_{13}(k)&=&\int d\left(\frac{L}{E}\right) \Phi(L/E)  P_{13}(L/E) e^{i\frac{kL}{E}}\\&=&\frac{\sigma\sqrt{\pi}\cp212}{4}\left(e^{\frac{\sigma^2}{4}\left(k+\frac{\mn31}{2}\right)^2}e^{i\left(k+\frac{\mn31}{2}\right)L\langle\frac{1}{E}\rangle}+(e^{\frac{\sigma^2}{4}\left(k.\frac{\mn31}{2}\right)^2}e^{i\left(k-\frac{\mn31}{2}\right)L\langle\frac{1}{E}\rangle}\right)\nonumber
\eea
where we have factored the $\spp2213$ out of the definition of $F_{13}$. This quantity has two peaks, one at $k=\mn31/2$ and one at $k=-\mn31/2$.  At each peak, one of the two terms in the parenthesis dominates, and the other is suppressed by a factor of order $e^{n^2/4}$, which is large enough that we will neglect the subdominant term.  Therefore, near the first peak
\beq
F_{13}(k)=\frac{\sigma\sqrt{\pi}\cp212}{4}e^{\frac{\sigma^2}{4}\left(k-\frac{\mn31}{2}\right)^2}e^{i\left(k-\frac{\mn31}{2}\right)L\langle\frac{1}{E}\rangle}
\eeq
The complex norm of $F$ has a maximum at $k=\mn31/2$, where $F$ is real.  The real part of $F$, corresponding to a cosine transform, is, within the validity of the approximations described above, symmetric about this maximum.  The imaginary part, corresponding to a sine transform, vanishes at this maximum and is antisymmetric about it.

The transform $F_{23}(k)$ can be calculated identically.  Near the positive $k$ maximum the sum is just
\bea
F(k)&=&\frac{\sigma\sqrt{\pi}}{4}\left[\cp212 e^{\frac{\sigma^2}{4}\left(k-\frac{\mn31}{2}\right)^2}e^{i\left(k-\frac{\mn31}{2}\right)L\langle\frac{1}{E}\rangle}+\sp212 e^{\frac{\sigma^2}{4}\left(k-\frac{\mn32}{2}\right)^2}e^{i\left(k-\frac{\mn32}{2}\right)L\langle\frac{1}{E}\rangle}\right]\nonumber\\
&=&\frac{\sigma\sqrt{\pi}}{4}\left[\cp212 e^{\frac{\sigma^2}{4}\left(k-\frac{\mn31}{2}\right)^2}+\sp212 e^{\frac{\sigma^2}{4}\left(k-\frac{\mn32}{2}\right)^2}e^{\pm i\frac{\m21 L}{2}\langle\frac{1}{E}\rangle}\right]e^{i\left(k-\frac{\mn31}{2}\right)L\langle\frac{1}{E}\rangle} \label{xform}
\eea
where the $+$ sign applies to the normal hierarchy.

The first term corresponds to the old peak, at $k=\mn31$ and the second to a new peak, which on its own would be $\tp212\sim 1/2$ as high as the first, as $k=\mn32$.  Of course, due to the phase difference of the two terms, for some choice of parameters the interference between the two terms implies that the second is not a local maximum of the norm of the Fourier transform.  However, each term is maximized when it is real, and so each term can be seen as a peak or a shoulder in the Fourier cosine transform.  The $k$ value of the peak then gives the corresponding mass difference.  Thus if the smaller peak is to the left, as smaller $k$, of the larger peak then $\mn32<\mn31$ and so one can conclude that there is a normal neutrino mass hierarchy \cite{hawaii}.

\subsection{Determining $\mn31$ at the 1-2 oscillation minimum}

We will refer to the baseline
\beq
L=\frac{2\pi}{\m21\langle\frac{1}{E}\rangle}
\eeq
as the 1-2 oscillation minimum, as it is roughly the baseline at which the largest fraction of neutrinos disappears as a result of $P_{12}$.   It is approximately 58 km.  At this distance the Fourier transform (\ref{xform}) simplifies as the two terms are precisely out of phase
\beq
F(k)=\frac{\sigma\sqrt{\pi}}{4}\left[\cp212 e^{\frac{\sigma^2}{4}\left(k-\frac{\mn31}{2}\right)^2}-\sp212 e^{\frac{\sigma^2}{4}\left(k-\frac{\mn32}{2}\right)^2}\right]e^{i\left(k-\frac{\mn31}{2}\right)\frac{2\pi}{\m21}} . \label{segno}
\eeq

The cosine transform is just the real part of $F$.  Up to $k$-independent factors, it is proportional to 
\beq
F_{\cos}(\tilde{k})=\left( e^{-\frac{\sigma^2}{4}\tilde{k}^2}-\tp212  e^{-\frac{\sigma^2}{4}\left(\tilde{k}\pm\frac{\m12}{2}\right)^2}\right)\cos\left(\frac{2\pi\tilde{k}}{\m21}\right) \label{coseq}
\eeq
where we have defined the distance to the $1-3$ peak to be
\beq
\tilde{k}=k-\frac{\mn31}{2}.
\eeq
The maxima of $F_{\cos}$ are found by setting to zero its derivative with respect to $\tilde{k}$, yielding the condition
\bea
&&\left(\frac{\sigma^2}{2}\tilde{k}+\frac{2\pi}{\m21}\tan\left(\frac{2\pi\tilde{k}}{\m21}\right)\right)e^{-\frac{\sigma^2}{4}\tilde{k}^2}\\&&\ \ \ \ \ \ \ \ \ \ \ \ \ =
\tp212\left(\frac{\sigma^2}{2}\left(\tilde{k}\pm\frac{\m21}{2}\right)+\frac{2\pi}{\m21}\tan\left(\frac{2\pi\tilde{k}}{\m21}\right)\right)e^{-\frac{\sigma^2}{4}\left(\tilde{k}\pm\frac{\m21}{2}\right)^2}\nonumber
\eea
where again the $+$ sign corresponds to the normal hierarchy and the $-$ sign to the inverted hierarchy.  The largest peak is the closest to $\tilde{k}=0$, the absolute maximum of the Fourier transform of $P_{13}$.  To find this peak we may expand $\tilde{k}$ about 0, at linear order we find
\beq
\tilde{k}=\pm\frac{\m21/2}{1+\frac{\pi^2}{2\beta}\left(e^\beta\ctp212-1\right)+2\beta} \label{ktilde}
\eeq
where we have defined the constant
\beq
\beta=\frac{\sigma^2\m21}{16}.
\eeq

As the denominator of (\ref{ktilde}) is much greater than 1, we learn that
\beq
|\tilde{k}|<<\frac{\m21}{2}
\eeq
and so the maximum of $F_{\cos}$ lies at approximately
\beq
k_{max}=\frac{\mn31}{2}.
\eeq
This means that if the detector is placed at the 1-2 oscillation minimum baseline then the peak of the cosine transform of the full neutrino spectrum lies essentially at the peak of $P_{13}$ alone.  This is not because because the frequencies of the $P_{13}$ and $P_{23}$ terms are similar, but because the absolute maximum of the Fourier transform of $P_{13}$, which corresponds to its frequency, happens to coincide with one of the minima of the cosine transform of $P_{23}$, which is not its frequency.  {\textbf{Thus, at the 1-2 oscillation minimum, the absolute maximum of the sum of two cosine transforms is coincident with that of $P_{13}$ alone, allowing for a direct and precise determination of $\mn31$}}.  This may be useful on its own even if the hierarchy has already been determined by NOvA of T2K.

\subsection{Determining the hierarchy at the 1-2 oscillation minimum}

Can this simplification also yield information about the hierarchy?  A precise determination of $\mn31$ combined with MINOS' best fit for $\mn32$ could give a 1$\sigma$ answer to this question, within 5 years MINOS+ may improve this to 2$\sigma$.  But with enough flux the peak structure Fourier transform alone yields some information about the hierarchy.  

\begin{figure} 
\begin{center}
\includegraphics[width=5.2in,height=2.5in]{FCT_58_D.pdf}
\caption{The cosine Fourier transform of $P_{12}$ (blue), $P_{23}$ (purple), $P_{13}$ (green) and $P_{ee}$ (yellow) at 58 km. {\textbf{This isn't quite true because our conventions for the P's are different, can we redo this figure with our conventions???}} Note that the $P_{13}$ and $P_{23}$ curves are just out of phase, so that the total extrema coincide with those of $P_{13}$.}
\label{cosfig}
\end{center}
\end{figure}

As can be seen in Fig. \ref{cosfig}, Just as the relative sign in (\ref{segno}) implied that the global maximum of the cosine transform of the total spectrum is essentially coincident with that of $P_{13}$, it also implies that the minima just to its left and right are coincident
\beq
k_{\min}^L=\frac{\mn31-\m21}{2}\hsp
k_{\min}^R=\frac{\mn31+\m21}{2}.
\eeq
These are minima for the simple reason that the cosine on the right of (\ref{coseq}) is equal to $-1$.  Substituting these values of $k$ into (\ref{coseq}) we find the values of the cosine transforms at the two minima
\beq
F_{\cos}(k_{min}^L)=-\left(e^{-\beta}-\tp212 e^{-\beta(-1\pm 1)^2}\right)\hsp
F_{\cos}(k_{min}^R)=-\left(e^{-\beta}-\tp212 e^{-\beta(1\pm 1)^2}\right).
\eeq
In the case of the normal hierarchy, the positive sign means that the second term is larger on the left, meaning that the minimum on the right is deeper.  In the case of the inverted hierarchy these two depths are interchanged, and so the peak on the left is deeper.  This is the criterion for determining the hierarchy from the cosine transform which was proposed in Ref. \cite{caojun}.

\begin{figure} 
\begin{center}
\includegraphics[width=5.2in,height=2.5in]{FST_58_D.pdf}
\caption{The sine Fourier transform of $P_{12}$ (blue), $P_{23}$ (purple), $P_{13}$ (green) and $P_{ee}$ (yellow) at 58 km. {\textbf{This isn't quite true because our conventions for the P's are different, can we redo this figure with our conventions???}} Note that the $P_{13}$ and $P_{23}$ curves are just out of phase, so that the total extrema coincide with those of $P_{13}$.}
\label{sinfig}
\end{center}
\end{figure}

A similar analysis can be applied to the imaginary part of $F(k)$, which is obtained via a sine transform.  As $F(\mn31/2)$ is real, the sine transform vanishes at the maximum of the global cosine transform.  The sine transform then has a maximum on the right and a minimum on the left.  The opposition of the phases of the Fourier transforms of $P_{13}$ and $P_{23}$ at the 1-2 oscillation minima again imply that these extrema of the sine transform of the full spectrum roughly coincide with the extrema of the sine transform of $P_{13}$ alone, as can be seen at 58 km in Fig. \ref{sinfig}.  They simply correspond to the values of $k$ for which the phase in (\ref{segno}) is $\pm\pi/2$
\beq
k^L=\frac{\mn31}{2}-\frac{\m21}{4}\hsp
k^R=\frac{\mn31}{2}+\frac{\m21}{4}.
\eeq
Defining the sine transform so as to take the imaginary part of $F$ with the same normalization as for the cosine transform one then finds
\beq
F_{\sin}(k^L)=-i\left(e^{-\beta/4}-\tp212 e^{-\frac{\beta}{4}(2\mp 1)^2}\right)\hsp
F_{\sin}(k^R)=i\left(e^{-\beta/4}-\tp212 e^{-\frac{\beta}{4}(2\pm 1)^2}\right)
\eeq
where the top sign corresponds to the normal hierarchy.  In the case of the normal hierarchy the second term in the minimum on the left is larger and so $|F_{\sin}(k^L)|<|F_{\sin}(k^R)|$, whereas in the case of the inverted hierarchy the maximum on the right is larger than the minimum on the left.  Thus we have reproduced the correlation between the hierarchy and the relative sizes of these extrema observed in Ref. \cite{caojun}.

Note that in the case of the normal hierarchy both extrema are more positive and in the case of the inverted hierarchy both are more negative, in other words {\textbf{near its maximum $F$ has a positive imaginary part in the case of a normal hierarchy and a negative imaginary part in the case of an inverted hierarchy}}.  This provides a new, third test which allows one to extrapolate the hierarchy from the Fourier transform of the survival probability.  Of course the optimal indicator of the hierarchy will be a weighted sum of these three tests, with weights that may be determined by series of simulations.  The ability to distinguish the hierarchies may also be improved by convoluting the observed spectrum with a slowly varying function of $(L/E)$ which weights neutrinos near the 1-2 minimum more heavily.  One can also use simulated data to determine which weighting functions are optimal for this task.

\subsection{Nonlinear Fourier transform}
The Fourier transform methods described above essentially work because the phase at the maximum is determined by the hierarchy, positive for the normal hierarchy and negative for the inverted hierarchy.  In the case of just $1-3$ oscillations, this phase is zero, since the corresponding oscillations are a cosine and the real part of the Fourier transform is determined by the cosine transform.  However when $P_{23}$ is included the peaks of the untransformed spectrum move according to Eq. (\ref{pichi}).  The distance between the untransformed peaks changes, which in general  moves the Fourier transformed peak.  

Critically, the untransformed $P_{13}+P_{23}$ also loses its mod $2/\mn31$ periodicity, as $\an$ is not a linear function of $n$.  A given detector is only sensitive to some of the peaks, corresponding to a certain region in Fig. \ref{anfig}.  Such regions are generally dominated by a domain in which $\an$ can be approximated not by a linear function, but by a linear function plus a constant offset.  This constant offset implies that the convolution of $P_{13}+P_{23}$ with the reactor neutrino spectrum is not of the form $\cos(\kappa L/E)$ but rather of the form $\cos(\kappa L/E+c)$ where the sign of $c$ is determined by the hierarchy.  This offset, $c$, leads to a translation in $L/E$ space which, after the Fourier transform, becomes a phase.  Thus the hierarchy determines an overall phase of the Fourier transform, which leads to the observable indicators described in the last section.

The Fourier transform method is robust since it sums multiple peaks together, and so it requires less neutrinos than a direct analysis of the positions of the peaks.  However it is inefficient because, as was just described, it works by approximating the function $\an$ to be affine in the energy range which is probed.  So one might ask if the performance would be increased by performing not an ordinary Fourier transform, but a Fourier transform with the nonlinearity of $\an$ built in.  The nonlinearity depends upon $\an$ which depends on the hierarchy and also weakly upon the neutrino mass matrix.  One may therefore attempt a nonlinear Fourier transform with both choices of hierarchy, and if desired, a weight function $g(L/E)$.  Such a nonlinear cosine transform is given by
\beq
F(k)=\int d\left(\frac{L}{E}\right)\Phi(L/E) P_{ee}(L/E) g(L/E) \cos\left(k\frac{L}{E}\pm 2\pi\alpha\left(\frac{k}{2\pi}\frac{L}{E}\right)\right)
\eeq
where $\alpha$ is a function which interpolates between the discrete values of $\an$ for the normal hierarchy and the positive sign corresponds to the inverted hierarchy.  This transform will add all of the peaks together with the same phase for the correct hierarchy whereas the peaks will be distorted by the other hierarchy.  Therefore the correct hierarchy can be determined from the fact that the corresponding nonlinear Fourier transform will have a larger absolute maximum.

\section{Interference effects} \label{intsez}

\subsection{Interference between reactors separated by 1 km} \label{unoproblema}

Consider two reactors of equal thermal power separated by a distance $D$.  If a detector is a distance $L>>D$ from the nearest and if the baseline makes an angle $\theta$ with respect to the line passing through both reactors, then the distance from the detector to the far reactor will be $L+d$ where $d=D\cos(\theta)$. 

Adding the flux from both reactors, one finds that the 1-3 oscillations interfere 
\beq
P_{13}\propto \cos\left(\frac{\mn31L}{2E}\right)+ \cos\left(\frac{\mn31(L+d)}{2E}\right)=2 \cos\left(\frac{\mn31d}{4E}\right)\cos\left(\frac{\mn31(L+d/2)}{2E}\right).
\eeq
The neutrinos from the two sources are not coherent, it is not the wavefunctions that add, but the probabilities.  And the result is that the amplitude of the oscillations at energy $E$ is suppressed by a factor of
\beq
\cos\left(\frac{\mn31d}{4E}\right)= \cos\left(3\frac{d/\mathrm{km}}{E/\mathrm{MeV}}\right).
\eeq
In particular they annihilate entirely when
\beq
\frac{d}{\mathrm{km}}=0.5 \frac{E}{\mathrm{MeV}}.
\eeq

Consider for example the pair of Daya Bay reactors and the two pairs of Ling Ao reactors which lie along a line at a distance of 0.8 and 1.3 km from the Daya Bay reactors.  The Daya Bay II reactor location suggested in Ref. \cite{caojunseminario} is at an angle of 15 degrees with respect to a continuation of this line, yielding $d=$1.2\ km for the nearest and furthest reactor pair.  As a result 1-3 oscillations at 2.4 MeV completely cancel between these two reactor pairs, leaving only the oscillations of the middle reactor, and so effectively damping the oscillation amplitude by a factor of 3, which implies that an equally precise measurement of a peak at that energy requires 9 times as much flux as it would have without interference.

At the peak energy, 3.6 MeV the annihilation is not complete, but is reduced by a factor of $\cos(1)=0.6$.  Thus the total amplitude of the peak from all three reactor pairs is reduced by about 30\%, and so twice the neutrino flux will be necessary to observe these peaks.

This problem can be avoided if the detector is equidistant from all of the reactors.  In the case of the Daya Bay and Ling Ao complex, in the case of RENO and somewhat trivially in the case of Double Chooz this is possible as all of the reactors essentially line along a line.  It means however that such a detector will not be equidistant from any reactor that may eventually be built at Haifeng, thus reducing the neutrino fluxes assumed in Ref. \cite{caojun} by a factor of 2.  Worse yet, the Haifeng reactor than provides a strong and undesirable background, as will be described in Subsec. \ref{centoproblema}.

\subsection{Interference between reactors separated by 100 km} \label{centoproblema}

The 1-2 oscillation minimum
\beq
L=\frac{2\pi E}{\mn31}=16\left(\frac{E}{\mathrm{MeV}}\right)\mathrm{km}
\eeq
provides an ideal baseline to determine the mass hierarchy for a number of reasons.  Among these is that while the $P_{13}+P_{23}$ oscillation amplitude of 
$(\cp212-\sp212)\spp2213$ is a factor of 3 less than its amplitude $\spp2213$ at the 1-2 maxima, the flux remaining after 1-2 oscillations is also smaller by a factor of $1/(1-\sp212\cp413)$ which is about 5.  Thus the 1-3 oscillations are a larger fraction of the total flux, making them easier to see above systematic errors, although not above statistical errors as $3>\sqrt{5}$.

However the smaller signal and smaller flux means that this part of the spectrum is particularly prone to interference from distant reactors, in particular those that may near the first 1-2 oscillation maximum
\beq
L=\frac{4\pi E}{\mn31}=32\left(\frac{E}{\mathrm{MeV}}\right)\mathrm{km}.
\eeq
In this case the addition factor of 5 in flux from the distant reactor outweighs the factor of 4 distance suppression, and so if both reactor complexes have the same strength than most of the flux at the 1-2 minimum is background.  This means that to obtain the same energy resolution at the 1-2 minimum peaks one will need more than 4 times as much neutrino flux.  

This is the case for example with a detector placed 58 km away from the Daya Bay/Ling Ao complex, perpendicular to the reactors.  It would be at the 1-2 maximum of the proposed Haifeng reactors and would also suffer significant contamination from the Taishan and Yanjiang reactor complexes, at each of which at least 3 reactors are already under construction.  Similarly a reactor placed 60 km from Daya Bay, Ling Ao and Haifeng would be at the 1-2 maximum of the proposed 17.4 GW thermal power reactor complex at Lufeng.

As this effect increases the 1-2 oscillation minimum flux, it is also a serious obstacle to an accurate measurement of $\theta_{12}$.  If a model of the background neutrino flux from distant reactors is wrong, the error in the total flux can be compensated for by an error in the determination of $\theta_{12}$.  Ideally this problem can be solved by using two medium baseline detectors instead of one, which would break this degeneracy.  However in a world limited by costs, it may instead be necessary to simply try to make the background from other reactors as small as possible, thus minimizing the potential error in the determination of the hierarchy and in the determination of $\theta_{12}$.

Unlike the short distance interference problem discussed in Subsec. \ref{unoproblema}, the fractional contamination caused by distance reactors depends on the baseline $L$ to the reactor whose neutrinos provide the signal.  The fractional contamination from undesired reactor neutrinos is inversely proportional to the desired signal strength, and so it is proportional to $L^2$.  Thus this problem is minimized by placing the detector as near to the reactor as possible.  If there is only one detector, and one wishes to use it to determine the hierarchy, then as explained in Subsec. \ref{cubicosez}, this minimum distance cannot be shorter than about 45 km.

\section{Conclusions}

The newly discovered high value of $\theta_{13}$ means that the determination of the neutrino mass hierarchy at a medium baseline reactor experiment is now practical.  Previous analysis of such experiments have assumed values of $\spp2213$ an order of magnitude or more below its true value.  At such low values, individual peaks of the electron antineutrino spectra could not be resolved and one instead needed to rely upon a Fourier analysis \cite{hawaii,caojun}.  

In this note we have reconsidered the determination of the neutrino mass hierarchy now that 1-3 oscillations are large and so individual 1-3 peaks in the spectrum can easily be observed.  We found that the position of each peak determines a particular combination of the neutrino mass differences, for example the first 10 peaks all determine the same difference $\cp212\mn31+\sp212\mn32$.  In particular this means that the first 10 peaks alone cannot be used to determine the hierarchy, as they are fit equally well by the wrong hierarchy model in which this mass difference is preserved.  On the other hand we found that the positions of the next 10 peaks depend on distinct combinations of the mass differences, and so combining the energies at different peaks with some beyond the $10$th can lead to a determination of the hierarchy.  We also estimated the detector resolution and neutrino flux which are needed for such a goal, although an accurate determination will be left for the simulations to be discussed in our companion paper.

The information about the hierarchy is therefore contained in the low energy peaks, which suffer the most from the effects of poor resolution,  low neutrino flux and interference.  While the individual peaks are somewhat difficult to resolve, the situation can nonetheless be improved with a Fourier transform.  We have analyzed the Fourier transform of the spectrum, deriving the phenomenological hierarchy indicators proposed in Refs. \cite{hawaii,caojun} and providing a new indicator, the complex phase at the peak of the Fourier transform, which we claim will be positive for the normal hierarchy and negative for the inverse hierarchy.  We also propose a new hierarchy-dependent nonlinear Fourier transform which will lead to a higher peak using the transform corresponding to the correct hierarchy.

The strength of the Fourier transform method is that it sums together all of the peaks to increase the strength of the signal with respect to the noise.  If the energy spectrum is shifted, this simply leads to an overall phase in the Fourier transform which does not seriously affect the analysis.  Likewise a uniform stretching of the spectrum simply shifts the peaks, which does not affect the determination of the hierarchy at all.  Any nonlinearity however poses a much more serious problem, as it can lead to an interference between the peaks in the spectrum which mutates or destroys the peak structure of the Fourier transform.  If the nature of the nonlinearity is known then one can adjust the analysis to correct for it \cite{petcov2010} however if the nonlinearity is known it could be corrected directly in the reading of the energy.  In general the nonlinearity of the response is only known at energies where the detector has been calibrated with radioactive sources.  It may therefore be desirable to modify the Fourier transform so as to weight the more reliable energies more heavily.  One may also wish to weight the energies near the 1-2 minimum more heavily, as it contributes few neutrinos but it is indispensable in a determination of the hierarchy.  These  weights can be optimized by testing various hierarchy determination algorithms against simulated data.

We also discussed two interference effects.  First of all, reactors in the same array are generally separated by distances of order 1 km, which means that neutrinos arriving at a detector from one reactor at a 1-3 maximum may be at the same energy as those arriving from another at a 1-3 minimum.  The result is that the 1-3 oscillation signal can be severely reduced at the low energies in which this oscillation can occur within a reactor complex.  These are just the low energies which are necessary for the determination of the hierarchy, and so this interference poses a serious problem.  One solution is to place the detectors orthogonal to arrays of reactors.

The second interference effect arises from the fact that while the 1-2 oscillation minimum is the most useful energy range at which to determine the hierarchy, it enjoys a much lower neutrino flux than the 1-2 oscillation maximum.  As a result at these energy ranges one can expect serious contamination from distant reactors at their 1-2 maximum.  This problem, as well as the degeneracy between the neutrino flux from distant reactors and $\theta_{12}$, is optimally solved by using two detectors at different baselines.  However a cheaper solution to the same problem is to use a single detector at a shorter baseline, such as 45-50 km.


\section* {Acknowledgement}

\noindent
JE is supported by the Chinese Academy of Sciences
Fellowship for Young International Scientists grant number
2010Y2JA01. EC and XZ are supported in part by the NSF of
China.  


\end{document}

\bibitem{lsnd}
A.~Aguilar-Arevalo {\it et al.}  [LSND Collaboration],
  ``Evidence for neutrino oscillations from the observation of anti-neutrino(electron) appearance in a anti-neutrino(muon) beam,''
  Phys.\ Rev.\ D {\bf 64} (2001) 112007
  [hep-ex/0104049].

\bibitem{minibooneanom}
A.~A.~Aguilar-Arevalo {\it et al.}  [The MiniBooNE Collaboration],
  ``A Search for electron neutrino appearance at the $\Delta m^{2} \sim 1$eV$^{2}$ scale,''
  Phys.\ Rev.\ Lett.\  {\bf 98} (2007) 231801
  [arXiv:0704.1500 [hep-ex]].
A.~A.~Aguilar-Arevalo {\it et al.}  [MiniBooNE Collaboration],
  ``Unexplained Excess of Electron-Like Events From a 1-GeV Neutrino Beam,''
  Phys.\ Rev.\ Lett.\  {\bf 102} (2009) 101802
  [arXiv:0812.2243 [hep-ex]].
 A.~A.~Aguilar-Arevalo {\it et al.}  [The MiniBooNE Collaboration],
  ``Event Excess in the MiniBooNE Search for $\bar \nu_\mu \rightarrow \bar \nu_e$ Oscillations,''
  Phys.\ Rev.\ Lett.\  {\bf 105} (2010) 181801
  [arXiv:1007.1150 [hep-ex]].

\bibitem{minosanom}
P.~Adamson {\it et al.}  [MINOS Collaboration],
  ``First direct observation of muon antineutrino disappearance,''
  Phys.\ Rev.\ Lett.\  {\bf 107} (2011) 021801
  [arXiv:1104.0344 [hep-ex]].

\bibitem{zichichi}
A.~Zichichi,
``Results from LVD-OPERA Combined Analysis: A Time-Shift in the OPERA Setup,"
available online at http://agenda.infn.it/getFile.py/access?resId=0\&materialId=slides\&confId=4896.

\bibitem{miniboonetuttobene}
E.~D.~Zimmerman [MiniBooNE Collaboration],
  ``Updated Search for Electron Antineutrino Appearance at MiniBooNE,''
  arXiv:1111.1375 [hep-ex].

\bibitem{minostuttobene}
P.~Adamson {\it et al.}  [MINOS Collaboration],
  ``An improved measurement of muon antineutrino disappearance in MINOS,''
  arXiv:1202.2772 [hep-ex].

\bibitem{nuovoflusso}
T.~.A.~Mueller, D.~Lhuillier, M.~Fallot, A.~Letourneau, S.~Cormon, M.~Fechner, L.~Giot and T.~Lasserre {\it et al.},
  ``Improved Predictions of Reactor Antineutrino Spectra,''
  Phys.\ Rev.\ C {\bf 83} (2011) 054615
  [arXiv:1101.2663 [hep-ex]].
P.~Huber,
  ``On the determination of anti-neutrino spectra from nuclear reactors,''
  Phys.\ Rev.\ C {\bf 84} (2011) 024617
   [Erratum-ibid.\ C {\bf 85} (2012) 029901]
  [arXiv:1106.0687 [hep-ph]].

\bibitem{reattoreanom}
G.~Mention, M.~Fechner, T.~.Lasserre, T.~.A.~Mueller, D.~Lhuillier, M.~Cribier and A.~Letourneau,
  ``The Reactor Antineutrino Anomaly,''
  Phys.\ Rev.\ D {\bf 83} (2011) 073006
  [arXiv:1101.2755 [hep-ex]].

\bibitem{smirnov}
P.~C.~de Holanda and A.~Y.~.Smirnov,
  ``Homestake result, sterile neutrinos and low-energy solar neutrino experiments,''
  Phys.\ Rev.\ D {\bf 69} (2004) 113002
  [hep-ph/0307266].
P.~C.~de Holanda and A.~Y.~.Smirnov,
  ``Solar neutrino spectrum, sterile neutrinos and additional radiation in the Universe,''
  Phys.\ Rev.\ D {\bf 83} (2011) 113011
  [arXiv:1012.5627 [hep-ph]].

\bibitem{icecube}
R.~Abbasi {\it et al.}  [IceCube Collaboration],
  ``Measurement of the atmospheric neutrino energy spectrum from 100 GeV to 400 TeV with IceCube,''
  Phys.\ Rev.\ D {\bf 83} (2011) 012001
  [arXiv:1010.3980 [astro-ph.HE]].

\bibitem{giuntireview}
  C.~Giunti and M.~Laveder,
  ``Implications of 3+1 Short-Baseline Neutrino Oscillations,''
  Phys.\ Lett.\ B {\bf 706} (2011) 200
  [arXiv:1111.1069 [hep-ph]].

\bibitem{sterilecosm}
J.~Hamann, S.~Hannestad, G.~G.~Raffelt and Y.~Y.~Y.~Wong,
  ``Sterile neutrinos with eV masses in cosmology: How disfavoured exactly?,''
  JCAP {\bf 1109} (2011) 034
  [arXiv:1108.4136 [astro-ph.CO]].

\bibitem{dayabay}
  F.~P.~An {\it et al.}  [DAYA-BAY Collaboration],
  ``Observation of electron-antineutrino disappearance at Daya Bay,''
  Phys.\ Rev.\ Lett.\  {\bf 108} (2012) 171803
  [arXiv:1203.1669 [hep-ex]].

\bibitem{piureattori}
L.~A.~Mikaelyan and V.~V.~Sinev,
  ``Neutrino oscillations at reactors: What next?,''
  Phys.\ Atom.\ Nucl.\  {\bf 63} (2000) 1002
   [Yad.\ Fiz.\  {\bf 63N6} (2000) 1077]
  [hep-ex/9908047].

\bibitem{neut2012}
D. Dwyer,
``Daya Bay Results," presented at Neutrino 2012 in Kyoto.
Soon to be available at http://neu2012.kek.jp/neu2012/programme.html.

\bibitem{doublechooz}
Y.~Abe {\it et al.}  [DOUBLE-Chooz Collaboration],
  ``Indication for the disappearance of reactor electron antineutrinos in the Double Chooz experiment,''
  Phys.\ Rev.\ Lett.\  {\bf 108} (2012) 131801
  [arXiv:1112.6353 [hep-ex]].

\bibitem{reno}
  J.~K.~Ahn {\it et al.}  [RENO Collaboration],
  ``Observation of Reactor Electron Antineutrino Disappearance in the RENO Experiment,''
  Phys.\ Rev.\ Lett.\  {\bf 108} (2012) 191802
  [arXiv:1204.0626 [hep-ex]].

\bibitem{nuturn}
``Observation of reactor neutrino disappearance at RENO," presented at $\nu$TURN 2012 under Gran Sasso.  Available at http://agenda.infn.it/contributionListDisplay.py?confId=4722.

\bibitem{globale1}
G.~L.~Fogli, E.~Lisi, A.~Marrone, A.~Palazzo and A.~M.~Rotunno,
  ``Evidence of $\theta_{13}$>0 from global neutrino data analysis,''
  Phys.\ Rev.\ D {\bf 84} (2011) 053007
  [arXiv:1106.6028 [hep-ph]].

\bibitem{globale2}
  T.~Schwetz, M.~Tortola and J.~W.~F.~Valle,
  ``Where we are on $\theta_{13}$: addendum to 'Global neutrino data and recent reactor fluxes: status of three-flavour oscillation parameters',''
  New J.\ Phys.\  {\bf 13} (2011) 109401
  [arXiv:1108.1376 [hep-ph]].

\bibitem{paloverde}
F.~Boehm, J.~Busenitz, B.~Cook, G.~Gratta, H.~Henrikson, J.~Kornis, D.~Lawrence and K.~B.~Lee {\it et al.},
  ``Final results from the Palo Verde neutrino oscillation experiment,''
  Phys.\ Rev.\ D {\bf 64} (2001) 112001
  [hep-ex/0107009].

\bibitem{chooz}
M.~Apollonio {\it et al.}  [Chooz Collaboration],
  ``Search for neutrino oscillations on a long baseline at the Chooz nuclear power station,''
  Eur.\ Phys.\ J.\ C {\bf 27} (2003) 331
  [hep-ex/0301017].

\bibitem{neutdarke}
X.~-J.~Bi, P.~-H.~Gu, X.~-l.~Wang and X.~-M.~Zhang,
  ``Thermal leptogenesis in a model with mass varying neutrinos,''
  Phys.\ Rev.\ D {\bf 69} (2004) 113007
  [hep-ph/0311022].
  R.~Takahashi and M.~Tanimoto,
  ``Model of mass varying neutrinos in SUSY,''
  Phys.\ Lett.\ B {\bf 633} (2006) 675
  [hep-ph/0507142].
  R.~Takahashi and M.~Tanimoto,
  ``Speed of sound in the mass varying neutrinos scenario,''
  JHEP {\bf 0605} (2006) 021
  [astro-ph/0601119].
  E.~Ciuffoli, J.~Evslin, J.~Liu and X.~Zhang,
  ``OPERA and a Neutrino Dark Energy Model,''
  arXiv:1109.6641 [hep-ph].

\bibitem{neal04}
  D.~B.~Kaplan, A.~E.~Nelson and N.~Weiner,
  ``Neutrino oscillations as a probe of dark energy,''
  Phys.\ Rev.\ Lett.\  {\bf 93} (2004) 091801
  [hep-ph/0401099].

\bibitem{tortola}
  M.~Tortola, J.~W.~F.~Valle and D.~Vanegas,
  ``Global status of neutrino oscillation parameters after recent reactor measurements,''
  arXiv:1205.4018 [hep-ph].

\bibitem{foglinuovo}
G.L. Fogli, E. Lisi, A. Marrone, D. Montanino, A. Palazzo and A.M. Rotunno,
``Global analysis of neutrino masses, mixings and phases: entering the era of leptonic CP violation searches,"
arXiv:1205.5254 [hep-ph].

\bibitem{bugey4}
Y.~Declais, H.~de Kerret, B.~Lefievre, M.~Obolensky, A.~Etenko, Y.~.Kozlov, I.~Machulin and V.~Martemyanov {\it et al.},
  ``Study of reactor anti-neutrino interaction with proton at Bugey nuclear power plant,''
  Phys.\ Lett.\ B {\bf 338} (1994) 383.

\bibitem{dayafeb}
F.~P.~An {\it et al.}  [Daya Bay Collaboration],
  ``A side-by-side comparison of Daya Bay antineutrino detectors,''
  arXiv:1202.6181 [physics.ins-det].

\bibitem{daya2007}
  X.~Guo {\it et al.}  [Daya-Bay Collaboration],
  ``A Precision measurement of the neutrino mixing angle theta(13) using reactor antineutrinos at Daya-Bay,''
  hep-ex/0701029.


\bibitem{tredueterm}
 A.~Melchiorri, O.~Mena, S.~Palomares-Ruiz, S.~Pascoli, A.~Slosar and M.~Sorel,
  ``Sterile Neutrinos in Light of Recent Cosmological and Oscillation Data: A Multi-Flavor Scheme Approach,''
  JCAP {\bf 0901} (2009) 036
  [arXiv:0810.5133 [hep-ph]].

\bibitem{neutrinoasym}
  S.~Hannestad, I.~Tamborra and T.~Tram,
  ``Thermalisation of light sterile neutrinos in the early universe,''
  arXiv:1204.5861 [astro-ph.CO].

\bibitem{baoscoperta}
  D.~J.~Eisenstein {\it et al.}  [SDSS Collaboration],
  ``Detection of the baryon acoustic peak in the large-scale correlation function of SDSS luminous red galaxies,''
  Astrophys.\ J.\  {\bf 633} (2005) 560
  [astro-ph/0501171].


\bibitem{bao}
W.~J.~Percival, S.~Cole, D.~J.~Eisenstein, R.~C.~Nichol, J.~A.~Peacock, A.~C.~Pope and A.~S.~Szalay,
  ``Measuring the Baryon Acoustic Oscillation scale using the SDSS and 2dFGRS,''
  Mon.\ Not.\ Roy.\ Astron.\ Soc.\  {\bf 381} (2007) 1053
  [arXiv:0705.3323 [astro-ph]].

\bibitem{cosmomc}
  A.~Lewis and S.~Bridle,
  ``Cosmological parameters from CMB and other data: A Monte Carlo approach,''
  Phys.\ Rev.\ D {\bf 66} (2002) 103511
  [astro-ph/0205436].

\bibitem{wmap}
E.~Komatsu {\it et al.}  [WMAP Collaboration],
  ``Seven-Year Wilkinson Microwave Anisotropy Probe (WMAP) Observations: Cosmological Interpretation,''
  Astrophys.\ J.\ Suppl.\  {\bf 192} (2011) 18
  [arXiv:1001.4538 [astro-ph.CO]].

\bibitem{Suzuki:2011hu}
  N.~Suzuki {\it et al.},
  ``The Hubble Space Telescope Cluster Supernova Survey: V. Improving the Dark
  Energy Constraints Above z>1 and Building an Early-Type-Hosted Supernova
  Sample,''
  arXiv:1105.3470 [astro-ph.CO].

\bibitem{h}
A.~G.~Riess, L.~Macri, S.~Casertano, H.~Lampeitl, H.~C.~Ferguson, A.~V.~Filippenko, S.~W.~Jha and W.~Li {\it et al.},
  ``A 3\% Solution: Determination of the Hubble Constant with the Hubble Space Telescope and Wide Field Camera 3,''
  Astrophys.\ J.\  {\bf 730} (2011) 119
   [Erratum-ibid.\  {\bf 732} (2011) 129]
  [arXiv:1103.2976 [astro-ph.CO]].

\bibitem{nessundivergenza}
J.~-Q.~Xia, G.~-B.~Zhao, B.~Feng, H.~Li and X.~Zhang,
  ``Observing dark energy dynamics with supernova, microwave background and galaxy clustering,''
  Phys.\ Rev.\ D {\bf 73} (2006) 063521
  [astro-ph/0511625].
G.~-B.~Zhao, J.~-Q.~Xia, M.~Li, B.~Feng and X.~Zhang,
  ``Perturbations of the quintom models of dark energy and the effects on observations,''
  Phys.\ Rev.\ D {\bf 72} (2005) 123515
  [astro-ph/0507482].

\bibitem{altroquintom}
  E.~Giusarma, M.~Archidiacono, R.~de Putter, A.~Melchiorri and O.~Mena,
  ``Sterile neutrino models and nonminimal cosmologies,''
  Phys.\ Rev.\ D {\bf 85} (2012) 083522
  [arXiv:1112.4661 [astro-ph.CO]].

\bibitem{unione2}
N.~Suzuki, D.~Rubin, C.~Lidman, G.~Aldering, R.~Amanullah, K.~Barbary, L.~F.~Barrientos and J.~Botyanszki {\it et al.},
  ``The Hubble Space Telescope Cluster Supernova Survey: V. Improving the Dark Energy Constraints Above z=1 and Building an Early-Type-Hosted Supernova Sample,''
  Astrophys.\ J.\  {\bf 746} (2012) 85
  [arXiv:1105.3470 [astro-ph.CO]].

\bibitem{quintom}
B.~Feng, M.~Li, Y.~-S.~Piao and X.~Zhang,
  ``Oscillating quintom and the recurrent universe,''
  Phys.\ Lett.\ B {\bf 634} (2006) 101
  [astro-ph/0407432].
  B.~Feng, X.~L.~Wang and X.~M.~Zhang, 
``Dark energy constraints from the cosmic age and supernova,''
 Phys.\ Lett.\  B {\bf 607} (2005) 35  
 [arXiv:astro-ph/0404224].
  X.~-F.~Zhang, H.~Li, Y.~-S.~Piao and X.~-M.~Zhang,
``Two-field models of dark energy with equation of state across -1,''
  Mod.\ Phys.\ Lett.\ A {\bf 21}, 231 (2006)
  [astro-ph/0501652].

\bibitem{neutrinoasymvecchio}
  R.~Foot and R.~R.~Volkas,
  ``Reconciling sterile neutrinos with big bang nucleosynthesis,''
  Phys.\ Rev.\ Lett.\  {\bf 75} (1995) 4350
  [hep-ph/9508275].

\bibitem{fr}
  H.~Motohashi, A.~A.~Starobinsky and J.~'i.~Yokoyama,
  ``Cosmology based on f(R) Gravity admits 1 eV Sterile Neutrinos,''
  arXiv:1203.6828 [astro-ph.CO].

\bibitem{robert}
R.~H.~Brandenberger, N.~Kaiser, D.~N.~Schramm and N.~Turok,
  ``Galaxy and Structure Formation with Hot Dark Matter and Cosmic Strings,''
  Phys.\ Rev.\ Lett.\  {\bf 59} (1987) 2371.
R.~H.~Brandenberger, A.~Mazumdar and M.~Yamaguchi,
  ``A Note on the robustness of the neutrino mass bounds from cosmology,''
  Phys.\ Rev.\ D {\bf 69} (2004) 081301
  [hep-ph/0401239].

\bibitem{monopoli}
J.~Evslin and S.~B.~Gudnason,
  ``High Q BPS Monopole Bags are Urchins,''
  arXiv:1111.3891 [hep-th].
J.~Evslin and S.~B.~Gudnason,
  ``Dwarf Galaxy Sized Monopoles as Dark Matter?,''
  arXiv:1202.0560 [astro-ph.CO].

\bibitem{lsndnonstandard}
  G.~Karagiorgi, M.~H.~Shaevitz and J.~M.~Conrad,
  ``Confronting the short-baseline oscillation anomalies with a single sterile neutrino and non-standard matter effects,''
  arXiv:1202.1024 [hep-ph].

\bibitem{envir}
  G.~Karagiorgi, M.~H.~Shaevitz and J.~M.~Conrad,
  ``Confronting the short-baseline oscillation anomalies with a single sterile neutrino and non-standard matter effects,''
  arXiv:1202.1024 [hep-ph].

\end{thebibliography}


\begin{thebibliography}{23}\setlength{\itemsep}{-2.3mm}



\bibitem{petcovidea}
  S.~T.~Petcov and M.~Piai,
  ``The LMA MSW solution of the solar neutrino problem, inverted neutrino mass hierarchy and reactor neutrino experiments,''
  Phys.\ Lett.\ B {\bf 533} (2002) 94
  [hep-ph/0112074]. S.~Choubey, S.~T.~Petcov and M.~Piai,
  ``Precision neutrino oscillation physics with an intermediate baseline reactor neutrino experiment,''
  Phys.\ Rev.\ D {\bf 68} (2003) 113006
  [hep-ph/0306017].

\bibitem{hawaii}
 M.~Batygov, S.~Dye, J.~Learned, S.~Matsuno, S.~Pakvasa and G.~Varner,
   J.~Learned, S.~T.~Dye, S.~Pakvasa and R.~C.~Svoboda,
  ``Determination of neutrino mass hierarchy and $\theta_{13}$ with a remote detector of reactor antineutrinos,''
  Phys.\ Rev.\ D {\bf 78} (2008) 071302
  [hep-ex/0612022].
 ``Prospects of neutrino oscillation measurements in the detection of reactor antineutrinos with a medium-baseline experiment,''
  arXiv:0810.2580 [hep-ph].
 
\bibitem{caojun}
 L.~Zhan, Y.~Wang, J.~Cao and L.~Wen,
  ``Determination of the Neutrino Mass Hierarchy at an Intermediate Baseline,''
  Phys.\ Rev.\ D {\bf 78} (2008) 111103
  [arXiv:0807.3203 [hep-ex]].
  L.~Zhan, Y.~Wang, J.~Cao and L.~Wen,
  ``Experimental Requirements to Determine the Neutrino Mass Hierarchy Using Reactor Neutrinos,''
  Phys.\ Rev.\ D {\bf 79} (2009) 073007
  [arXiv:0901.2976 [hep-ex]].
 
\bibitem{daya}
  F.~P.~An {\it et al.}  [DAYA-BAY Collaboration],
  ``Observation of electron-antineutrino disappearance at Daya Bay,''
  Phys.\ Rev.\ Lett.\  {\bf 108} (2012) 171803
  [arXiv:1203.1669 [hep-ex]].

\bibitem{reno}
  J.~K.~Ahn {\it et al.}  [RENO Collaboration],
  ``Observation of Reactor Electron Antineutrino Disappearance in the RENO Experiment,''
  Phys.\ Rev.\ Lett.\  {\bf 108} (2012) 191802
  [arXiv:1204.0626 [hep-ex]].

\bibitem{doublechooz}
Y.~Abe {\it et al.}  [DOUBLE-Chooz Collaboration],
  ``Indication for the disappearance of reactor electron antineutrinos in the Double Chooz experiment,''
  Phys.\ Rev.\ Lett.\  {\bf 108} (2012) 131801
  [arXiv:1112.6353 [hep-ex]].

\bibitem{globale}
G.~L.~Fogli, E.~Lisi, A.~Marrone, A.~Palazzo and A.~M.~Rotunno,
  ``Evidence of $\theta_{13}$>0 from global neutrino data analysis,''
  Phys.\ Rev.\ D {\bf 84} (2011) 053007
  [arXiv:1106.6028 [hep-ph]].
  T.~Schwetz, M.~Tortola and J.~W.~F.~Valle,
  ``Where we are on $\theta_{13}$: addendum to 'Global neutrino data and recent reactor fluxes: status of three-flavour oscillation parameters',''
  New J.\ Phys.\  {\bf 13} (2011) 109401
  [arXiv:1108.1376 [hep-ph]].

\bibitem{caojunseminario}
J. Cao,
``Observation of $\overline{\nu}_e$ Disappearance at Daya Bay," presented at $\nu$Turn under Gran Sasso.
Available at http://agenda.infn.it/contributionListDisplay.py?confId=4722.

\bibitem{renonuturn}
``Observation of reactor neutrino disappearance at RENO," presented at $\nu$TURN 2012 under Gran Sasso.  Available at http://agenda.infn.it/contributionListDisplay.py?confId=4722.


\bibitem{yifangseminario}
Y. Wang,
``Daya Bay II: The Next Generation Reactor Neutrino Experiment" presented at NuFact in Williamsburg, Virginia.
Available at https://www.jlab.org/indico/conferenceTimeTable.py?confId=0\#20120725.detailed.

\bibitem{vogelengel}
  P.~Vogel and J.~Engel,
  ``Neutrino Electromagnetic Form-Factors,''
  Phys.\ Rev.\ D {\bf 39} (1989) 3378.

\bibitem{nuovoflusso}
T.~.A.~Mueller, D.~Lhuillier, M.~Fallot, A.~Letourneau, S.~Cormon, M.~Fechner, L.~Giot and T.~Lasserre {\it et al.},
  ``Improved Predictions of Reactor Antineutrino Spectra,''
  Phys.\ Rev.\ C {\bf 83} (2011) 054615
  [arXiv:1101.2663 [hep-ex]].
P.~Huber,
  ``On the determination of anti-neutrino spectra from nuclear reactors,''
  Phys.\ Rev.\ C {\bf 84} (2011) 024617
   [Erratum-ibid.\ C {\bf 85} (2012) 029901]
  [arXiv:1106.0687 [hep-ph]].

\bibitem{reattoreanom}
G.~Mention, M.~Fechner, T.~.Lasserre, T.~.A.~Mueller, D.~Lhuillier, M.~Cribier and A.~Letourneau,
  ``The Reactor Antineutrino Anomaly,''
  Phys.\ Rev.\ D {\bf 83} (2011) 073006
  [arXiv:1101.2755 [hep-ex]].

\bibitem{noiunokm}
  E.~Ciuffoli, J.~Evslin and H.~Li,
  ``The Reactor Anomaly after Daya Bay and RENO,''
  arXiv:1205.5499 [hep-ph].
 
\bibitem{huber2004}
  P.~Huber and T.~Schwetz,
  ``Precision spectroscopy with reactor anti-neutrinos,''
  Phys.\ Rev.\ D {\bf 70} (2004) 053011
  [hep-ph/0407026].
 
\bibitem{cartabianca}
  K.~N.~Abazajian, M.~A.~Acero, S.~K.~Agarwalla, A.~A.~Aguilar-Arevalo, C.~H.~Albright, S.~Antusch, C.~A.~Arguelles and A.~B.~Balantekin {\it et al.},
  ``Light Sterile Neutrinos: A White Paper,''
  arXiv:1204.5379 [hep-ph].

\bibitem{sezionedurto}
  P.~Vogel and J.~F.~Beacom,
  ``Angular distribution of neutron inverse beta decay, anti-neutrino(e) + p ---> e+ + n,''
  Phys.\ Rev.\ D {\bf 60} (1999) 053003
  [hep-ph/9903554].

 
\bibitem{gando}
  A.~Gando {\it et al.}  [KamLAND Collaboration],
  ``Constraints on $\theta_{13}$ from A Three-Flavor Oscillation Analysis of Reactor Antineutrinos at KamLAND,''
  Phys.\ Rev.\ D {\bf 83} (2011) 052002
  [arXiv:1009.4771 [hep-ex]].

\bibitem{minosneut2012}
R. Nichol,
``Final MINOS Results," presented at Neutrino 2012 in Kyoto.
Soon to be available at http://neu2012.kek.jp/neu2012/programme.html.


\bibitem{petcov2010}
  P.~Ghoshal and S.~T.~Petcov,
  ``Neutrino Mass Hierarchy Determination Using Reactor Antineutrinos,''
  JHEP {\bf 1103} (2011) 058
  [arXiv:1011.1646 [hep-ph]].



\end{thebibliography}

\begin{thebibliography}{23}\setlength{\itemsep}{-2.3mm}



\bibitem{dayabay}
  F.~P.~An {\it et al.}  [DAYA-BAY Collaboration],
  ``Observation of electron-antineutrino disappearance at Daya Bay,''
  Phys.\ Rev.\ Lett.\  {\bf 108} (2012) 171803
  [arXiv:1203.1669 [hep-ex]].

\bibitem{neut2012}
D. Dwyer,
``Daya Bay Results," presented at Neutrino 2012 in Kyoto.
Available at http://neu2012.kek.jp/neu2012/programme.html.


\bibitem{reno}
  J.~K.~Ahn {\it et al.}  [RENO Collaboration],
  ``Observation of Reactor Electron Antineutrino Disappearance in the RENO Experiment,''
  Phys.\ Rev.\ Lett.\  {\bf 108} (2012) 191802
  [arXiv:1204.0626 [hep-ex]].


\bibitem{petcovidea}
  S.~T.~Petcov and M.~Piai,
  ``The LMA MSW solution of the solar neutrino problem, inverted neutrino mass hierarchy and reactor neutrino experiments,''
  Phys.\ Lett.\ B {\bf 533} (2002) 94
  [hep-ph/0112074]. S.~Choubey, S.~T.~Petcov and M.~Piai,
  ``Precision neutrino oscillation physics with an intermediate baseline reactor neutrino experiment,''
  Phys.\ Rev.\ D {\bf 68} (2003) 113006
  [hep-ph/0306017].

\bibitem{caojunseminario}
J. Cao,
``Observation of $\overline{\nu}_e$ Disappearance at Daya Bay," presented at $\nu$Turn under Gran Sasso.
Available at http://agenda.infn.it/contributionListDisplay.py?confId=4722.

\bibitem{renonuturn}
``Observation of reactor neutrino disappearance at RENO," presented at $\nu$TURN 2012 under Gran Sasso.  Available at http://agenda.infn.it/contributionListDisplay.py?confId=4722.

\bibitem{yifangseminario}
Y. Wang,
``Daya Bay II: The Next Generation Reactor Neutrino Experiment" presented at NuFact in Williamsburg, Virginia.
Available at https://www.jlab.org/indico/conferenceTimeTable.py?confId=0\#20120725.detailed.

\bibitem{parke2007}
  H.~Minakata, H.~Nunokawa, S.~J.~Parke and R.~Zukanovich Funchal,
  ``Determination of the neutrino mass hierarchy via the phase of the disappearance oscillation probability with a monochromatic anti-electron-neutrino source,''
  Phys.\ Rev.\ D {\bf 76} (2007) 053004
   [Erratum-ibid.\ D {\bf 76} (2007) 079901]
  [hep-ph/0701151].

\bibitem{oggi}
  X.~Qian, D.~A.~Dwyer, R.~D.~McKeown, P.~Vogel, W.~Wang and C.~Zhang,
  ``Mass Hierarchy Resolution in Reactor Anti-neutrino Experiments: Parameter Degeneracies and Detector Energy Response,''
  PRD, 87, {\bf 033005} (2013)
  [arXiv:1208.1551 [physics.ins-det]].

\bibitem{noiteor}
  E.~Ciuffoli, J.~Evslin and X.~Zhang,
  ``The Neutrino Mass Hierarchy at Reactor Experiments now that theta13 is Large,''
  JHEP {\bf 1303} (2013) 016
  [arXiv:1208.1991 [hep-ex]].

\bibitem{hawaii}
 M.~Batygov, S.~Dye, J.~Learned, S.~Matsuno, S.~Pakvasa and G.~Varner,
   J.~Learned, S.~T.~Dye, S.~Pakvasa and R.~C.~Svoboda,
  ``Determination of neutrino mass hierarchy and $\theta_{13}$ with a remote detector of reactor antineutrinos,''
  Phys.\ Rev.\ D {\bf 78} (2008) 071302
  [hep-ex/0612022].
 ``Prospects of neutrino oscillation measurements in the detection of reactor antineutrinos with a medium-baseline experiment,''
  arXiv:0810.2580 [hep-ph].


\bibitem{caojun}
 L.~Zhan, Y.~Wang, J.~Cao and L.~Wen,
  ``Determination of the Neutrino Mass Hierarchy at an Intermediate Baseline,''
  Phys.\ Rev.\ D {\bf 78} (2008) 111103
  [arXiv:0807.3203 [hep-ex]].

\bibitem{caojun2}
  L.~Zhan, Y.~Wang, J.~Cao and L.~Wen,
  ``Experimental Requirements to Determine the Neutrino Mass Hierarchy Using Reactor Neutrinos,''
  Phys.\ Rev.\ D {\bf 79} (2009) 073007
  [arXiv:0901.2976 [hep-ex]].

\bibitem{minos}
R. Nichol,
``Final MINOS Results," presented at Neutrino 2012 in Kyoto.
Available at http://neu2012.kek.jp/neu2012/programme.html.
P.~Adamson {\it et al.}  [MINOS Collaboration],
  ``Measurements of atmospheric neutrinos and antineutrinos in the MINOS Far Detector,''
  arXiv:1208.2915 [hep-ex].

\bibitem{parke2005}
 H.~Nunokawa, S.~J.~Parke and R.~Zukanovich Funchal,
  ``Another possible way to determine the neutrino mass hierarchy,''
  Phys.\ Rev.\ D {\bf 72} (2005) 013009
  [hep-ph/0503283].

\bibitem{nuovoflusso}
T.~.A.~Mueller, D.~Lhuillier, M.~Fallot, A.~Letourneau, S.~Cormon, M.~Fechner, L.~Giot and T.~Lasserre {\it et al.},
  ``Improved Predictions of Reactor Antineutrino Spectra,''
  Phys.\ Rev.\ C {\bf 83} (2011) 054615
  [arXiv:1101.2663 [hep-ex]].

\bibitem{huber}
P.~Huber,
  ``On the determination of anti-neutrino spectra from nuclear reactors,''
  Phys.\ Rev.\ C {\bf 84} (2011) 024617
   [Erratum-ibid.\ C {\bf 85} (2012) 029901]
  [arXiv:1106.0687 [hep-ph]].

\bibitem{reattoreanom}
G.~Mention, M.~Fechner, T.~.Lasserre, T.~.A.~Mueller, D.~Lhuillier, M.~Cribier and A.~Letourneau,
  ``The Reactor Antineutrino Anomaly,''
  Phys.\ Rev.\ D {\bf 83} (2011) 073006
  [arXiv:1101.2755 [hep-ex]].

\bibitem{noiunokm}
  E.~Ciuffoli, J.~Evslin and H.~Li,
  ``The Reactor Anomaly after Daya Bay and RENO,''
  arXiv:1205.5499 [hep-ph].
 


\bibitem{sezionedurto}
  P.~Vogel and J.~F.~Beacom,
  ``Angular distribution of neutron inverse beta decay, anti-neutrino $e^-+p\rightarrow e^++n$,''
  Phys.\ Rev.\ D {\bf 60} (1999) 053003
  [hep-ph/9903554].

\bibitem{uflusso}
  K.~Schreckenbach, G.~Colvin, W.~Gelletly and F.~Von Feilitzsch,
  ``Determination Of The Anti-neutrino Spectrum From U-235 Thermal Neutron Fission Products Up To 9.5-mev,''
  Phys.\ Lett.\ B {\bf 160} (1985) 325.

\bibitem{pflusso}
  A.~A.~Hahn, K.~Schreckenbach, G.~Colvin, B.~Krusche, W.~Gelletly and F.~Von Feilitzsch,
  ``Anti-neutrino Spectra From Pu-241 And Pu-239 Thermal Neutron Fission Products,''
  Phys.\ Lett.\ B {\bf 218} (1989) 365.

\bibitem{quadflusso}
  P.~Vogel and J.~Engel,
  ``Neutrino Electromagnetic Form-Factors,''
  Phys.\ Rev.\ D {\bf 39} (1989) 3378.

\bibitem{noisim}
  E.~Ciuffoli, J.~Evslin and X.~Zhang,
  ``Mass Hierarchy Determination Using Neutrinos from Multiple Reactors,''
  JHEP {\bf 1212} (2012) 004
  [arXiv:1209.2227 [hep-ph]].

\end{thebibliography}

\begin{thebibliography}{23}\setlength{\itemsep}{-2.3mm}


\bibitem{caojun}
 L.~Zhan, Y.~Wang, J.~Cao and L.~Wen,
  ``Determination of the Neutrino Mass Hierarchy at an Intermediate Baseline,''
  Phys.\ Rev.\ D {\bf 78} (2008) 111103
  [arXiv:0807.3203 [hep-ex]].

\bibitem{caojun2}
  L.~Zhan, Y.~Wang, J.~Cao and L.~Wen,
  ``Experimental Requirements to Determine the Neutrino Mass Hierarchy Using Reactor Neutrinos,''
  Phys.\ Rev.\ D {\bf 79} (2009) 073007
  [arXiv:0901.2976 [hep-ex]].
 

\bibitem{dayabay}
  F.~P.~An {\it et al.}  [DAYA-BAY Collaboration],
  ``Observation of electron-antineutrino disappearance at Daya Bay,''
  Phys.\ Rev.\ Lett.\  {\bf 108} (2012) 171803
  [arXiv:1203.1669 [hep-ex]].






\bibitem{reno}
  J.~K.~Ahn {\it et al.}  [RENO Collaboration],
  ``Observation of Reactor Electron Antineutrino Disappearance in the RENO Experiment,''
  Phys.\ Rev.\ Lett.\  {\bf 108} (2012) 191802
  [arXiv:1204.0626 [hep-ex]].


\bibitem{doublechooz}
Y.~Abe {\it et al.}  [DOUBLE-Chooz Collaboration],
  ``Indication for the disappearance of reactor electron antineutrinos in the Double Chooz experiment,''
  Phys.\ Rev.\ Lett.\  {\bf 108} (2012) 131801
  [arXiv:1112.6353 [hep-ex]].


\bibitem{neut2012}
D. Dwyer,
``Daya Bay Results," presented at Neutrino 2012 in Kyoto.
Available at http://neu2012.kek.jp/neu2012/programme.html.

\bibitem{petcovidea}
  S.~T.~Petcov and M.~Piai,
  ``The LMA MSW solution of the solar neutrino problem, inverted neutrino mass hierarchy and reactor neutrino experiments,''
  Phys.\ Lett.\ B {\bf 533} (2002) 94
  [hep-ph/0112074]. S.~Choubey, S.~T.~Petcov and M.~Piai,
  ``Precision neutrino oscillation physics with an intermediate baseline reactor neutrino experiment,''
  Phys.\ Rev.\ D {\bf 68} (2003) 113006
  [hep-ph/0306017].

\bibitem{parke2007}
  H.~Minakata, H.~Nunokawa, S.~J.~Parke and R.~Zukanovich Funchal,
  ``Determination of the neutrino mass hierarchy via the phase of the disappearance oscillation probability with a monochromatic anti-electron-neutrino source,''
  Phys.\ Rev.\ D {\bf 76} (2007) 053004
   [Erratum-ibid.\ D {\bf 76} (2007) 079901]
  [hep-ph/0701151].

\bibitem{sezionedurto}
  P.~Vogel and J.~F.~Beacom,
  ``Angular distribution of neutron inverse beta decay, anti-neutrino(e) + p ---> e+ + n,''
  Phys.\ Rev.\ D {\bf 60} (1999) 053003
  [hep-ph/9903554].

\bibitem{caojunseminario}
J. Cao,
``Observation of $\overline{\nu}_e$ Disappearance at Daya Bay," presented at $\nu$Turn under Gran Sasso.
Available at http://agenda.infn.it/contributionListDisplay.py?confId=4722.

\bibitem{parke2005}
 H.~Nunokawa, S.~J.~Parke and R.~Zukanovich Funchal,
  ``Another possible way to determine the neutrino mass hierarchy,''
  Phys.\ Rev.\ D {\bf 72} (2005) 013009
  [hep-ph/0503283].

\bibitem{noi}
  E.~Ciuffoli, J.~Evslin and X.~Zhang,
  ``The Neutrino Mass Hierarchy at Reactor Experiments now that theta13 is Large,''
  arXiv:1208.1991 [hep-ex].

\bibitem{petcov2010}
  P.~Ghoshal and S.~T.~Petcov,
  ``Neutrino Mass Hierarchy Determination Using Reactor Antineutrinos,''
  JHEP {\bf 1103} (2011) 058
  [arXiv:1011.1646 [hep-ph]].

\bibitem{oggi}
X. Qian,  D. A. Dwyer,  R. D. McKeown, P. Vogel, W. Wang, and C. Zhang,
``Mass Hierarchy Resolution in Reactor Anti-neutrino Experiments: Parameter Degeneracies and Detector Energy Response,"
arXiv:1208.1551 [hep-ex].


\bibitem{hawaii}
 M.~Batygov, S.~Dye, J.~Learned, S.~Matsuno, S.~Pakvasa and G.~Varner,
   J.~Learned, S.~T.~Dye, S.~Pakvasa and R.~C.~Svoboda,
  ``Determination of neutrino mass hierarchy and $\theta_{13}$ with a remote detector of reactor antineutrinos,''
  Phys.\ Rev.\ D {\bf 78} (2008) 071302
  [hep-ex/0612022].
 ``Prospects of neutrino oscillation measurements in the detection of reactor antineutrinos with a medium-baseline experiment,''
  arXiv:0810.2580 [hep-ph].

\bibitem{reattoreanom}
G.~Mention, M.~Fechner, T.~.Lasserre, T.~.A.~Mueller, D.~Lhuillier, M.~Cribier and A.~Letourneau,
  ``The Reactor Antineutrino Anomaly,''
  Phys.\ Rev.\ D {\bf 83} (2011) 073006
  [arXiv:1101.2755 [hep-ex]].

\bibitem{huber2011}
  P.~Huber,
  ``On the determination of anti-neutrino spectra from nuclear reactors,''
  Phys.\ Rev.\ C {\bf 84} (2011) 024617
   [Erratum-ibid.\ C {\bf 85} (2012) 029901]
  [arXiv:1106.0687 [hep-ph]].

\bibitem{renofrancia}
  T.~Lasserre, G.~Mention, M.~Cribier, A.~Collin, V.~Durand, V.~Fischer, J.~Gaffiot and D.~Lhuillier {\it et al.},
  ``Comment on Phys. Rev. Lett. 108, 191802 (2012): 'Observation of Reactor Electron Antineutrino Disappearance in the RENO Experiment,''
  arXiv:1205.5626 [hep-ex].

\bibitem{weihai}
M. Weng,
``Observation of Electron Antineutrino Disappearance at Daya Bay,''
presented at ICFPC 2012 in Weihai.

\bibitem{yifangseminario}
Y. Wang,
``Daya Bay II: The Next Generation Reactor Neutrino Experiment" presented at NuFact in Williamsburg, Virginia.
Available at https://www.jlab.org/indico/conferenceTimeTable.py?confId=0\#20120725.detailed.




\end{thebibliography}
\end{document}